\documentclass[fleqn,usenatbib]{mnras}

\usepackage{newtxtext,newtxmath}

\usepackage[T1]{fontenc}
\usepackage{ae,aecompl}

\usepackage{graphicx}	
\usepackage{amsmath}	
\usepackage{mathtools}

\usepackage{etoolbox} 

\usepackage{physics}
\usepackage{siunitx}
\DeclareSIUnit \h {\ensuremath{\mathit{h}}}
\DeclareSIUnit \parsec {pc}
\DeclareSIUnit \deg {deg}
\DeclareSIUnit \solarmass {\ensuremath{\mathit{M}_{\odot}}}
\usepackage{bm}
\usepackage[normalem]{ulem}
\usepackage[capitalise]{cleveref}
\usepackage{multirow}
\usepackage[dvipsnames]{xcolor}
\usepackage{xspace}
\usepackage{booktabs}
\usepackage{verbatim}
\usepackage{tikz}

\newcommand{\cd}[1]{{#1}\xspace}

\newcommand{\treecorr}{\textsc{TreeCorr}\xspace}
\newcommand{\namaster}{\textsc{NaMaster}\xspace}
\newcommand{\multinest}{\textsc{MultiNest}\xspace}
\newcommand{\halofit}{\textsc{Halofit}\xspace}
\newcommand{\cosmosis}{\textsc{CosmoSIS}\xspace}
\newcommand{\numcosmo}{\textsc{NumCosmo}\xspace}
\newcommand{\healpix}{\textsc{HEALPix}\xspace}
\newcommand{\healpy}{\textsc{healpy}\xspace}

\newcommand{\Cl}{{C_{\ell}}}
\newcommand{\xip}{{\xi_{+}}}
\newcommand{\xim}{{\xi_{-}}}
\newcommand{\xipm}{{\xi_{\pm}}}
\newcommand{\ellmin}{{\ell_{\rm min}}}
\newcommand{\ellmax}{{\ell_{\rm max}}}
\newcommand{\tmin}{{\theta_{\rm min}}}
\newcommand{\tmax}{{\theta_{\rm max}}}
\newcommand{\kmax}{{k_{\rm max}}}
\newcommand{\nside}{{N_{\rm side}}}
\newcommand{\fsky}{{f_{\rm sky}}}
\DeclareMathOperator*{\cov}{Cov}

\newcommand{\Om}{{\Omega_{\rm m}}}
\newcommand{\sqrtOm}{\sqrt{\Omega_{\rm m}}}
\newcommand{\Ob}{{\Omega_{\rm b}}}
\newcommand{\Onu}{{\Omega_{\nu}}}

\newcommand{\ns}{{n_{\rm s}}}
\newcommand{\Aia}{{A_{\rm IA}}}
\newcommand{\aia}{{\alpha_{\rm IA}}}

\newcommand{\SeightCl}{\widehat{S_8}\vert_{\Cl}}
\newcommand{\Seightxi}{\widehat{S_8}\vert_{\xipm}}

\newcommand{\eg}{\emph{e.g.}\xspace}
\newcommand{\ie}{\emph{i.e.}\xspace}
\newcommand{\vs}{\emph{vs}\xspace}

\title[Cosmic shear in harmonic \vs real space]{Consistency of cosmic shear analyses in harmonic and real space}

\author[Doux, Chang, Jain et al.]{
\parbox{\textwidth}{
\Large
C.~Doux,$^{1\star}$
C.~Chang,$^{2,3\dag}$
B.~Jain,$^{1}$
J.~Blazek,$^{4,5}$
H.~Camacho,$^{6,7}$
X.~Fang,$^{8}$
M.~Gatti,$^{9}$
E.~Krause,$^{8}$
N.~MacCrann,$^{10}$
S.~Samuroff,$^{11}$
L.~F.~Secco,$^{1}$
M.~A.~Troxel,$^{12}$
J.~Zuntz,$^{13}$
M.~Aguena,$^{14,7}$
S.~Allam,$^{15}$
A.~Amon,$^{16}$
S.~Avila,$^{17}$
D.~Bacon,$^{18}$
E.~Bertin,$^{19,20}$
D.~Brooks,$^{21}$
D.~L.~Burke,$^{16,22}$
A.~Carnero~Rosell,$^{23,7,24}$
M.~Carrasco~Kind,$^{25,26}$
J.~Carretero,$^{9}$
A.~Choi,$^{4}$
M.~Costanzi,$^{27,28,29}$
M.~Crocce,$^{30,31}$
L.~N.~da Costa,$^{7,32}$
M.~E.~S.~Pereira,$^{33}$
T.~M.~Davis,$^{34}$
J.~P.~Dietrich,$^{35}$
P.~Doel,$^{21}$
I.~Ferrero,$^{36}$
A.~Fert\'e,$^{37}$
P.~Fosalba,$^{30,31}$
J.~Garc\'ia-Bellido,$^{17}$
E.~Gaztanaga,$^{30,31}$
D.~W.~Gerdes,$^{38,33}$
D.~Gruen,$^{39,16,22}$
R.~A.~Gruendl,$^{25,26}$
J.~Gschwend,$^{7,32}$
G.~Gutierrez,$^{15}$
W.~G.~Hartley,$^{40}$
S.~R.~Hinton,$^{34}$
D.~L.~Hollowood,$^{41}$
D.~Huterer,$^{33}$
D.~J.~James,$^{42}$
K.~Kuehn,$^{43,44}$
N.~Kuropatkin,$^{15}$
M.~A.~G.~Maia,$^{7,32}$
J.~L.~Marshall,$^{45}$
F.~Menanteau,$^{25,26}$
R.~Miquel,$^{46,9}$
R.~Morgan,$^{47}$
A.~Palmese,$^{15,3}$
F.~Paz-Chinch\'{o}n,$^{25,48}$
A.~A.~Plazas,$^{49}$
A.~Roodman,$^{16,22}$
E.~Sanchez,$^{50}$
M.~Schubnell,$^{33}$
S.~Serrano,$^{30,31}$
I.~Sevilla-Noarbe,$^{50}$
M.~Smith,$^{51}$
M.~Soares-Santos,$^{33}$
E.~Suchyta,$^{52}$
G.~Tarle,$^{33}$
C.~To,$^{39,16,22}$
T.~N.~Varga,$^{53,54}$
J.~Weller,$^{53,54}$
and R.D.~Wilkinson$^{55}$
\begin{center} (DES Collaboration) \end{center}
}
}

\date{Accepted XXX. Received YYY; in original form ZZZ}

\pubyear{2015}

\usepackage{eso-pic}

\AddToShipoutPictureBG*{%
  \AtPageUpperLeft{%
    \hspace{0.75\paperwidth}%
    \raisebox{-3.5\baselineskip}{%
      \makebox[0pt][l]{\textnormal{DES-2020-0564}}
}}}%

\AddToShipoutPictureBG*{%
  \AtPageUpperLeft{%
    \hspace{0.75\paperwidth}%
    \raisebox{-4.5\baselineskip}{%
      \makebox[0pt][l]{\textnormal{FERMILAB-PUB-20-550-AE}}
}}}%

\begin{document}
\label{firstpage}
\pagerange{\pageref{firstpage}--\pageref{lastpage}}
\maketitle

\begin{abstract}
Recent cosmic shear studies have reported  discrepancies of up to $1\sigma$ on the parameter ${S_{8}=\sigma_{8}\sqrt{\Om/0.3}}$ between the analysis of shear power spectra and two-point correlation functions, derived from the same shear catalogs. It is not \textit{a priori} clear whether the measured discrepancies are consistent with statistical fluctuations. In this paper, we investigate this issue in the context of the forthcoming analyses from the third year data of the Dark Energy Survey (DES-Y3). We analyze DES-Y3 mock catalogs from Gaussian simulations with a fast and accurate importance sampling pipeline. We show that the methodology for determining matching scale cuts in harmonic and real space is the key factor that contributes to the scatter between constraints derived from the two statistics. We compare the published scales cuts of the KiDS, Subaru-HSC and DES surveys, and find that the correlation coefficients of posterior means range from over 80\% for our proposed cuts, down to 10\% for cuts used in the literature.  We then study the interaction between scale cuts and systematic uncertainties arising from multiple sources: non-linear power spectrum, baryonic feedback, intrinsic alignments, uncertainties in the point-spread function, and redshift distributions. We find that, given DES-Y3 characteristics and proposed cuts, these uncertainties affect the two statistics similarly; the differential biases are below a third of the statistical uncertainty, with the largest biases arising from intrinsic alignment and baryonic feedback. While this work is aimed at DES-Y3, the tools developed can be applied to Stage-IV surveys where statistical errors will be much smaller.
\end{abstract}



\makeatletter
\def \blfootnote{\xdef\@thefnmark{}\@footnotetext}
\makeatother

\blfootnote{$^{\star}$ E-mail: cdoux@sas.upenn.edu}
\blfootnote{$^{\dag}$ E-mail: chihway@kicp.uchicago.edu}


\section{Introduction}
\label{sec:intro}

Weak gravitational lensing, the apparent distortion of galaxy shapes due to the intervening dark matter distribution in the line-of-sight, is one of the most powerful tools for constraining cosmological parameters at low redshift \citep{2006astro.ph..9591A}. In particular, the derived parameter $S_{8}=\sigma_{8} \sqrt{\Omega_{m}/0.3}$ points roughly to the most constraining direction of weak lensing datasets \citep{1997ApJ...484..560J}, where $\sigma_{8}$ is the amplitude of structure in the Universe, parametrized as the standard deviation of the linear overdensity fluctuations in \SI{8}{\per\h\mega\parsec} spheres at present time, and $\Omega_{\rm m}$ is the density parameter of matter at present time. The primary cosmological constraints from weak lensing analyses come from two-point statistics of weak lensing distortion, referred to as cosmic shear, which may be evaluated in real (or configuration), harmonic or other spaces. The most commonly used two-point statistics in real space are the shear two-point angular correlation functions $\xipm(\theta)$ \citep{ 2017MNRAS.465.1454H,2018PhRvD..98d3528T,2020PASJ...72...16H}, whereas in harmonic space, the shear power spectrum $\Cl$ of lensing E-modes \citep[see e.g.][and \cref{sec:theory} for a review of formalism]{2001PhR...340..291B} is often used \citep{2017MNRAS.471.4412K,2019PASJ...71...43H}. Both present advantages and drawbacks. $\xipm$ measurements are insensitive to the survey geometry and straightforward to compute \citep{2004MNRAS.352..338J}, but they are heavily correlated across scales (\ie the covariance has large off-diagonal contributions). $\Cl$ measurements, on the other hand, are typically based on a pixelization of the shear field estimated with galaxy shapes. They must be deconvolved from the survey mask \citep{2019MNRAS.484.4127A} and have their shape-noise contributions subtracted, but are almost uncorrelated across multipoles for large observed areas of the sky \cd{(alternatively, one may choose to forward model the effect of the mask)}. Fast theoretical predictions are available for both, a requirement to perform Bayesian analysis.

Even though different statistics compress the information in different ways, one may expect both statistics---$\xipm$ and $\Cl$---to return reasonably close constraints on cosmological parameters when applied to the same data set. However,
both the Kilo-Degree Survey \citep[KiDS,][]{2013Msngr.154...44D,2015MNRAS.454.3500K} and the Hyper Suprime-Cam survey \citep[HSC,][]{2018PASJ...70S...4A,2018PASJ...70S...8A} have released cosmological results of the same weak lensing data using both real and harmonic statistics,
and found discrepancies at the level of 0.5 to 1.5$\sigma$.
In \citet{2017MNRAS.465.1454H}, the authors derived the constraint $S_{8}=0.745 \pm 0.039$ based on cosmic shear measured with $\xipm$ in the 450 deg$^{2}$ KiDS dataset, while \citet{2017MNRAS.471.4412K} carried out an analysis using instead the $\Cl$ statistic and obtained $S_{8}=0.651 \pm 0.058$. Similarly, with the first year of about $\SI{137}{deg\squared}$ of HSC data, \citet{2020PASJ...72...16H} found $S_{8}=0.804^{+0.032}_{-0.029}$ using $\xipm$, while \citet{2019PASJ...71...43H} found $S_{8}=0.780^{+0.030}_{-0.033}$ based on $\Cl$.
These discrepancies are concerning and make the comparison of weak lensing with other types of cosmological data ambiguous.
Therefore, in light of the observed tensions between the value of $\sigma_8$ (or $S_8$) inferred from the cosmic microwave background and from large-scale structure at low redshift \citep[e.g.][]{2015MNRAS.451.2877M,2017MNRAS.465.1454H,2018PhRvD..98d3526A}, it is crucial
to understand the limitations and consistency of standard statistics such as $\xipm$ and $\Cl$ used in weak lensing.
This is important for ongoing weak lensing surveys, such as the Dark Energy Survey \citep[DES,][]{2005IJMPA..20.3121F}, the Kilo-Degree Survey \citep[KiDS,][]{2013Msngr.154...44D,2015MNRAS.454.3500K}, the Hyper Suprime-Cam survey \citep[HSC,][]{2018PASJ...70S...4A,2018PASJ...70S...8A}, and even more so for upcoming ones, such as the Vera Rubin Observatory Legacy Survey of Space and Time \citep[LSST,][]{2019ApJ...873..111I}, the ESA satellite Euclid \citep{2012SPIE.8442E..0TL} and the NASA's Nancy Grace Roman Space Telescope \citep{2019arXiv190205569A}.

In this paper we will show that scale cuts play a significant role in terms of the consistency between the two statistics.
To illustrate this, we imagine, as a thought experiment, that we know the true shear field. If we were able to measure two-point statistics with infinite resolution---that is, for arbitrarily large multipoles $\ell$, and arbitrarily small separation angles $\theta$ measured in infinitesimal bins---and provided we can fully characterize the likelihood of those estimators, both would capture the full {\it Gaussian} information. Therefore, we would expect identical, cosmic variance-limited posteriors on $S_8$ and other parameters, and equal estimators, denoted generically $\Seightxi$ and $\SeightCl$---\eg, from the mean or mode of the respective posteriors. Now, if we consider a catalog of galaxy shapes sampled from the shear field, this step adds noise to our measurements, but the results should also perfectly agree between the $\xipm$ and $\Cl$ measurements. That is, we still expect ${\Seightxi=\SeightCl}$.

\begin{figure}
    \centering
    \includegraphics[scale=0.6]{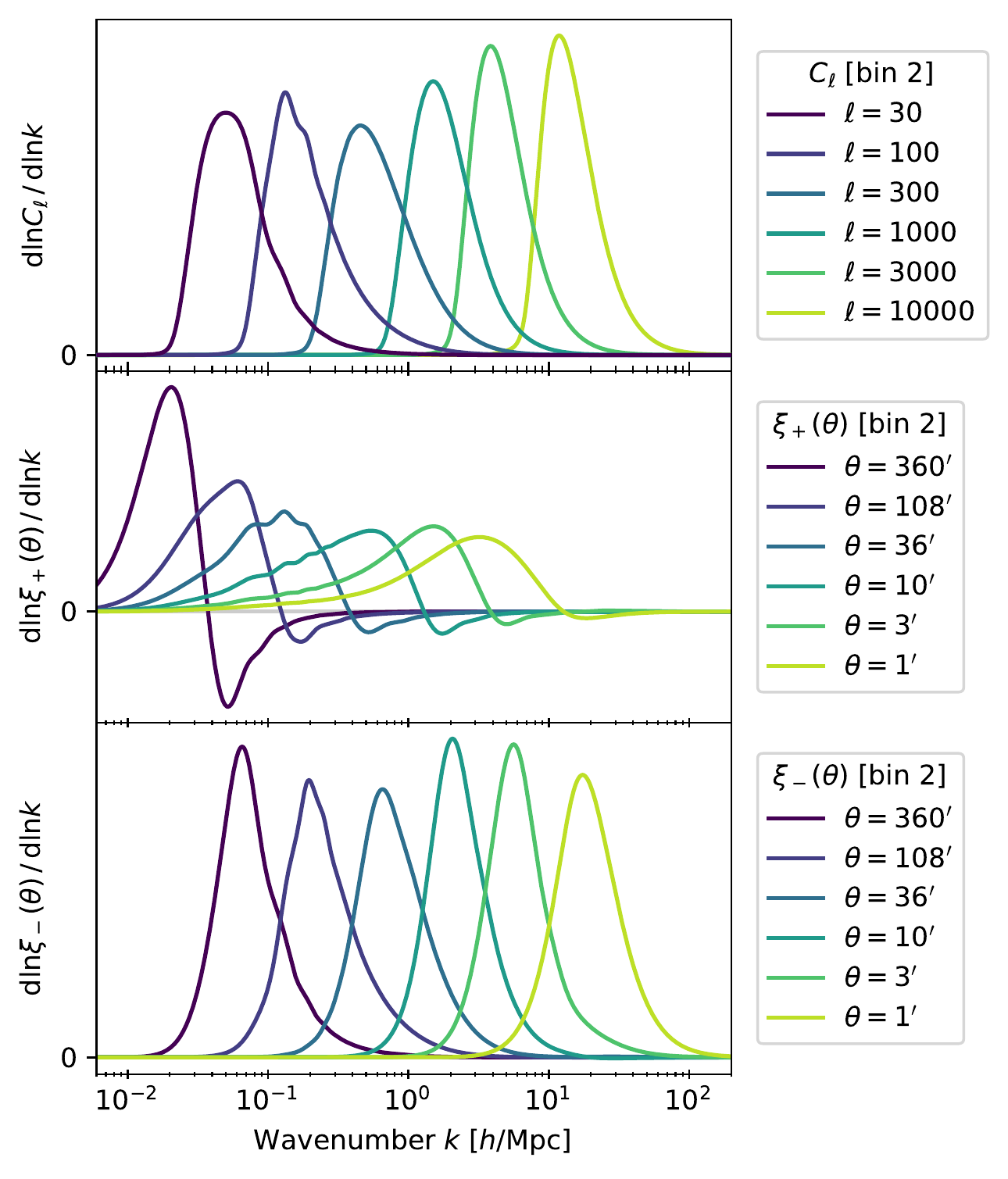}
    \caption{Contributions of physical $k$-modes to the shear power spectrum $\Cl$ and two-point functions $\xipm(\theta)$, for different multipoles $\ell$ and angular separations $\theta$ in the ranges used in cosmic shear analyses. These curves correspond to the integrand of the Limber formula, \cref{eq:Cl}, with a change of variable ${z \rightarrow k=\flatfrac{\qty(\ell+1/2)}{\chi(z)}}$.
    We show $\dv*{\ln X}{\ln k}$ for ${X=\Cl,\xipm(\theta)}$ on a logarithmic scale in $k$, such that each curve is normalized to have area unity under the curve. Here, the auto-correlation for redshift bin 2 is shown (see redshift distributions and the broad lensing efficiency functions in \cref{fig:dndz_footprint}).
    }
    \label{fig:Cl_xi_vs_lnk}
\end{figure}

In practice, however, neither of the previous scenarios are realistic. First, the finite survey area and the density of observed (and selected) galaxies introduce respectively large- and small-scale cut-offs. Furthermore, cosmic shear analyses are inherently limited by theoretical uncertainties (\eg, baryonic effects, intrinsic alignments) and observational effects that restrict the use of two-point statistic measurements at small scales---though accessible in the data---to derive constraints on cosmological parameters.
The most straightforward solution is to decide on hard cuts, \ie using only
real-space angular bins between certain $\tmin$ and $\tmax$ for $\xipm$, and, similarly, only multipoles between $\ellmin$ and  $\ellmax$ for $\Cl$. 
The $\theta$ and $\ell$ cuts cannot, however, be directly translated because both statistics are related through a Bessel integral, see \cref{eq:xipm} below, such that a hard scale cut introduced in real space induces an {\it oscillatory} cut in harmonic space, and vice versa \citep[see, \eg,][]{2002PhRvD..65f3001H}.
Therefore, realistic cosmic shear analyses must exclude different information for each statistic and we do not expect $\Seightxi$ and $\SeightCl$ to be equal because of the necessary scale cuts.
In addition, cosmic shear two-point statistics at a fixed angular scale receive contributions from a wide range of physical scales---or Fourier modes~$k$---stemming from the projection of the shear field along the line of sight that sums up distortions sourced by the matter density field from the sources to the observer.
This is illustrated in \cref{fig:Cl_xi_vs_lnk}, where
the three panels show $\dv*{\ln X}{\ln k}$ for $X=\xipm(\theta),\Cl$, \ie the normalized integrand of $\xipm$ and $\Cl$ statistics as a function of scales $k$ (for details, see \cref{sec:theory}).
Due to this mixing of $k$-modes, and for a given choice of scale cuts, the estimators $\Seightxi$ and $\SeightCl$ from cosmic shear analyses are placing different weights on accessible $k$-modes.
The question of consistency thus pertains, to a large extent, to the corresponding choice of scale cuts in harmonic and real space.

The problem was recently studied, in particular by the HSC and KiDS collaborations.
\citet{2020PASJ...72...16H}, investigated the origin of the difference in their cosmological constraints from $\xipm$ and $\Cl$, noting that for some parameters, such as $\Omega_{\rm m}$, the difference could be significant. To understand the problem, the authors performed the same real and harmonic-space analyses on 100 $N$-body simulations and found significant scatter between the constrains from $\xipm$ and $\Cl$ \citep[see Fig. 19 of][]{2020PASJ...72...16H}. They concluded that one potential explanation of the difference in the constraints from \citet{2020PASJ...72...16H} and \citet{2019PASJ...71...43H} is that the scales used in the two analyses do not match, such that the observed discrepancy could be explained by a statistical fluctuation at the $\sim1.4\sigma$ level, based on discrepancies found in simulations.
Recently, KiDS-1000 proposed and compared different statistics \citep{2020arXiv200715633A}, in addition to the real-space two-point functions, namely band powers and COSEBIs \citep[Complete Orthogonal Sets of E-/B-mode Integrals, see][]{2010A&A...520A.116S}. The latter are based on linear combinations of thinly-binned measurements of $\xipm$ weighted with specific window functions that amount to apply effective soft cuts. They similarly analyzed a number of simulations to quantify the expected differences. Finally, \citet{2019MNRAS.490.5033L} investigated a similar discrepancy found in cosmic shear data from the Canada-France-Hawaii Telescope Lensing Survey  \citep{2013MNRAS.430.2200K,2015PhRvD..91f3507L} and identified excess power at very small scales ($\ell\gtrsim5000$) to drive the larger value of the matter fluctuation amplitude inferred from the power spectrum.

In this paper, our goal is to
examine conditions for consistency of the DES Y3 cosmic shear analysis in harmonic space and that performed in real space.
For a given data set and choices of scales used in each analysis,
the standard deviation of ${\Delta\widehat{S_8}\equiv\Seightxi-\SeightCl}$ over a large number of realizations of the shear field, denoted $\sigma(\Delta\widehat{S_8})$, is fixed. The more common information the two statistics probe, the smaller $\sigma(\Delta\widehat{S_8})$ is, and vice versa.
We will measure the distribution of $(\Seightxi,\SeightCl)$ and $\sigma(\Delta\widehat{S_8})$ from simulations of mock DES Y3 surveys, under different assumptions of noise and scale cuts. 
These simulations are required to accurately estimate ${\Delta\widehat{S_8}}$. They allow us to include the effects of cosmic variance, shape noise, and any effects coming from the particular geometry of the DES footprint.
In addition, we will compare the estimated scatter of ${\Delta\widehat{S_8}}$ to the biases in ${S_8}$ predicted for a set of systematic effects and modeling uncertainties.
Indeed, observational systematics and theoretical uncertainties in modeling are liable to impact $\Cl$ and $\xipm$ measurements differently depending on whether the effect has more compact support in real or harmonic space, or neither (\eg for a $k$-dependent effect). Therefore, it is useful to measure the differential impact of systematics to the spread expected from statistical fluctuations.
In order to analyze hundreds of simulations for several choices of scale cuts, we implement (and validate) an importance sampling pipeline that provides fast estimates of $\widehat{S_8}$. Finally, we will also evaluate the equivalent quantities for $\sigma_8$ and $\Om$.

This exercise is meant to provide guidance to determine scale cuts for the DES Y3 cosmic shear analysis in harmonic space.
In the context of multiprobe analysis \citep{2018MNRAS.474.4894J,2018MNRAS.476.4662V,2018PhRvD..98d3526A,2020arXiv200715632H}, where one analyzes cosmic shear measurements in conjunction with galaxy clustering and galaxy-galaxy lensing measurements---\ie the cross-correlation of galaxy positions and shapes---we restrict our analysis to cosmic shear. The reason is that one expects less $k$-mode mixing for the other two probes (at least for $\Cl$) since the integral over redshift (or $k$) has limited support, given by the width of the redshift distributions of clustering galaxy samples.

The paper is structured as follows. In \Cref{sec:model}, we provide a brief overview of the theoretical background and we describe the baseline model used in this analysis, as well as alternative modeling choices and systematic effects for which we will examine the differential impact on harmonic \vs real space. In \Cref{sec:scales} we describe several strategies in choosing scale cuts in real and harmonic space. In \Cref{sec:sims} we describe the generation and validation of the simulations used in this work. In \Cref{sec:is} we present and validate the use of importance sampling to obtain fast estimators $\widehat{S_8}$ in both harmonic and real space. We present our estimation of ${\Delta\widehat{S_8}}$ in \Cref{sec:results}, estimate $\widehat{S_8}_{\xipm}-\widehat{S_8}_{\Cl}$ for various systematics and discuss discrepancies found in the literature. Finally, we summarize our findings in \Cref{sec:summary} and discuss the implications for the forthcoming DES Y3 analysis and future surveys.

\section{Formalism and modeling}
\label{sec:model}

\begin{figure*}
    \centering
    \includegraphics[scale=0.6]{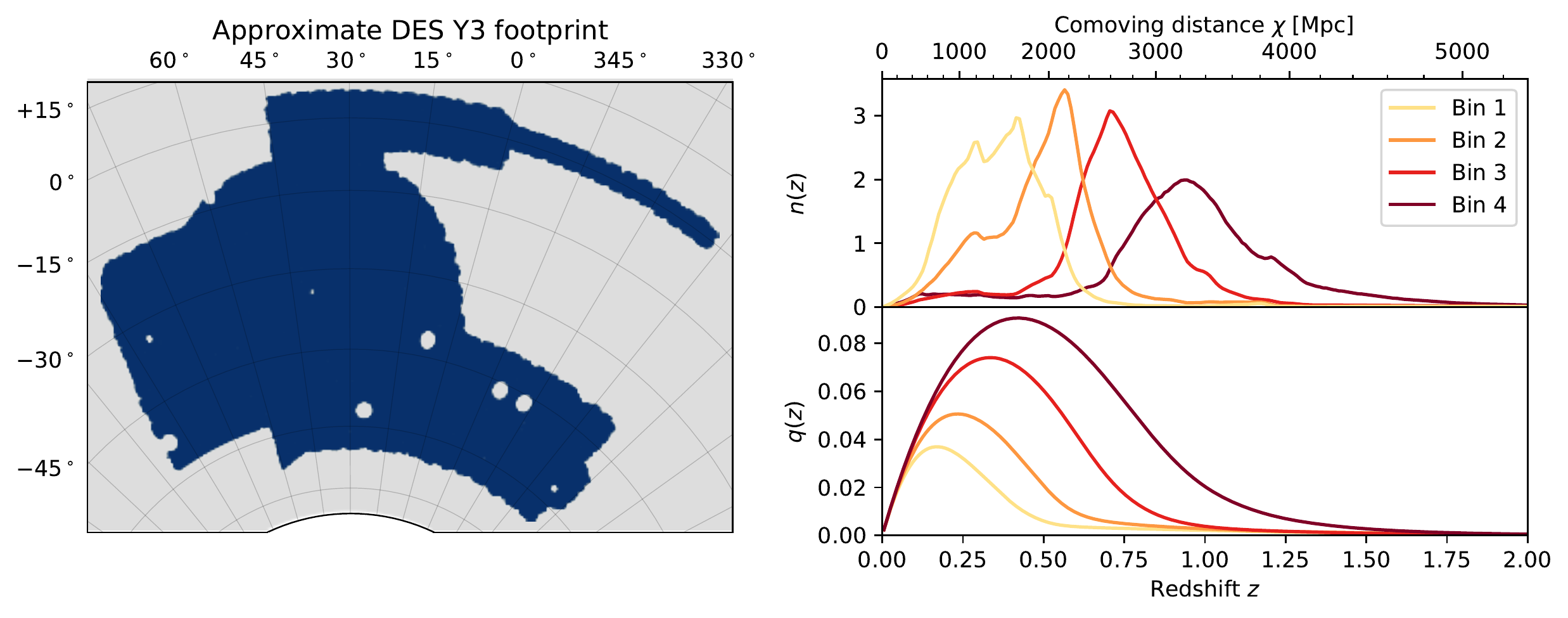}
    \caption{\textit{Left}: DES Y3 survey footprint used in this work. \textit{Right}: Normalized redshift distributions $n(z)$ from DES Y1 (top) and corresponding lensing efficiency functions $q(z)$ at the fiducial cosmology (bottom).}
    \label{fig:dndz_footprint}
\end{figure*}

Our modeling of $\xipm$ and $\Cl$ follows closely the framework used in \citet{2017arXiv170609359K} and the cosmological analysis of the first year (Y1) data of DES \citep{2018PhRvD..98d3528T,2018PhRvD..98d3526A}. For DES Y3, there were several improvements to this pipeline though, for the purpose of this study, the Y1 pipeline is sufficient as our simulations do not contain the higher-order corrections. We will adopt the approximate Y3 footprint with Y1 redshift distributions, shown in \cref{fig:dndz_footprint}.

\subsection{Theory}
\label{sec:theory}

Here we provide the basic theoretical framework associated with the harmonic- and real-space cosmic shear two-point statistics, the power spectrum $\Cl$ and two-point correlation functions $\xipm$.

Under the Limber approximation \citep{1953ApJ...117..134L,1992ApJ...388..272K,1998ApJ...498...26K,2008PhRvD..78l3506L} and in a spatially flat Universe,
the lensing power spectrum encodes cosmological information through 
\begin{equation}
C^{ij}_\ell = \int_{0}^{\chi_{\rm H}} \dd{\chi} \frac{q^{i}(\chi) q^{j}(\chi)}{\chi^2} P_{\rm NL}\qty( k=\frac{\ell + 1/2}{\chi}, \chi ),
\label{eq:Cl}
\end{equation}
where $\chi$ is the radial comoving distance, $\chi_{\rm H}$ is the distance to the horizon, $P_{\rm NL}$ is the non-linear matter power spectrum, and $q(\chi)$ is the lensing efficiency defined via
\begin{equation}
q^{i}(\chi) = \frac{3}{2} \Omega_{\rm m} \left( \frac{H_{0}}{c}\right)^{2} \frac{\chi}{a(\chi)} \int_{\chi}^{\chi_{\rm H}} \dd{\chi '} n^{i}(\chi') \frac{\chi' - \chi}{\chi'},
\label{eq:lensing_efficiency}
\end{equation}
where $\Omega_{\rm m}$ is the matter density today, $H_{0}$ is the Hubble parameter today, $a$ is the scale factor, and $n^{i}(\chi)$ is the normalized redshift distribution of the galaxy sample $i$.

Assuming the flat-sky approximation \citep{1992ApJ...388..272K,1998ApJ...498...26K}, $\xipm$ and $\Cl$ are connected via
\begin{equation}
\xi^{ij}_{\pm}(\theta) = \int_0^\infty \frac{\ell \dd{\ell}}{2\pi} J_{0/4}(\theta \ell) \, C^{ij}(\ell),
\label{eq:xipm}
\end{equation}
where $J_{n}$ is the $n$th-order spherical Bessel function of the first kind, with $n=0$ ($n=4$) for $\xip$ ($\xim$), and $C(\ell)$ is an interpolation of $\Cl$ for non-integer $\ell$ \citep[see][for discussions of this approximation]{2017MNRAS.469.2737K,2017JCAP...05..014L}.

Both quantities receive contributions from the matter power spectrum over a range of physical $k$-modes. By applying the change of variables ${k=\flatfrac{\qty(\ell+1/2)}{\chi(z)}}$ in \cref{eq:Cl} and using \cref{eq:xipm}, we can write both $\Cl$ and $\xipm$ as integrals over ${\ln k}$. We show the corresponding (normalized) integrands in \cref{fig:Cl_xi_vs_lnk} for different values of $\ell$ and $\theta$. Information from different physical $k$-modes in $P_{\rm NL}$ contribute to a given $\ell$ in $\Cl$, and since $\xipm$ is a Fourier transform of $\Cl$, information in different $k$-modes get further redistributed into different $\theta$ scales. 

\subsection{Baseline model}

Following \citet{2017arXiv170609359K}, we compute the non-linear power spectrum $P_{\rm NL}$ using the Boltzmann code CAMB \citep{2000ApJ...538..473L,2012JCAP...04..027H} with the \halofit extension to non-linear scales \citep{2003MNRAS.341.1311S} with updates from \citet{2012ApJ...761..152T}.
Later in \cref{sec:sys}, we investigate the effect of alternative prescription for the non-linear matter power spectrum from either an emulator \citep{2017ApJ...847...50L} or the introduction of baryonic effects based on hydrodynamical simulations \citep{2010MNRAS.402.1536S}. Consistent with the DES Y3 analysis, we vary six parameters of the $\Lambda$CDM model, namely the total matter density parameter $\Om$, the amplitude of structure $\sigma_{8}$, the baryon density parameter $\Ob$, the Hubble parameter $h$ (where ${H_0=\SI{100}{\h \kilo\meter\per\second\per\mega\parsec}}$), the spectral index of the primordial curvature power spectrum $\ns$ and the neutrino physical density parameter $\Onu h^2$. Throughout this paper we assume the Planck 2018 \citep{2020A&A...641A...6P} best-fit cosmology derived from TT,TE,EE+lowE+lensing+BAO data. The list of parameters, their definition, fiducial values used as inputs for simulations (see \cref{sec:sims}) and priors are shown in \cref{tab:params}.

In addition, our baseline model includes a number of observational and astronomical systematic effects, parametrized by nuisance parameters that we marginalize over.
\begin{itemize}
    \item \textit{Shear calibration bias.} To account for uncertainties in shear calibration, we model the observed shear $\gamma_{\rm obs}^i$ in redshift bin $i$ from the true shear $\gamma^i$ by
    \begin{equation}
    \gamma_{\rm obs}^i=(1+m_{i})\gamma^i,
    \end{equation}
    where $m_{i}$ is the multiplicative shear bias, constant within each redshift bin $i$ \citep{2006MNRAS.366..101H,2006MNRAS.368.1323H}. These biases act as an overall rescaling for each redshift bin pair, such that theoretical predictions are scaled as
    \begin{flalign}
        C_\ell^{ij} & \rightarrow (1+m_i)(1+m_j)C_\ell^{ij}, \\
        \xi_\pm^{ij}(\theta) & \rightarrow (1+m_i)(1+m_j)\xi_\pm^{ij}(\theta).
    \end{flalign}
    The posterior on these parameters is typically strongly dominated by the tight prior, a centered Gaussian of standard deviation 0.005 (see \cref{tab:params}). Therefore, we approximate the posterior by marginalizing analytically over these nuisance parameters by a simple modification of the covariance matrix, as explained \cref{sec:covariance}.
    This procedure allows us to reduce the parameter space and hence decrease the variance of importance sampling estimators (see \cref{sec:is}). \cd{We use the same priors on $m_i$'s for both statistics, but we do not require these multiplicative bias corrections to be strictly equal. In fact, they may slightly differ since biases are redshift dependent and contributions along the line of sight are mixed into angular scales differently.}
    \item \textit{Bias in the redshift distributions.} To account for uncertainty in the means of the redshift distributions, we model the estimated $n_{\rm obs}(z)$ to be shifted from the true $n(z)$. That is, we have
    \begin{equation}
        n_{\rm obs}^i(z) = n^i(z-\Delta z_{i}),
        \label{eq:nz_bias}
    \end{equation} 
    with a bias $\Delta z_{i}$ for each redshift bin $i$. 
    \item \textit{Intrinsic alignment (IA).}
    The intrinsic shapes of galaxies are correlated through various mechanisms, such that the cosmic shear power spectrum receives contribution from the auto-correlation of intrinsic shapes, $C_{\ell,{\rm II}}^{ij}$, and the cross-correlation of intrinsic shapes with the shear field, $C_{\ell,{\rm \gamma I}}^{ij}$.
    For our baseline model, we use the so-called Non-Linear Alignment (NLA) model, proposed by \citet{2004PhRvD..70f3526H,2007NJPh....9..444B},
    to compute $C_{\ell,{\rm II}}^{ij}$ and $C_{\ell,{\rm \gamma I}}^{ij}$, which is mathematically equivalent to
    replacing the lensing efficiency $q(\chi)$ in \cref{eq:Cl} in the following way
    \begin{equation}
        q^i(\chi) \rightarrow q^i(\chi) - A(z(\chi)) n^i(\chi),
    \end{equation}
    with
    \begin{equation}
        A(z) = - \Aia \bar{C}_1 \rho_{c} \frac{\Om}{D(z)} \qty(\frac{1+z}{1+z_0})^\aia,
    \end{equation}
    where ${\rho_{c}=3 H_0^2/8 \pi G}$ is the critical density.
    Here $D(z)$ is the linear growth factor, $\bar{C}_1$ is a normalization constant set to $\SI{5e-14}{\per\solarmass\per\h\squared\mega\parsec\cubed}$ \citep{2002MNRAS.333..501B} and we set the pivot redshift at $z_0 = 0.62$ as done in \citet{2018PhRvD..98d3528T,2019MNRAS.489.5453S}.
    The amplitude $\Aia$ and power-law scaling with redshift, $\aia$, are treated as free parameters of the model with uniform priors over the range $[-5,+5]$ for both.
\end{itemize}
The full modeling pipeline described above is implemented in the software package \cosmosis \citep{2015A&C....12...45Z}.

\begin{table*}
    \centering
    \begin{tabular}{l l l l l}
        Parameters & Symbols & Fiducial values & Priors & IS proposal distributions \\
        \hline
        Total matter density            & $\Om$ & 0.3111 & $\mathcal{U}(0.1,0.6)$ & \multirow{2}{*}{Uniform in $(S_8,\sqrtOm)$, see \cref{sec:IS_q}.} \\
        Density fluctuation amplitude   & $\sigma_8$ & 0.8076 & $\mathcal{U}(0.5,1.3)$ & \\
        Baryon density                  & $\Ob$ & 0.04897 & $\mathcal{U}(0.03,0.12)$ & $\mathcal{U}(0.03,0.12)$ \\
        Hubble parameter                & $h$ & 0.6766 & $\mathcal{U}(0.55,0.91)$ & $\mathcal{U}(0.55,0.91)$ \\
        Spectral index                  & $\ns$ & 0.9665 & $\mathcal{U}(0.87,1.07)$ & $\mathcal{U}(0.87,1.07)$ \\
        Physical neutrino density       & $\Onu h^2$ & 0.00083 & $\mathcal{U}(0.0006, 0.01)$ &  $\mathcal{U}(0.0006, 0.01)$ \\
        \hline
        Intrinsic alignment amplitude   & $\Aia$ & 0 & $\mathcal{U}(-5,5)$ & $\mathcal{N}(0, 1.5)$ \\
        Intrinsic alignment redshift dependence & $\aia$ & 0 & $\mathcal{U}(-5,5)$ & $\mathcal{U}(-5,5)$ \\
        \hline
        Photo-$z$ shift in bin $i$ ($i=1,2,3,4$) & $\Delta z_i$ & 0 & $\mathcal{N}(0, 0.005)$ & $\mathcal{N}(0, 0.005)$ \\
        Shear bias in bin $i$ ($i=1,2,3,4$) & $m_i$ & 0 & $\mathcal{N}(0, 0.005)$ & None (analytical marginalization) \\

    \end{tabular}
    \caption{Cosmological and nuisance parameters in the baseline model. Uniform distributions in the range $[a,b]$ are denoted $\mathcal{U}(a,b)$ and Gaussian distributions with mean $\mu$ and standard deviation $\sigma$ are denoted $\mathcal{N}(\mu,\sigma)$.}
    \label{tab:params}
\end{table*}
 
\subsection{Covariance and likelihood}
\label{sec:covariance}

Both sets of point statistics, $C_\ell^{ij}$ and $\xi_\pm^{ij}$, measured from data at multipoles $\ell$ and angular separation $\theta$, are stacked into two data vectors which are modeled as multivariate Gaussian variables with expected values described above by \cref{eq:Cl,eq:xipm}.
We use Gaussian analytic covariance matrices computed with \cosmosis for all the cosmological inference in this work. The covariance for the power spectra $\Cl$ can then be written as
\begin{equation}
    \cov\qty(C_\ell^{ij}, C_{\ell^\prime}^{i^{\prime}j^{\prime}})  \approx \delta_{\ell \ell^{\prime}} \frac{ D_{\ell}^{ii^{\prime}} D_{\ell}^{jj^{\prime}}+D_{\ell}^{ij^{\prime}} D_{\ell}^{ji^{\prime}}}{(2 \ell + 1) \fsky} 
\label{eq:cov_cl}
\end{equation}
where $i,j$ and $i^{\prime},j^{\prime}$ denotes the redshift bin pairs associated with the two considered $\Cl$'s, ${D_\ell^{ij}\equiv C_\ell^{ij}+N_\ell^{ij}}$ is the sum of the signal and noise power spectra, $\delta_{\ell \ell^{\prime}}$ is the Kronecker delta function and $\fsky$ is the fractional sky coverage set to $\fsky=0.1181$ in this work \citep[corresponding to \SI{4872}{\deg\squared}, the approximate DES Y3 area when including small, isolated regions rejected in the Gold catalog as presented in][]{y3-gold}. The shape noise contribution $N_\ell^{ij}$ is zero for cross-correlation and $\sigma_{e,i}^2/\bar{n}_i$ for auto-correlations, where $\sigma_{e,i}$ is the standard deviation of the measured galaxy shapes and $\bar{n}_i$ is the effective mean density of galaxies per unit sky area in redshift bin~$i$. In practice, we estimate power spectra $\hat{C}_L^{ij}$ averaged over band powers $L$ defined by $\ell_{\min}^L \leq \ell < \ell_{\max}^L$
and their covariance matrix is obtained by averaging \cref{eq:cov_cl} accordingly,
\begin{equation}
    \cov\qty(C_L^{ij}, C_{L^\prime}^{i^{\prime}j^{\prime}}) = \frac{1}{\Delta_L} \frac{1}{\Delta_{L^\prime}} \sum_{\ell \in L} \sum_{\ell^{\prime} \in L^{\prime}} \cov\qty(C_\ell^{ij}, C_{\ell^\prime}^{i^{\prime}j^{\prime}}),
\end{equation}
where $\Delta_L=\ell_{\max}^L-\ell_{\min}^L$ is the number of multipoles in band $L$.

The covariance for the correlation functions $\xipm$ is essentially a Fourier transform of \cref{eq:cov_cl}, and can be written as
\begin{align}
    &\cov\qty(\xipm^{ij}(\theta), \xipm^{i^{\prime}j^{\prime}}(\theta^{\prime}))  \approx \notag \\
    & \quad \int \frac{\ell\dd{\ell}}{2\pi} J_{n}(\ell \theta) \int \frac{\ell' \dd{\ell'}}{2\pi}  J_{n}(\ell' \theta') \cov\qty(C_\ell^{ij}, C_{\ell^\prime}^{i^{\prime}j^{\prime}}),
\label{eq:cov_xipm}
\end{align}
where spherical Bessel functions $J_{n}$ are of order $n=0$ ($n=4$) for $\xip$ ($\xim$).

As mentioned in the previous section, we do not vary shear calibration biases in the cosmological analysis in order to reduce the dimension of the parameter space and thus improve the accuracy of importance sampling (see \cref{sec:is}). However, we do account for the uncertainty due to those nuisance parameters by marginalizing analytically, following the procedure laid out in \citet{2002MNRAS.335.1193B} and extended in \citet{2010MNRAS.408..865T}. Assuming independent Gaussian priors, and to first order in the shear biases $m_i$, the marginalized likelihood remains Gaussian with a marginalized covariance receiving an extra term\footnote{This derivation starts from equation~(24) in \citet{2010MNRAS.408..865T}, noting that the prior covariance is diagonal with coefficients $\sigma^2_m$ and that the mean is $(1+m_i)(1+m_j)C_\ell^{ij}$. Taking derivatives of the mean with respect to $m_k$ gives $(\delta_{ki}+\delta_{kj})C_\ell^{ij}$, to first order in $m_k$, and summing over indices $k$ leads to \cref{eq:shear_marg}. Note that this expression slightly differs from that found in \citet{2017MNRAS.465.1454H} as our parametrization is different (we allow all shear biases to vary independently).}
\begin{align}
    \cov\qty(C_L^{ij}, C_{L^\prime}^{i^{\prime}j^{\prime}}) \rightarrow & \cov\qty(C_L^{ij}, C_{L^\prime}^{i^{\prime}j^{\prime}}) \nonumber \\ & + \sigma^2_m C_L^{ij} C_{L^\prime}^{i^{\prime}j^{\prime}} \qty(\delta_{ii'} + \delta_{ij'} + \delta_{ji'} + \delta_{jj'}),
    \label{eq:shear_marg}
\end{align}
where $\delta_{\alpha\beta}$ is the Kronecker symbol and $\sigma_m=0.005$ is the standard deviation of the Gaussian prior on shear biases. The real-space covariance matrix is modified in exactly the same way, by replacing $C_L$'s by $\xip(\theta)$ or $\xim(\theta)$.

We now comment on approximations made in the covariance matrix.
First, we do not account for the survey geometry in either harmonic or real space as it has little impact for the DES Y3 footprint \citep{2018MNRAS.479.4998T,y3-covariances}.
Second, we note that the choice of Gaussian covariance (and Gaussian likelihood) is an approximation that matches the choice of Gaussian simulations. However, as shown in \citet{2018JCAP...10..053B}, the Gaussian covariance is largely sufficient even for non-Gaussian simulations. The next leading term is the so-called super-sample covariance term, accounting for correlations with $k$-modes larger than the survey footprint, but even this term is largely subdominant for a survey like DES Y3, as shown in \citet{y3-covariances}.
This means that, for the purposes of this work targetted at DES Y3, we can safely employ Gaussian simulations and analyze them with a Gaussian covariance.

\subsection{Contaminated data vectors: systematic effects and alternative modeling}
\label{sec:sys_th}

The baseline model described above matches that in the DES Y1 analysis. However, there are known physical and instrumental effects that impact the measured $\Cl$ and $\xipm$\, potentially differently, which may impact cosmological constraints depending on whether they affect scales that are used for the analysis. We therefore compute alternative theoretical data vectors, including one effect at a time, either modifying modeling or including additional biases in the data vectors from systematic effects.

\begin{itemize}
    \item \textit{Non-linear matter power spectrum.} Our fiducial model is based on the \halofit prescription \citep{2012ApJ...761..152T} to model the impact of non-linear gravitational evolution of the large-scale structure at small scales, \ie $k \gtrsim \SI{0.2}{\h\per\mega\parsec}$. However, it is known to be accurate only up to 5\% for $k \leq \SI{5}{\h\per\mega\parsec}$ and degrading for smaller scales. We recompute the fiducial data vectors using the matter power spectrum emulator from \citet{2017ApJ...847...50L}, a Gaussian process interpolator based on the Mira-Titan Universe simulations \citep{2016ApJ...820..108H}, leaving the rest of the pipeline unchanged.
    With respect to \halofit, the emulator predicts a power spectrum roughly 5\% lower in the range \SIrange{0.2}{2}{\h\per\mega\parsec} with damped acoustic oscillation features. The cosmic shear power spectra and two-point functions are respectively reduced by about 5\% at most for multipoles $\ell\sim1000$ and angular separations $\theta\sim\SI{10}{\arcmin}$, which is about 10\% (respectively 40\%) of the error bars for the auto-correlation of redshift bin 1 (redshift bin 4) at these scales.

    \item \textit{Baryonic feedback.} Baryonic processes within dark matter haloes redistribute matter and therefore impact the matter power spectrum at small scales. Energy injection from active galactic nuclei causes a small suppression of the matter power spectrum in the range $k \sim \SIrange{1}{10}{\h\per\mega\parsec}$, and cooling as well as star formation enhance it at smaller scales \citep{2018MNRAS.480.3962C,2019JCAP...03..020S}.
    In order to model the impact of baryons on the fiducial data vectors, we rescale the non-linear matter power spectrum by the ratio of the matter power spectra measured in the OWLS simulations \citep{2011MNRAS.415.3649V} with dark matter only, $P_{\rm DM}(k,z)$, and with AGN feedback, $P_{\rm AGN}(k,z)$,
    such that
    \begin{equation}
        P_{\rm NL}(k,z) \rightarrow P_{\rm NL}(k,z) \frac{P_{\rm AGN}(k,z)}{P_{\rm DM}(k,z)},
        \label{eq:baryon_Pk_ratio}
    \end{equation}
    as was done for the real space analysis of DES Y1, \citet{2018PhRvD..98d3528T}.
    Note that we also derive scale cuts from this modified power spectrum in \cref{sec:scales}.
    
    \item \textit{Intrinsic alignments from tidal torquing.} The NLA model accounts for tidal alignment (TA) mechanisms but not for tidal torquing (TT) ones that were proposed by \citet{2001MNRAS.320L...7C,2001ApJ...559..552C,2002MNRAS.332..788M} as extra contributions in the observed shear power spectrum. These contributions were unified, including cross terms, into a single model (TATT) in \citet{2019PhRvD.100j3506B}, following a perturbation theory expansion of the tidal field. The TATT model was applied to DES Y1 data in \citet{2019MNRAS.489.5453S}. 
    Here, we adopt the same model and measure biases on data vectors including (part or all of) TA and TT contributions with respective amplitudes $A_1$ and $A_2$, and redshift dependence parametrized by power-law $\alpha_1$ and $\alpha_2$.
    We follow \citet{2019MNRAS.489.5453S} and fix the source bias to $b_g^{\rm src}=1$, accounting for the density-tidal field term in the TA component (which is not included in the NLA model). Note that NLA is a special case of TATT with $A_2=0$ and $b_g^{\rm src}=0$.

    \item \textit{Point-spread function leakage.} The point-spread function (PSF) needs to be estimated and accounted for when measuring the ellipticities of galaxies. It is usually measured at the positions of stars and interpolated to the positions of galaxies, while modeling residuals can be estimated from a fraction of stars reserved for this purpose \citep*{2018MNRAS.481.1149Z,2016MNRAS.460.2245J}. PSF ellipticity residuals leak directly into cosmic shear measurements and may introduce biases. In particular, the PSF does not have the symmetries of gravitational lensing and has roughly equal E- and B-mode signals. We use the measurements from DES Y3 data presented in \cite{y3-piff} and a similar parametrization of the bias in the two-point functions $\xipm(\theta)$ (although for simplicity we only keep the dominant term and fix $\beta=1$), given by
    \begin{equation}
        \Delta \xipm(\theta) = \rho_1^{\pm}(\theta) - \alpha \rho_2^{\pm}(\theta).
    \end{equation}
    Here, $\rho_1^{\pm}$ is the auto-correlation function of PSF ellipticity residuals, $\rho_2^{\pm}$ is the cross-correlation between the model and residual ellipticities and $\alpha$ is the leakage coefficient. It is, however, difficult to directly evaluate the equivalent of $\rho_{1/2}$ in harmonic space directly because of the bias created by the noise power spectrum of the residuals (which is \emph{not} a simple shape-noise). Instead, for the purpose of this work, we treat this bias as a small perturbation to the fiducial data vectors. Given the cross-covariance between $\xipm$ and $\Cl$, we can compute the expectation value of the harmonic-space bias $\Delta \Cl$ conditioned on the real-space bias $\Delta \xipm(\theta)$. More precisely, we combine \cref{eq:xipm} and \cref{eq:cov_cl} to compute an approximate, analytic Gaussian cross-covariance $\mathbf{C}_{\ell \pm} \equiv \cov\qty(\Cl,\xipm(\theta))$ and the bias is estimated by
    \begin{equation}
        \mqty[\Delta \Cl] = \mathbf{C}_{\ell \pm} \vdot \mathbf{C}_{\pm}^{-1} \vdot \mqty[\Delta \xipm(\theta)],
    \end{equation}
    where $\mathbf{C}_{\pm}$ is the covariance matrix of $\xipm(\theta)$ given by \cref{eq:cov_xipm}. We used brackets to indicate data vectors and $\vdot$ for matrix-matrix and matrix-vector products. We find the biases induced by the PSF, $\Delta \Cl$ and $\Delta \xipm(\theta)$, to be very small for DES Y3, in agreement with \citet{y3-piff}, even for a leakage as high as 10\% (\ie $\alpha=0.1$), which is excluded by measurements presented in \citet*{y3-shapecatalog}.

    \item \textit{Width of the redshift distributions.} In our fiducial pipeline, the uncertainty in the redshift distributions is solely encoded by a coherent shift, as shown in \cref{eq:nz_bias}, capturing the principal mode of uncertainty. Here, we additionally probe the effect of underestimating the width of the redshift distribution. To do so, we convolve the redshift distribution with a Gaussian kernel of width $\sigma_z=0.1$. Given that the widths of the redshift distributions are of order \numrange{0.2}{0.3}, this convolution increases the width by about \numrange{5}{10}\%, consistent with typical width uncertainties found with self-organizing maps methods for DES Y3 \citep*{2019MNRAS.489..820B}.
\end{itemize}

\section{Scale cuts}
\label{sec:scales}

As discussed in \cref{sec:intro}, the scale cuts imposed on the data vectors determine the physical $k$-modes of the cosmic shear field that are accessible through the two-point functions. In particular, since a hard cut in multipole space $\ell$ is not a hard cut in real space $\theta$ (and vice-versa), there is no straightforward translation between the two spaces, and a hard cut in $k$ corresponds to neither. We explore below several methods to establish an approximate correspondence between scale cuts in harmonic and real space.

Methodologies employed for determining scale cuts in the literature are usually based on balancing the trade-off between systematic uncertainty and statistical uncertainty \citep{2017MNRAS.465.1454H,2018PhRvD..98d3528T,2019PASJ...71...43H}. The scale cuts are usually chosen to be such that the systematic uncertainty introduced by errors in the modeling are subdominant to the statistical uncertainty determined by the survey characteristics (\eg area, density of galaxies).
In particular, the modeling of intrinsic alignments \citep{2017arXiv170609359K,2019MNRAS.489.5453S} and the impact of baryons at small scales \citep{2018MNRAS.480.3962C,2019JCAP...03..020S}, both connected to small-scale, non-linear, astrophysical processes, typically drives small-scale cuts, while observational considerations, such as PSF residuals and shear calibration uncertainty, determine the large-scale cuts.
For the real space analysis of DES Y1, \citet{2018PhRvD..98d3528T} chose small-scale cuts for each redshift bin such that differences between $\xipm$ data vectors, with or without the effect of baryons, is less than 2\%. The impact of baryons was modeled from the ratio of the matter power spectra measured in the OWLS simulations, as in \cref{eq:baryon_Pk_ratio}. No additional scale cuts were applied for IA modeling uncertainties, given robustness tests presented in \citet{2017arXiv170609359K}.
The DES Y1 scale cuts are listed in \cref{tab:all_cuts}. We will present and discuss scale cuts used in HSC Y1 and KiDS-450 analyses in \cref{sec:res_hsc_kids}.

\begin{figure}
    \centering
    \includegraphics[scale=0.6]{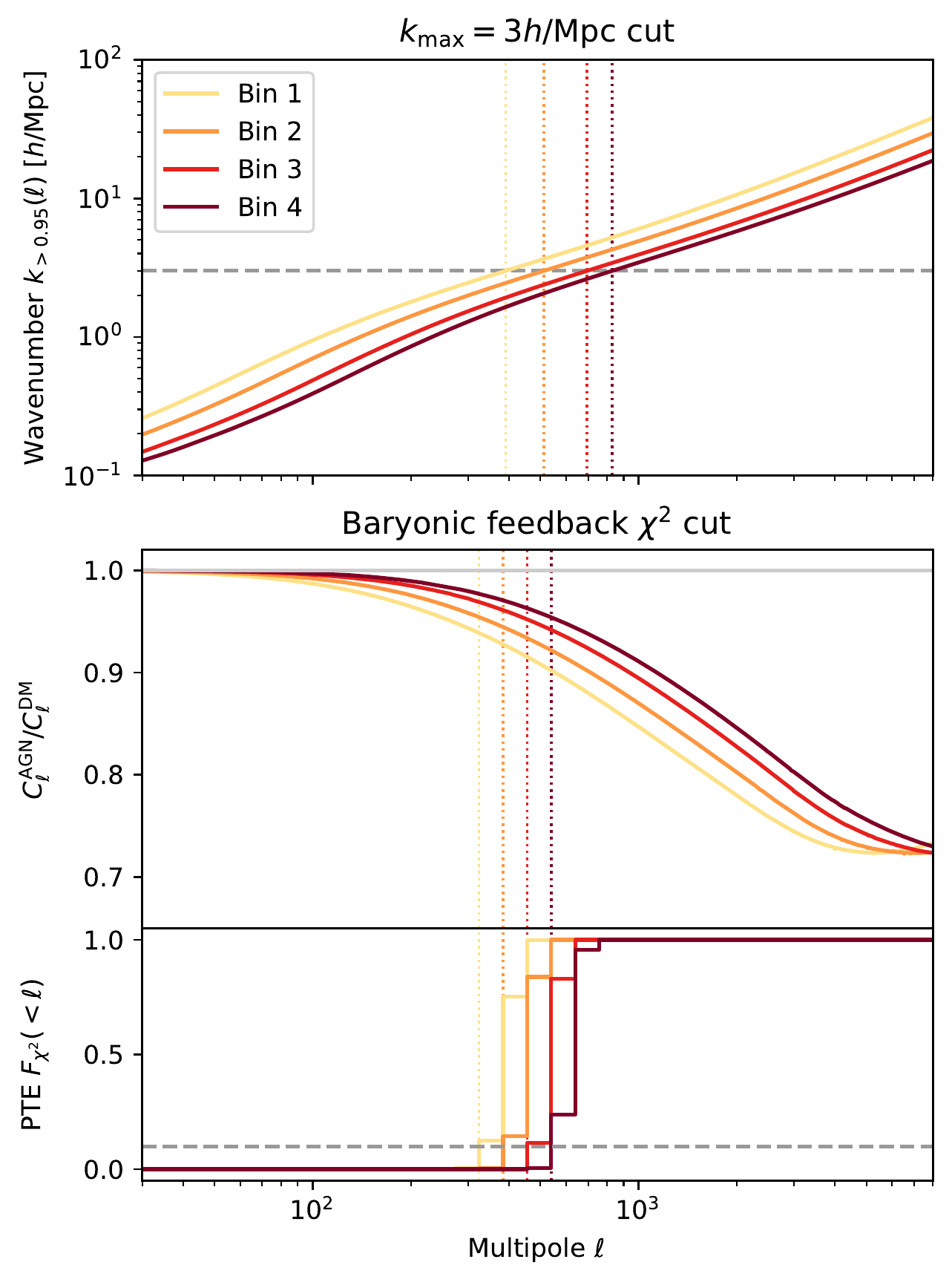}
    \caption{
    Harmonic space scale cuts derived from $\kmax$ and baryonic feedback cut-offs. In both plots, we show the curves corresponding to auto-correlations for bins 1 (yellow) through 4 (red) and derived $\ellmax$ cuts as the vertical dotted lines. We do not show cross-correlations for readability, although we apply the same method to derive cuts for those bins, which we report in \cref{tab:all_cuts}.
    \textit{Top:}~Scale cuts derived from physical mode cut-off at $\kmax=\SI{3}{\h\per\mega\parsec}$ (grey dashed line). For a given multipole $\ell$, we compute the $k$-mode at which the Limber integral, \cref{eq:Cl}, reaches 95\% of its total value, $k_{>0.95}(\ell)$. We exclude multipoles with ${k_{>0.95}(\ell)>\kmax}$, \ie those that receive more than 5\% of their signal from scales beyond $\kmax$.
    \textit{Bottom:}~Scale cuts derived from baryonic feedback (from OWLS).
    The top panel shows the ratio of predicted $\Cl$ with and without baryonic feedback. The bottom panel shows the probability-to-exceed $F_{\chi^2}(>\ell)$ of the $\chi^2$ statistics computed between $\Cl$ data vectors computed with baryonic feedback to that without (for the binning used in Gaussian simulations, see \cref{sec:sims}) when including all multipole bins below $\ell$. We exclude multipoles with a $\chi^2$ above its tenth percentile (grey dashed line).
    }
    \label{fig:kmax_baryons_cls}
\end{figure}

\begin{figure}
    \centering
    \includegraphics[scale=0.6]{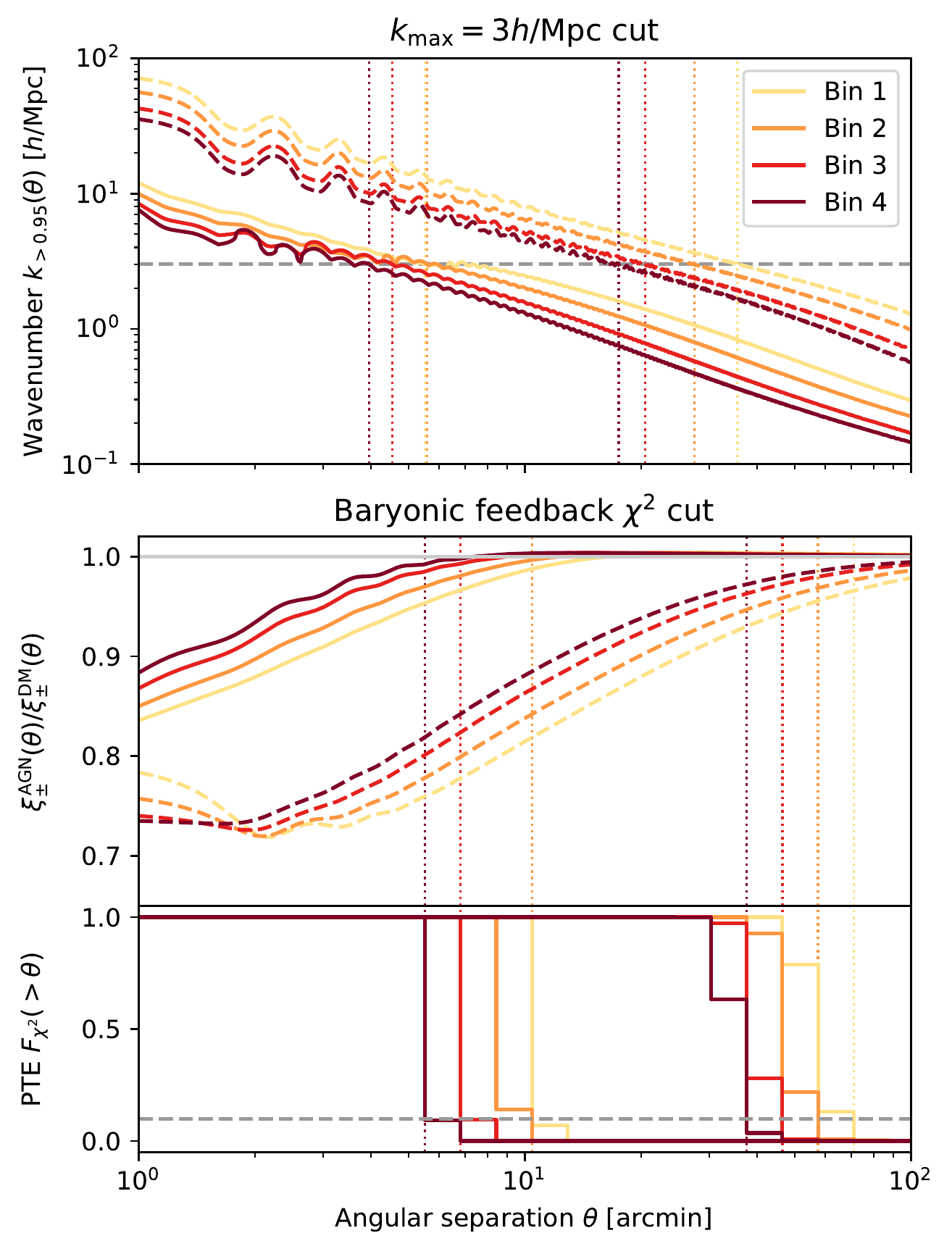}
    \caption{
    Real space scale cuts derived from $\kmax$ and baryonic feedback cut-offs, similar to \cref{fig:kmax_baryons_cls} for real-space two-point functions $\xipm(\theta)$. We use solid (respectively dashed) curves for $\xip$ (respectively $\xim$). The wiggles in $k_{>0.95}(\theta)$ are due to acoustic features and sign flipping of $\dv*{\ln \xipm}{\ln k}$ visible in \cref{fig:Cl_xi_vs_lnk}. The bottom panel in the bottom plot should be read from large scales (right) to small scales (left), as it shows the probability-to-exceed of the $\chi^2$ statistic, $F_{\chi^2}(>\theta)$ computed for bins with a separation angle above $\theta$.
    }
    \label{fig:kmax_baryons_xis}
\end{figure}

For this analysis, we fix the large-scale cut at $\tmax=\SI{10}{\degree}$ and ${\ellmin=\pi/\tmax=18}$ (because signal-to-noise ratio is low in the largest-scale bin)
and focus on different methods to derive the small-scale cuts.
To derive the scale cuts we need to specify a number of quantities associated with the survey properties, such as the redshift distribution, the survey area and the number density of galaxies. We have chosen to use numbers that match the DES Y3 dataset. However, we note that the framework we developed here can be easily adapted for a different survey.  
As an illustrative exercise, we will present results in \cref{sec:results} that use all scales measured in the simulations described in \cref{sec:sims}. In order to maintain consistency between sections, we therefore include this option before presenting three realistic methods. 
\begin{enumerate}
    \item \label{it:sc_i} \textit{No scale cuts.} In \cref{sec:results}, we will perform measurements using all available scales for which we have measurements.
    
    \item \label{it:sc_ii} \textit{$\ell \sim \flatfrac{\pi}{\theta}$ relation.} If cuts are available in one space, they can be very approximately translated to the other space using the relation $\ell \sim \flatfrac{\pi}{\theta}$. In particular, we will use DES Y1 scale cuts in real space and approximately match them in harmonic space. To do so, we choose to use the geometric mean of the $\xip$ and $\xim$ small-scale cuts, respectively $\theta_{\rm min}^+$ and $\theta_{\rm min}^-$, \ie we set $\ellmax={\pi/\sqrt{ \theta_{\rm min}^+ \theta_{\rm min}^-}}$, which we find to preserve signal-to-noise ratio (as opposed to using either $\theta_{\rm min}^+$ or $\theta_{\rm min}^-$ to do the conversion). This is explained by the comparable marginal signal-to-noise ratios of $\xip$ and $\xim$ measurements at their respective scale cuts.

    \item \label{it:sc_iii} \textit{Physical mode cut-off $\kmax$.} \label{item:kmax} The next option we consider is motivated by \cref{fig:Cl_xi_vs_lnk}. We pick a small-scale physical mode cut, $\kmax$, and determine an effective corresponding $\ell_{\rm max}$ and $\theta_{\rm min}$ from $\dv*{\ln C_\ell}{\ln k}$ and $\dv*{\ln \xi_\pm}{\ln k}$. To do so, we write the power spectra $\Cl$ and correlation functions $\xipm(\theta)$ as integrals over wavenumber $k$, using the Limber formula and the change of variables $k=\flatfrac{\qty(\ell+1/2)}{\chi(z)}$, and compute the corresponding scale $k_{>\alpha}$---which is a function of $\ell$ or $\theta$---at which the integral reaches a fraction $\alpha$ of its total value, \emph{i.e.}
    \begin{equation}
        \int_{- \infty}^{\ln k_{>\alpha}} \dd{\ln k} \abs{\dv{\ln X}{\ln k}} = \alpha,
    \end{equation}
    where $X$ is either $\Cl$, $\xip(\theta)$ or $\xim(\theta)$. Since $\xip$ receives negative contributions for a range of $k$-modes (especially at small scales), we consider the absolute value of the integrand to determine this cut\footnote{We tested both with and without the absolute value, and found that including it yielded cuts with closer signal-to-noise ratio between harmonic and real spaces.}. We then compute the value $\ellmax$ (respectively $\tmin$)
    for which $k_{>\alpha}(\ellmax)=\kmax$ (respectively $k_{>\alpha}(\tmin)=\kmax$).
    Here, we use a fraction of ${\alpha=\num{0.95}}$, \ie we keep scales for which 95\% of the signal comes from modes under $\kmax$. In other words, theoretical uncertainties beyond $\kmax$ may only affect 5\% of the smallest scales included in the analysis. We will vary $\kmax$ from 1 to \SI{5}{\h\per\mega\parsec}, which is the scale above which errors in the \halofit model exceed 10\% \citep{2012ApJ...761..152T}. The results are shown in \cref{fig:kmax_baryons_cls} for harmonic space and in \cref{fig:kmax_baryons_xis} for real space, for $\kmax=\SI{3}{\h\per\mega\parsec}$, which we use as our fiducial value, in the middle of the range \SIrange{1}{5}{\h\per\mega\parsec}. 
    
    \item \label{it:sc_iv} \textit{Impact of baryons.} \label{item:baryons} Following the DES Y1 method, we use baryonic feedback models to compare two-point data vectors with and without modeling baryons (using the same OWLS AGN model). The DES Y1 analysis set a fixed threshold for the fractional difference between the data vector with and without baryons. This threshold is, however, somewhat arbitrary. We improve on the method by instead requiring the $\chi^2$ distance between the two data vectors, which incorporates correlations between elements of the data vectors, not to exceed the percentile corresponding to a fixed probability-to-exceed (PTE). We use a covariance matrix without shape-noise to preserve the theoretical motivation for the cut, alleviating the dependence on survey depth (except for the area through the $\fsky$ factor). Given a binning scheme (see \cref{sec:sims}), we compute, separately for each redshift bin pair, $\chi^2$ distances between data vectors, starting from the largest-scale bin and progressively including smaller-scale bins.
    At each step, we then compute the corresponding PTE for a $\chi^2$ distribution with a number of degrees of freedom equal to the number of aggregated bins, denoted $F_{\chi^2}(<\ell)$. We set a threshold at the tenth percentile, \ie we discard small-scale bins where $F_{\chi^2}(<\ell)>0.1$. We plot $F_{\chi^2}(<\ell)$ in the lowest panel of \cref{fig:kmax_baryons_cls} as a piecewise constant function matching bin edges.
    The procedure works similarly in real space, where we instead compute $\chi^2(\theta)$ and $F_{\chi^2}(>\theta)$, discarding small-angle bins $\theta$ where $F_{\chi^2}(>\theta)>0.1$.
    This procedure allows us to obtain theoretically motivated cuts in both spaces with relatively little dependence on the threshold choice, as shown by the sharp transition 
    in the lower panels of \cref{fig:kmax_baryons_cls,fig:kmax_baryons_xis}.
    Note that this method is applicable to any kind of comparison between a fiducial and contaminated model.
\end{enumerate}
We list all scale cuts used in the analysis (results presented in \cref{fig:S8_r_vs_F} and thereafter) in \cref{tab:all_cuts} and plot them in \cref{fig:sims_validation_Cl,fig:sims_validation_xi} for comparison.

\begin{table*}
    \centering
    \begin{tabular}{l c c c c c c c c c c c c}
        & & & \multicolumn{10}{c}{Redshift bin pairs}\\
        \cmidrule(r){4-13}
        Scale cut & Fiducial $S/N$ & $\Cl$/$\xipm$ cut & 1-1 & 1-2 & 1-3 & 1-4 & 2-2 & 2-3 & 2-4 & 3-3 & 3-4 & 4-4 \\
    \midrule
    \multirow{3}{*}{No scale cut} & 69.8 ($\Cl$)
        & $\ellmax$ &  8192 & 8192 & 8192 & 8192 & 8192 & 8192 & 8192 & 8192 & 8192 & 8192 \\
    & \multirow{2}{*}{62.8 ($\xipm$)}
        & $\theta_{\rm min}^+$ &  \SI{1.0}{\arcminute} & \SI{1.0}{\arcminute} & \SI{1.0}{\arcminute} & \SI{1.0}{\arcminute} & \SI{1.0}{\arcminute} & \SI{1.0}{\arcminute} & \SI{1.0}{\arcminute} & \SI{1.0}{\arcminute} & \SI{1.0}{\arcminute} & \SI{1.0}{\arcminute} \\
        & & $\theta_{\rm min}^-$ &  \SI{10.0}{\arcminute} & \SI{10.0}{\arcminute} & \SI{10.0}{\arcminute} & \SI{10.0}{\arcminute} & \SI{10.0}{\arcminute} & \SI{10.0}{\arcminute} & \SI{10.0}{\arcminute} & \SI{10.0}{\arcminute} & \SI{10.0}{\arcminute} & \SI{10.0}{\arcminute} \\
    \midrule
    \multirow{3}{*}{DES Y1 ($\ell\sim\pi/\theta$)} & 54.8 ($\Cl$)
        & $\ellmax$ &  423 & 474 & 532 & 532 & 670 & 670 & 752 & 844 & 844 & 947 \\
    & \multirow{2}{*}{43.8 ($\xipm$)}
        & $\theta_{\rm min}^+$ &  \SI{7.2}{\arcminute} & \SI{7.2}{\arcminute} & \SI{5.7}{\arcminute} & \SI{5.7}{\arcminute} & \SI{4.5}{\arcminute} & \SI{4.5}{\arcminute} & \SI{4.5}{\arcminute} & \SI{3.6}{\arcminute} & \SI{3.6}{\arcminute} & \SI{3.6}{\arcminute} \\
        & & $\theta_{\rm min}^-$ &  \SI{90.6}{\arcminute} & \SI{72.0}{\arcminute} & \SI{72.0}{\arcminute} & \SI{72.0}{\arcminute} & \SI{57.2}{\arcminute} & \SI{57.2}{\arcminute} & \SI{45.4}{\arcminute} & \SI{45.4}{\arcminute} & \SI{45.4}{\arcminute} & \SI{36.1}{\arcminute} \\
    \midrule
    \multirow{3}{*}{$\kmax=\SI{3}{\h\per\mega\parsec}$} & 53.0 ($\Cl$)
        & $\ellmax$ &  391 & 441 & 491 & 513 & 512 & 586 & 619 & 694 & 751 & 830 \\
    & \multirow{2}{*}{50.1 ($\xipm$)} 
        & $\theta_{\rm min}^+$ &  \SI{5.5}{\arcminute} & \SI{5.0}{\arcminute} & \SI{6.0}{\arcminute} & \SI{6.0}{\arcminute} & \SI{5.6}{\arcminute} & \SI{5.1}{\arcminute} & \SI{5.0}{\arcminute} & \SI{4.5}{\arcminute} & \SI{4.1}{\arcminute} & \SI{3.9}{\arcminute} \\
        & & $\theta_{\rm min}^-$ &  \SI{35.8}{\arcminute} & \SI{31.6}{\arcminute} & \SI{28.5}{\arcminute} & \SI{27.1}{\arcminute} & \SI{27.1}{\arcminute} & \SI{24.0}{\arcminute} & \SI{22.9}{\arcminute} & \SI{20.5}{\arcminute} & \SI{18.6}{\arcminute} & \SI{17.5}{\arcminute} \\
    \midrule
    \multirow{3}{*}{Baryonic feedback $\chi^2$} & 47.5 ($\Cl$)
        & $\ellmax$ &  341 & 364 & 410 & 429 & 397 & 432 & 456 & 485 & 507 & 532 \\
    & \multirow{2}{*}{40.3 ($\xipm$)} 
        & $\theta_{\rm min}^+$ &  \SI{11.5}{\arcminute} & \SI{11.4}{\arcminute} & \SI{10.5}{\arcminute} & \SI{9.3}{\arcminute} & \SI{10.0}{\arcminute} & \SI{9.2}{\arcminute} & \SI{9.0}{\arcminute} & \SI{7.6}{\arcminute} & \SI{7.4}{\arcminute} & \SI{6.1}{\arcminute} \\
        & & $\theta_{\rm min}^-$ &  \SI{67.7}{\arcminute} & \SI{62.3}{\arcminute} & \SI{57.7}{\arcminute} & \SI{51.3}{\arcminute} & \SI{58.6}{\arcminute} & \SI{50.9}{\arcminute} & \SI{49.6}{\arcminute} & \SI{48.3}{\arcminute} & \SI{43.4}{\arcminute} & \SI{40.9}{\arcminute} \\
    \end{tabular}
    \caption{Small-scale cuts used in this work (in particular the results shown in \cref{fig:S8_r_vs_F}) and the signal-to-noise ratio $S/N$ computed from the the fiducial data vector. The large-scale cuts are $\ellmin=18$ and $\theta_{\max}^\pm=\pi/\ellmin=\SI{600}{\arcminute}=\SI{10}{\degree}$ for all bins.}
    \label{tab:all_cuts}
\end{table*}

\section{Simulations}
\label{sec:sims}

Simulations are essential in this work as they allow us to realistically capture the correlated information that is used by the harmonic and real space statistics, and by representing the survey geometry, galaxy density and noise level expected in the real survey. In particular, simulations allow us to generate pairs of $\Cl$ and $\xipm$ data vectors with the correct cross-covariance, which is challenging to compute analytically with good accuracy, especially when accounting for survey geometry.

Our fiducial analysis is targeted towards the DES Y3 cosmic shear analysis, which motivates the choice of tomographic redshift bins, redshift distributions, number density and shape noise. At the time of completing this analysis the DES Y3 shear catalog and redshift distribution were not finalized. As a result we only approximately match final DES Y3 choices. We use four tomographic redshift bins with redshift distributions taken from the DES Y1 dataset, as shown in \cref{fig:dndz_footprint}. The number density for each redshift bin is fixed to $\bar{n}=\SI{1.5}{gal \per arcmin \squared}$, with shape noise of $\sigma_{e}=0.3$ per component in the fiducial analysis, and $\sigma_{e}=0.3/\sqrt{2}$ for the \textit{low noise} analysis (see \cref{sec:posterior_shifts}). We use an approximate DES Y3 footprint, also shown in \cref{fig:dndz_footprint}, with a survey area of \SI{4872}{\deg\squared}. We note that the main conclusions of this study should be robust even though these numbers do not match exactly that in DES Y3.

We provide here a brief description of our simulations. Starting from fiducial power spectra $C_\ell^{ij}$ for redshift bins $i$ and $j$, we generate a full-sky realization of the four correlated shear fields in \healpix\footnote{\url{http://healpix.sf.net}} \citep{2005ApJ...622..759G} maps of resolution ${\nside=4096}$ (with an approximate resolution of \SI{0.86}{\arcminute}). 
To do so, we first generate the harmonic coefficients of the E-mode of the shear fields, $E_{\ell m}^{i}$, for multipoles up to $\ellmax=3\nside-1$. These coefficients are Gaussian random variables with covariance $\expval*{E_{\ell m}^{i} E_{\ell^\prime m^\prime}^{j}} = \delta_{\ell\ell^\prime} \delta_{mm^\prime} C_\ell^{ij}/2w_\ell^2$, where we have included the \healpix pixel window function $w_\ell$. These variables are independent for different $\ell$ and $m$ indices, such that they can be sampled in parallel. At fixed $\ell,m$, the Cholesky decomposition of the covariance matrix (indexed by $i,j$) is used to generate $E_{\ell m}^{i}$ coefficients from four standard random variables, following a standard procedure to sample multivariate Gaussian variables. We then use the \texttt{alm2map} function of \healpy \citep{2019JOSS....4.1298Z} in polarization mode, with $T_{\ell m}^i=B_{\ell m}^i=0$, to generate the four correlated, true shear maps.
Then, we draw random positions of galaxies within the DES Y3 footprint\footnote{Galaxies are drawn independently, therefore there is no clustering of source galaxies, which may cause B-mode patterns, beyond the scope of this paper. We also do not account for blending here, supposedly included in the shear biases.} in each redshift bin with density $\bar{n}$ and compute the values $\bm{\gamma}_i$ of the shear field at the positions of the galaxies\footnote{The shear field is thus sampled at fixed effective redshift and not at the redshift of the galaxies.} (indexed by $i$ here). We then draw random intrinsic ellipticities of the galaxies $\vb{e}_i$ from a zero-mean normal distribution with variance $\sigma_e^2$ and finally compute the observed ellipticity, given by
\begin{equation}
    \vb{e}_i^{\rm obs} = \frac{\bm{\gamma}_i+\vb{e}_i}{1+\bm{\gamma}_i^\ast \vb{e}_i}.
\end{equation}

The next step is to compute the two-point data vectors from these mock catalogs both in harmonic and real space. In harmonic space, we use a pseudo-$\Cl$ estimator computed with \namaster \citep{2019MNRAS.484.4127A}. To do so, we first compute ellipticity maps, for both components, by averaging ellipticities in each pixel. We use the count map as the inverse-variance weighting mask. We use \namaster to measure mask-deconvolved, binned power spectra $\hat{C}_\ell^{ij}$ in 36 logarithmically spaced bins between $\ellmin=18$ and ${\ellmax=2\nside=8192}$. The measurements are corrected for the pixel window function introduced above (by multiplying raw $\Cl$ by $w_\ell^2$) and we correct for the noise power spectrum bias in each bin by applying random rotations to ellipticities and subtracting the mean power spectrum of 16 realizations to the measured auto power spectra. In real space, we use the package \treecorr \citep{2004MNRAS.352..338J} to estimate $\xipm$ from the mock catalogs in 30 log-spaced bins between $\tmin=\SI{1}{\arcminute}$ and $\tmax=\SI{600}{\arcminute}$. For both harmonic and real space measurements, the large-scale cut corresponds to \SI{10}{\degree}, which is of the order of the largest scale that can be measured well in the DES footprint, and the small-scale cut corresponds to the resolution of the simulations.

\begin{figure*}
    \centering
    \includegraphics[scale=0.6]{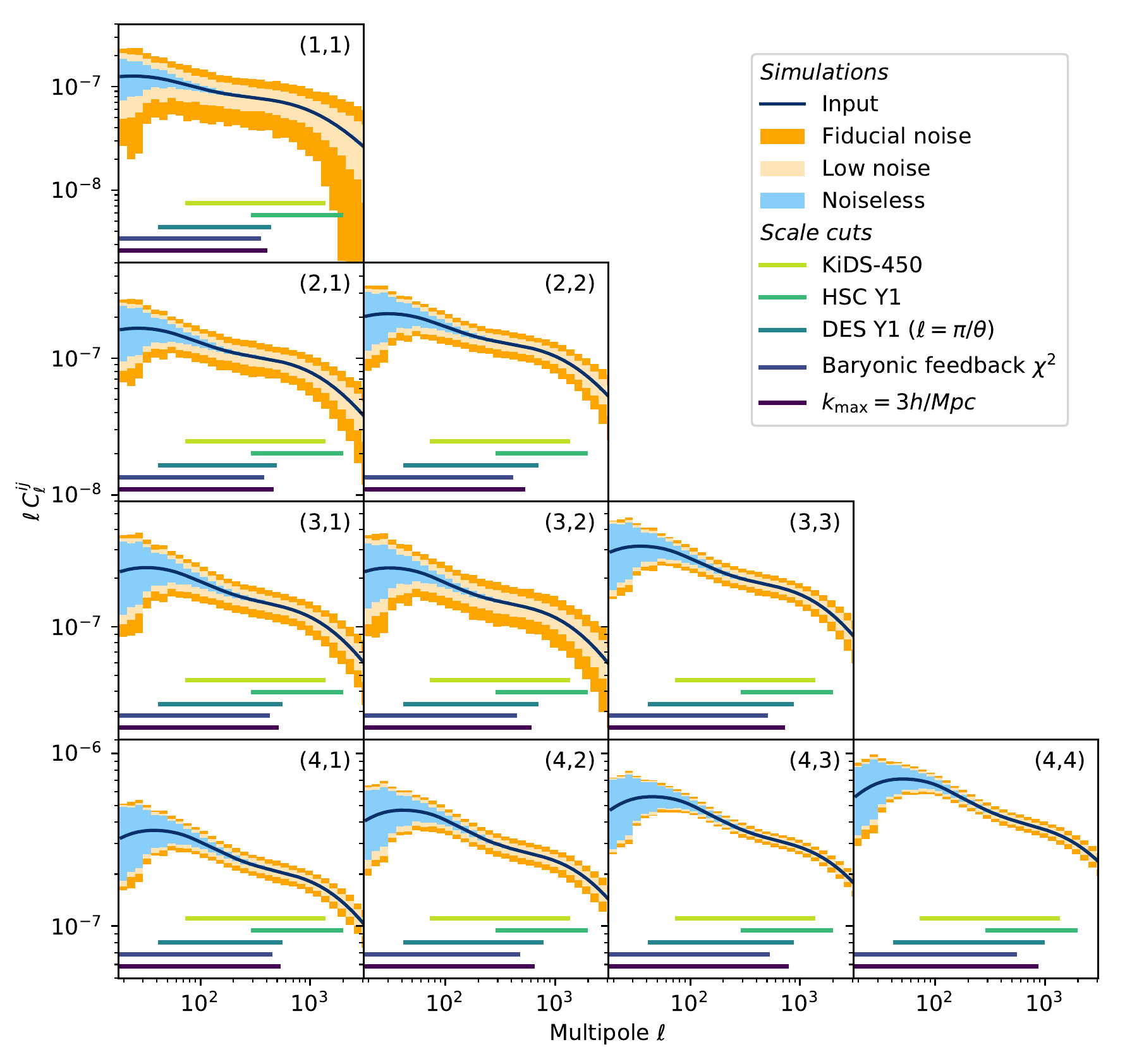}
    \caption{Power spectra $\Cl$ measured from Gaussian simulations described in \cref{sec:sims} and scale cuts used in the analysis in \cref{sec:results}. The input power spectrum are plotted in dark blue. Boxes show the measured mean and standard deviation in each $\ell$ bin, in orange for the fiducial noise level ($\sigma_e=0.3$), in moccasin for the low noise level ($\sigma_e=0.3/\sqrt{2}$) and in light blue for the noiseless case (where each galaxy shape corresponds to the shear field value). The ranges of scale cuts used in the analysis are shown as solid horizontal lines.}
    \label{fig:sims_validation_Cl}
\end{figure*}

\begin{figure*}
    \centering
    \begin{tikzpicture}
    \node(a){\includegraphics[scale=0.6]{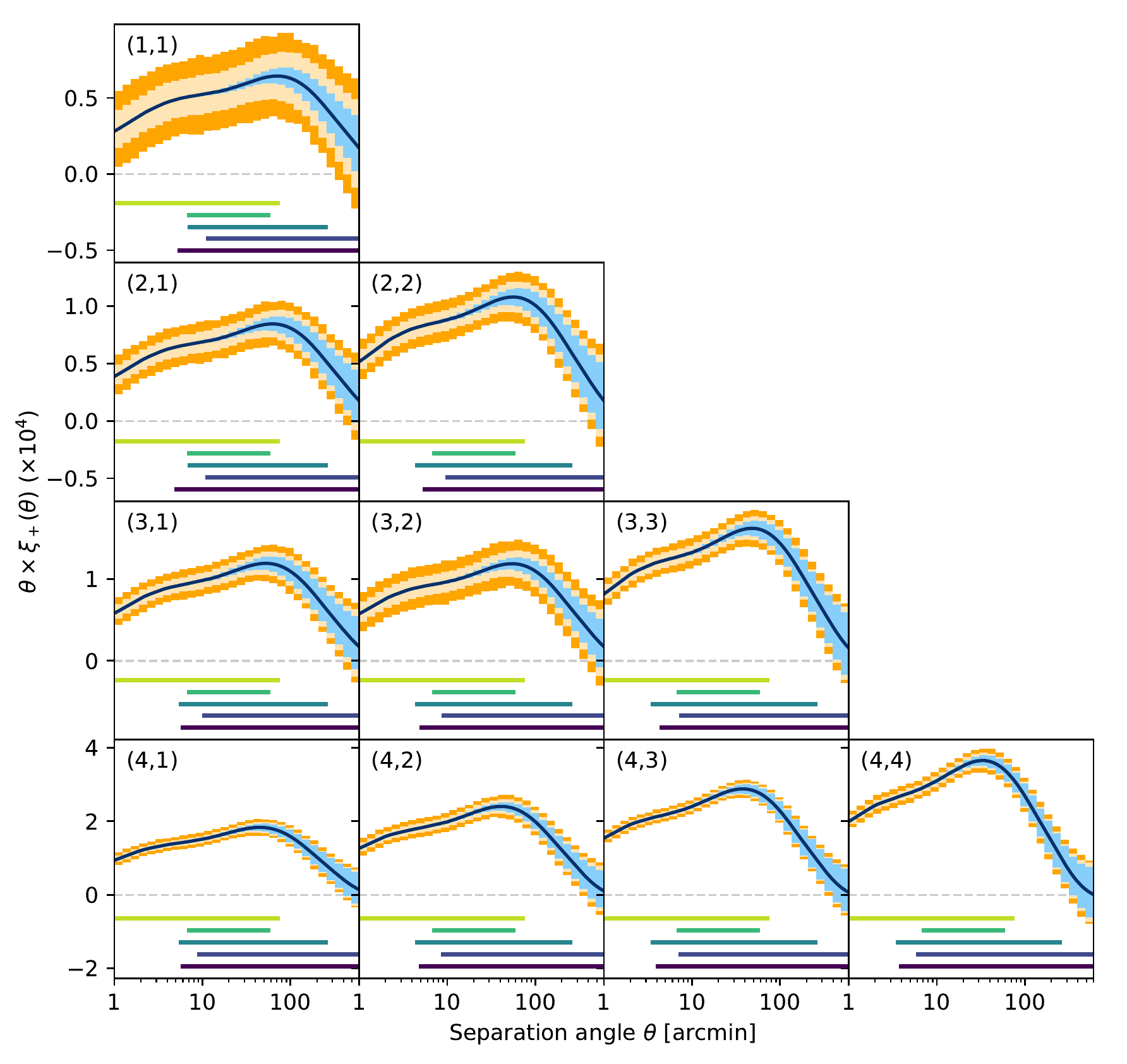}};
    \node(b) at (a.north east)[anchor=center, xshift=-37mm, yshift=-13mm]{\includegraphics[scale=0.6]{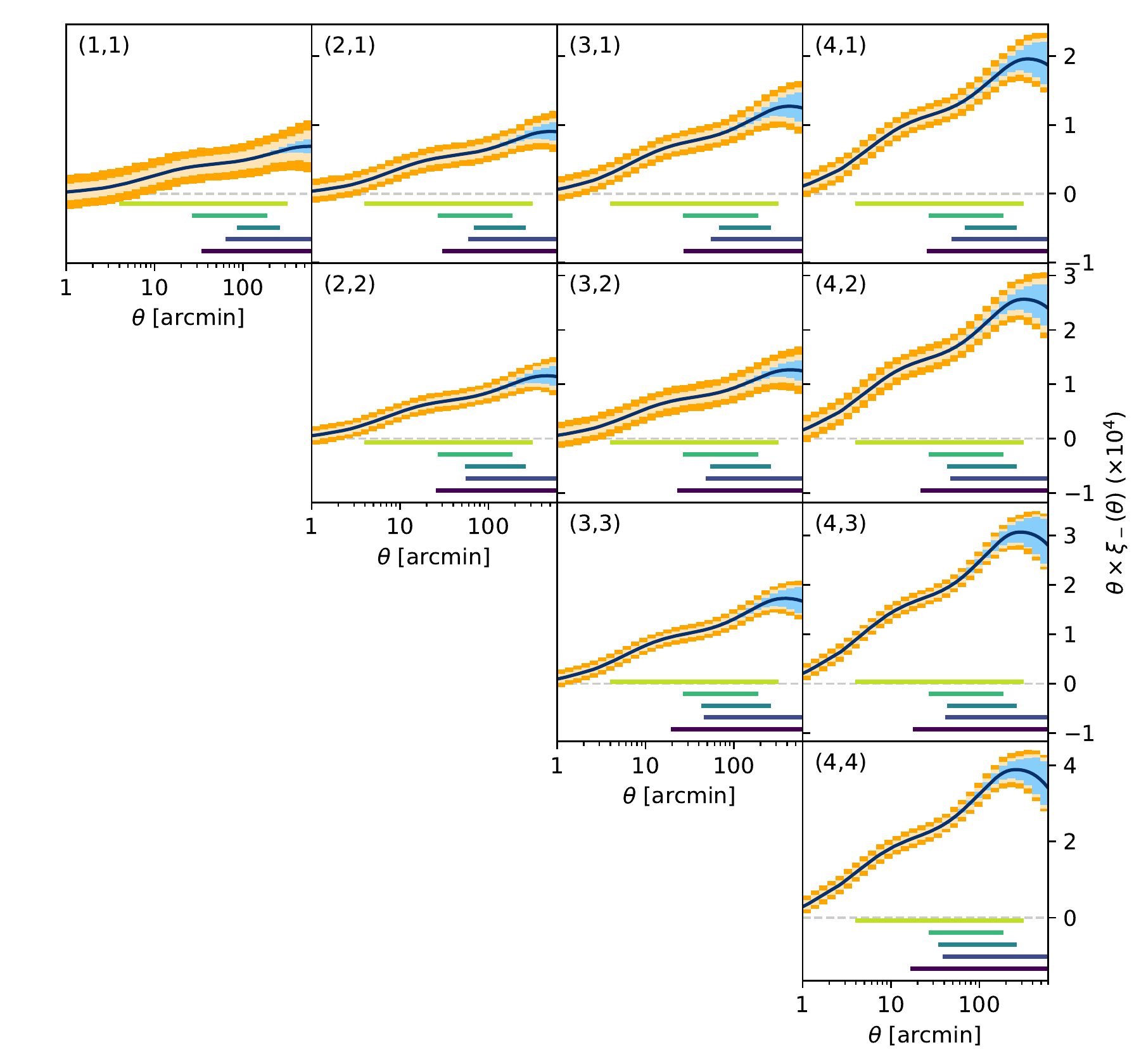}};
    \node at (b.east)[xshift=19mm, yshift=-15mm]{\includegraphics[scale=0.6]{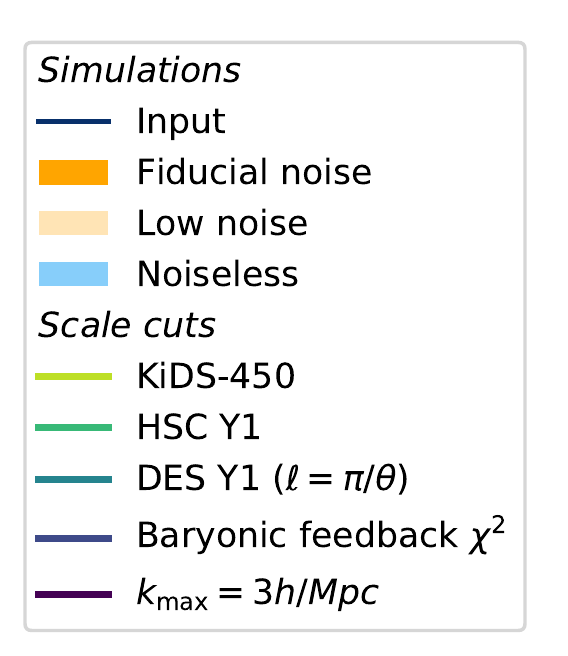}};
    \end{tikzpicture}
    \caption{Two-point functions $\xipm(\theta)$ measured from Gaussian simulations described in \cref{sec:sims} and scale cuts used in the analysis in \cref{sec:results}, similar to \cref{fig:sims_validation_Cl}. The input two-point functions are plotted in dark blue. Boxes show the measured mean and standard deviation in each $\theta$ bin, in orange for the fiducial noise level ($\sigma_e=0.3$), in moccasin for the low noise level ($\sigma_e=0.3/\sqrt{2}$) and in light blue for the noiseless case (where each galaxy shape corresponds to the shear field value). The ranges of scale cuts used in the analysis are shown as solid horizontal lines.}
    \label{fig:sims_validation_xi}
\end{figure*}

\Cref{fig:sims_validation_Cl,fig:sims_validation_xi} show two-point data vectors measured from the 500 simulations we use for the analysis. The input $\Cl$ power spectra are recovered well within 5\% of the error bars at all considered scales\footnote{Note that the sampling noise creates a small bias at very large multipoles $\ell \gtrsim 4000$, likely due to the choice of resolution parameter $\nside=4096$ with our fiducial galaxy density, leading to many empty pixels.}. Corresponding real-space two-point functions are recovered within 5\% and 10\%, respectively, for $\xip$ and $\xim$, for scales larger than \SI{10}{\arcminute}, which is accurate enough for our purposes.
At scales below \SI{10}{\arcminute}, resolution effects of at most 20\% of the error bars for $\xip$ are visible and a simple correction is applied (see below). For $\xim$, these effects are larger, leading us to discard these scales for $\xim$, although these are already cut off in all scale cuts we consider. We therefore apply an additive correction to each simulation corresponding to the difference between the fiducial data vector and the mean measured data vector to correct for small resolution and binning effects, which we approximate to be realization independent. This allows us to use the fiducial data vector as a reference point while not changing the variance of measured data vectors. We verified that the shifts in cosmological parameters arising from this difference are indeed negligible, thus validating \textit{a posteriori} the simulations and this correction.

\section{Fast derivation of mean posterior point using Importance Sampling}
\label{sec:is}

For a given choice of scale cuts, we wish to measure the difference in the posterior of cosmological parameters, for each simulations presented in the previous section, between harmonic and real space two-point statistics measurements. Running two full Markov Chain Monte Carlo (MCMC) analyses for each simulation and each choice of scale cuts is computationally unfeasible. As a result, we employ importance sampling (IS) in this work to rapidly compute, for each simulation, a point estimator of $S_8$ in both harmonic and real space, which we denote $\SeightCl$ and $\Seightxi$ (or simply $\widehat{S_8}$ to mention both). We choose to use \textit{the mean of the posterior} as point estimator, as it is less noisy than the mode of the posterior (or maximum a posteriori) when evaluated from a fixed sample.
Since the likelihood is Gaussian, we can generate a library of theoretical data vectors (for $\Cl$ and $\xipm$) for a sample of cosmological parameters representative of the full
prior space, and use them to rapidly compute importance weights for all simulations and scale cuts.
The IS pipeline and its validation are detailed in the following subsections. We show the distributions of $\widehat{S_8}$ computed from simulations in \cref{fig:is_validation_posterior} and compare it to the posteriors derived from a single (noiseless) fiducial data vector without scale cuts. While expectedly close, these distributions are not mathematically equal, \ie the spread of $\widehat{S_8}$ does not necessarily trace the width of the posterior. Moreover, $\SeightCl$ and $\Seightxi$ are correlated variables---the two-dimensional distribution of which we will study in \cref{sec:results}, see \eg \cref{fig:S8_r_vs_F}.

\subsection{Improved weighted importance sampling}
\label{sec:iwis}

We now provide a brief introduction to the theory of importance sampling. Given a sample of size $n$ of parameters $\qty{\theta_i}_{1 \leq i \leq n}$ from a proposal distribution with density $q$, one can estimate the expectation value of a function $\expval{f(\theta)}_p$ under a target distribution with density $p$ with the estimator
\begin{equation}
    \hat{f}_n= \frac{1}{n}\sum_{1 \leq i \leq n} w_i f(\theta_i),
    \label{eq:is_standard}
\end{equation}
where $w_i \equiv \flatfrac{p(\theta_i)}{q(\theta_i)}$ are ratios of the densities, called importance weights.
For our study, the target distribution is the posterior $p(\theta) \propto \mathcal{L}(X|\theta) \, \pi(\theta)$, where $X$ is either $\Cl$ or $\xipm$ measured from simulations, $\mathcal{L}$ is the likelihood and $\pi$ is the prior. However, the posterior, computed this way, has unknown normalization. Therefore, one needs to normalize the importance weights such that they sum to 1 and then use the weighted average estimator instead of the standard estimator \cref{eq:is_standard}. This operation introduces an order $\order{1/n}$ bias which can be reduced to $\order*{1/n^2}$ by using the improved weighted importance sampling (IWIS) estimator from \citet{Skare:2003kv}, where weights are modified to $ w_i^\prime = \flatfrac{w_i}{S_{-i}}$ with $S_{-i}=\sum_{j \neq i} w_j$ and then normalized, providing final IS weights $\tilde{w}_i=w_i^\prime / \sum_{1 \leq i \leq n} w_i^\prime$.

The efficiency of importance sampling strongly depends on the choice of proposal distribution $q$, \ie how the parameter sample is generated. The effective number of samples is given by ${N^{\rm IS}_{\rm eff}=1/\sum_{1 \leq i \leq n} \tilde{w}_i^2}$. It is bounded by the total number of samples $n$, corresponding to the case where the proposal distribution is equal to the target distribution and all weights are equal to $1/n$. In practice, the closer the proposal distribution $q$ is to the target distribution $p$, the higher the effective number of samples will be, and therefore the more accurate the IWIS estimator is.

\subsection{Cosmological parameter sampling}

For the purposes of this work, the proposal distribution $q$ of cosmological parameters needs to efficiently cover
all regions in parameter space where the posteriors corresponding to each simulations have support.
We detail the choice of proposal distribution in the next section, motivated by posteriors obtained with MCMC from few random realizations of the data vectors.

\subsubsection{Choice of proposal distribution}
\label{sec:IS_q}

We choose to sample all parameters independently, except for $\sigma_8$ and $\Om$. Therefore, the proposal distribution can be factorized into the product of distributions for $(\sigma_8,\Om)$ and all the other parameters.
We adopt a uniform distribution in the $\qty(\sigma_8,\Om)$-plane within a band along the degeneracy observed in weak lensing experiments, which better constrain the combination $S_8\propto\sigma_8\sqrtOm$. To do so, we note that the Jacobian of the transformation ${\qty(\sigma_8,\Om) \rightarrow (S_8,\sqrtOm)}$ is constant, such that we can uniformly sample over a rectangle in the $(S_8,\sqrtOm)$-plane in order to obtain the desired distribution.
We draw $S_8$ in the range $\qty[0.7,0.9]$ and $\sqrtOm$ within $[\sqrt{0.1},\sqrt{0.6}]$ (\ie the prior range of $\Om$). 
We sample over other cosmological parameters ($\Ob$, $h$, $\ns$ and ${\Onu h^2}$) uniformly within their prior bounds specified in \cref{tab:params}.
For intrinsic alignment parameters, we opt for a uniform distribution for the intrinsic alignment tilt parameter $\aia$; however,
we sample the amplitude of intrinsic alignments $\Aia$ with a Gaussian proposal distribution of standard deviation 1.5 and centered at zero, as the prior range is much broader than observed posterior distributions. Finally, we adopt the Gaussian priors over redshift biases $\Delta z_i$'s for the proposal distribution. As explained in \cref{sec:covariance}, we dot not sample shear biases as they are marginalized analytically.
Given these choices, summarized in the fourth column of \cref{tab:params}, the proposal distribution is therefore proportional to the product of redshift bias priors and the proposal distribution of $\Aia$, the other marginal distributions being uniform within their support.

\subsubsection{Sample generation}
\label{sec:lhs}

In order to generate a sample with good space-filling properties, we use optimized Latin Hypercube Sampling (LHS).
More precisely, we first generate $10^6$ samples within the unit hypercube $[0,1]^{12}$ with standard LHS and then optimize its design using the Enhanced Stochastic Evolutionary algorithm\footnote{We use the implementation from the Surrogate Modeling Toolbox python library \citep{SMT2019}, available at \url{https://smt.readthedocs.io/en/latest/_src_docs/sampling_methods/lhs.html}.} \citep{JIN2005268}. This technique operates by exchanging coordinates of points to make the sample closer to uniform (which is formally quantified by a \emph{discrepancy criterion}). This reduces the variance of the IS estimator while maintaining its convergence properties and leaving it unbiased \citep[see][for a derivation]{Packham:2015jf}.
Finally, we apply to the LHS sample the inverse cumulative distribution function of the proposal distribution $q$ to generate the sample of cosmological parameters.

\subsection{Fast derivation of posterior mean}

For each simulation, indexed $j$, from which we obtained a measurement $X_j$ (where $X$ is either $\Cl$ or $\xipm$), we compute the IWIS estimator of the mean under the posterior given by 
\begin{equation}
    \widehat{\theta}_j = \frac{1}{n}\sum_{1 \leq i \leq n} \tilde{w}_{ij} \theta_i,
\end{equation}
where the normalized weights $\tilde{w}_{ij}$ are computed as explained in \cref{sec:iwis} from unnormalized weights, given by
\begin{equation}
    w_{ij} = \frac{\mathcal{L}(X_j|\theta_i) \pi(\theta_i)}{q(\theta_i)}.
\end{equation}
This computation can be accelerated and parralelized by noting that
\begin{equation}
    \log w_{ij} = -\frac{1}{2} \norm{\mathbf{L}X_j - \mathbf{L}X(\theta_i)}^2 + \log \pi(\theta_i) - \log q(\theta_i) + c,
\end{equation}
up to an irrelevant constant $c$, where $\mathbf{L}$ is the Cholesky decomposition of the covariance matrix, ${\mathbf{C}=\mathbf{L}^{\intercal}\mathbf{L}}$. We first compute $\mathbf{L}X_j$ and $\mathbf{L}X(\theta_i)$ for all $j$ and $i$ (slow but parallelizable operations), and then we compute the first term for all pairs $i$ and $j$ (fast operations). The computation thus becomes linear in the number of samples plus the number of simulations---instead of the product---and we can analyze all 500 simulations in under a minute on a single 28-core node.

\label{sec:is_validation}

\begin{figure}
    \centering
    \includegraphics[scale=0.6]{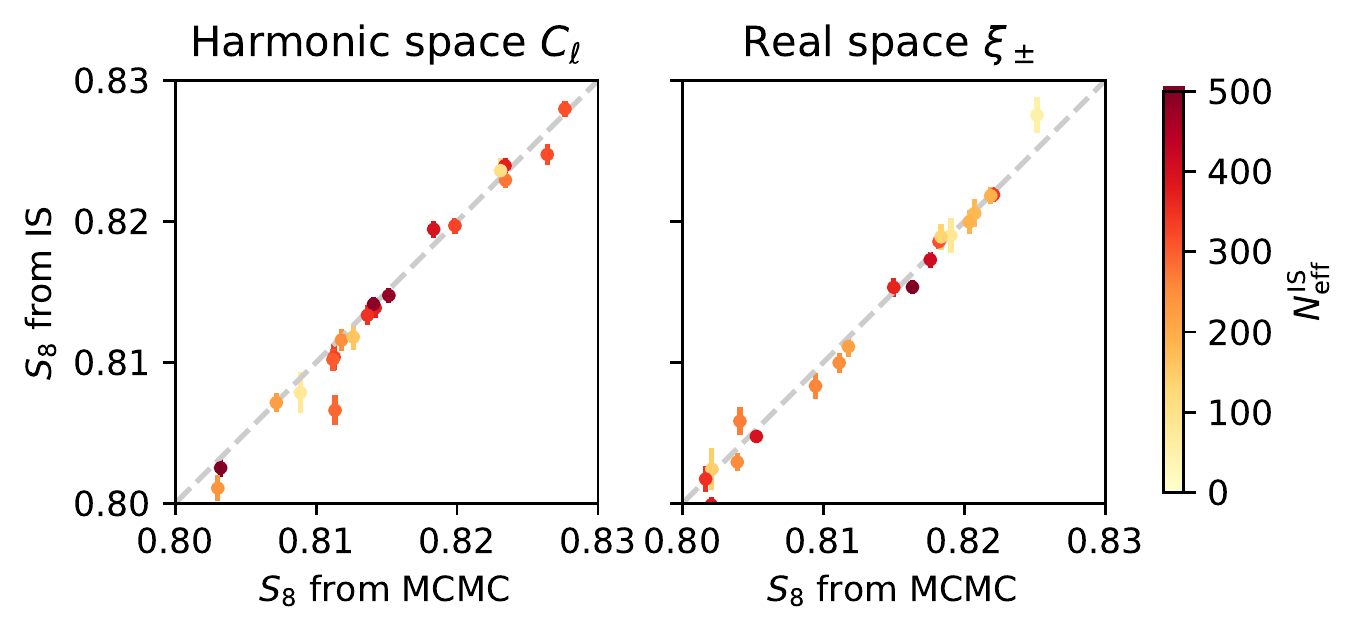}
    \caption{Validation of the importance sampling (IS) pipeline by comparison of estimated mean $S_8$ obtained with standard nested sampling of the posterior (horizontal axis) and with importance sampling (vertical axis) for 20 noisy data vectors. The error bars show the error on the IS estimator, which depends on the effective number of samples, shown by the color of the error bars. This test uses no scale cuts: applying scale cuts increases the effective number of samples and further reduces errors.}
    \label{fig:is_validation}
\end{figure}

To validate the importance sampling pipeline and the choice of proposal distribution, we draw 20 noisy data vector realizations in harmonic and real space (uncorrelated), which were obtained by independently sampling the likelihoods at the fiducial model. We then run a standard nested sampling analysis with \multinest \citep{2009MNRAS.398.1601F} and compare the mean of the parameter posteriors obtained from the nested sampling to those obtained with importance sampling. \Cref{fig:is_validation} shows the comparisom in the case where we use all the measured scales described in \cref{sec:sims}, which represents the most stringent test. In the plot, the points are colored by their effective number of samples. We obtain biases well below 0.5\% for most realizations as seen in \cref{fig:is_validation}, with effective number of samples typically in the few hundreds. Note that for all other cuts, we obtain much higher effective number of samples, typically few thousands for $\kmax=\SI{5}{\h\per\mega\parsec}$ to order $10^4$ for $\kmax=\SI{1}{\h\per\mega\parsec}$, making errors on the posterior means negligible for our purposes.

\begin{figure}
    \centering
    \includegraphics[scale=0.6]{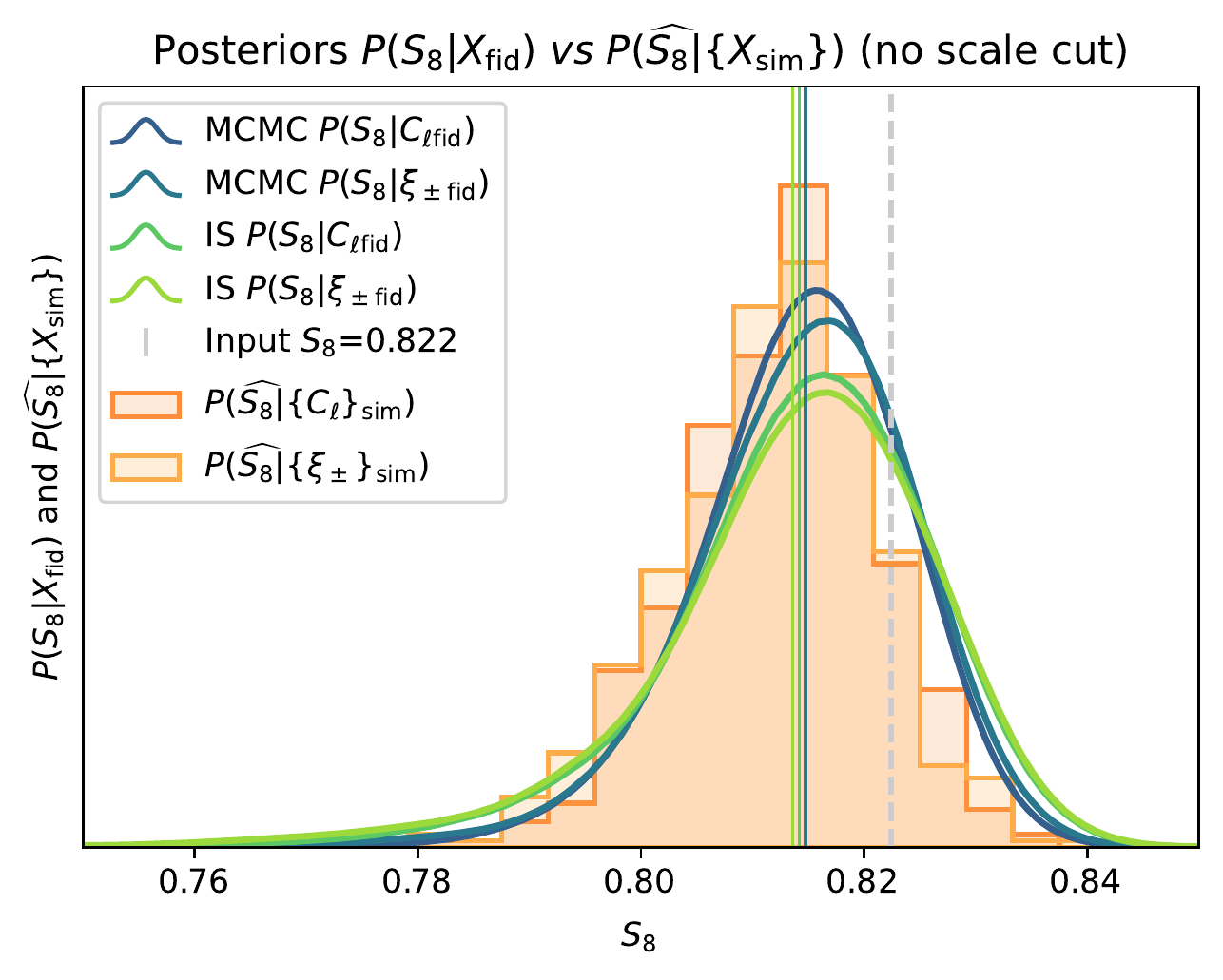}
    \caption{Comparison of marginal posterior distributions on $S_8$ on the fiducial data vectors $P(S_8|X_{\rm fid})$ in harmonic ($X=\Cl$) and real space ($X=\xipm$) without scale cuts, derived from standard MCMC sampling (blue) and IS (green). Colored vertical lines (mostly overlapping) indicate the mean under each posterior and the black vertical line shows the input $S_8$ value.
    In orange, we show the histogram of mean posterior points $\widehat{S_8}$ derived from $\Cl$ and $\xipm$ measurements for 500 simulations, see \cref{sec:results}. The gray dashed line show the input $S_8$ value.}
    \label{fig:is_validation_posterior}
\end{figure}

As a complementary test, we also compare the full shape of the marginalized posterior derived from nested sampling and importance sampling, shown in \cref{fig:is_validation_posterior} for the fiducial, noiseless data vector, in both spaces and without scale cuts. In this case, the importance sampling pipeline recovers the mean with an accuracy of 0.1\% and the width within 10\%\cd{, with very similar results in harmonic and real space}. For comparison, we also show the distribution of mean posterior points evaluated from simulations (orange histograms).

\section{Results}
\label{sec:results}

This section presents our main results. In \cref{sec:posterior_shifts}, we probe the joint distribution of $S_8$ estimators, $\SeightCl$ and $\Seightxi$, derived from measurements of two-points statistics in harmonic and real space, on 500 simulated mock DES Y3 surveys, for different scale cuts. In particular, we measure the Pearson correlation coefficients of these estimators and the scatter of their difference to gauge the expected consistency of harmonic and real space analyses. In \cref{sec:sys}, we compute the biases ${\Delta\widehat{S_8}=\Seightxi-\SeightCl}$ due to various potential theoretical, astrophysical and observational residual systematic uncertainties and compare it with the scatter measured from statistical fluctuations, $\sigma(\Delta\widehat{S_8})$. Finally, we apply scale cuts from published analyses of HSC and KiDS-450 data and discuss observed discrepancies in $S_8$ in \cref{sec:res_hsc_kids}.
Although we partially focus our analysis on $S_8$, we also report results on $\sigma_8$ and $\Om$, with figures in \cref{sec:app_s8_Om}.

Before describing our results, we draw attention to important features of the marginalized $S_8$ posterior. As can be seen in \cref{fig:is_validation_posterior}, it is slightly asymmetric towards lower values, and both the mode and the mean are biased low, with respect to the input $S_8$ value, on a fiducial data vector. This \textit{projection effect} is expected when a high-dimensional posterior, with its associated degeneracies and prior boundaries, is projected onto one dimension. Similar trends were observed with other cosmic shear analyses \citep[\eg,][]{2020arXiv200701844J}. As a consequence, our measurements of posterior means generally appear to be biased low with respect to the input $S_8$ parameter (similarly, $\sigma_8$ is biased low and $\Om$ high). However, this problem pertains to the choice of point estimate---\eg, the mode, median, or, like here, the mean of the marginal posterior---and we find similar trends in both harmonic and real space. Moreover, the truth value consistently lies within the $\sim1\sigma$ interval of the posterior. Therefore, this apparent bias does not interfere with the question of the consistency between analyses, as long as we relate to the mean of the posterior evaluated from the fiducial data vector as a reference point, and shifts thereof.

\begin{figure*}
    \centering
    \begin{tikzpicture}
    \node(a){
        \includegraphics[scale=0.6, trim=0 0 0 0, clip]{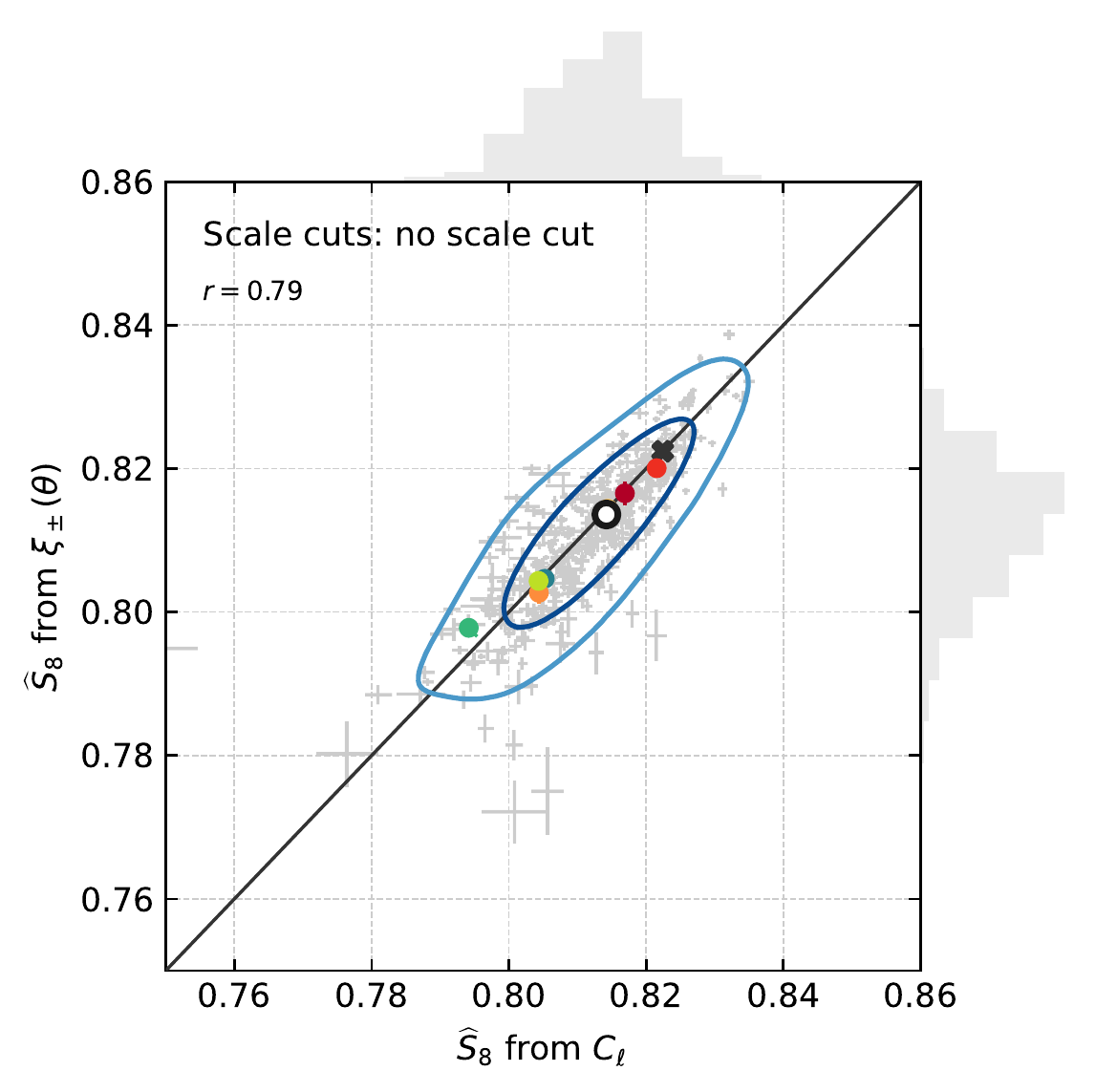}
    };
    \node at (a.north east) [anchor=north west] {
        \includegraphics[scale=0.6, trim=0 0 0 0, clip]{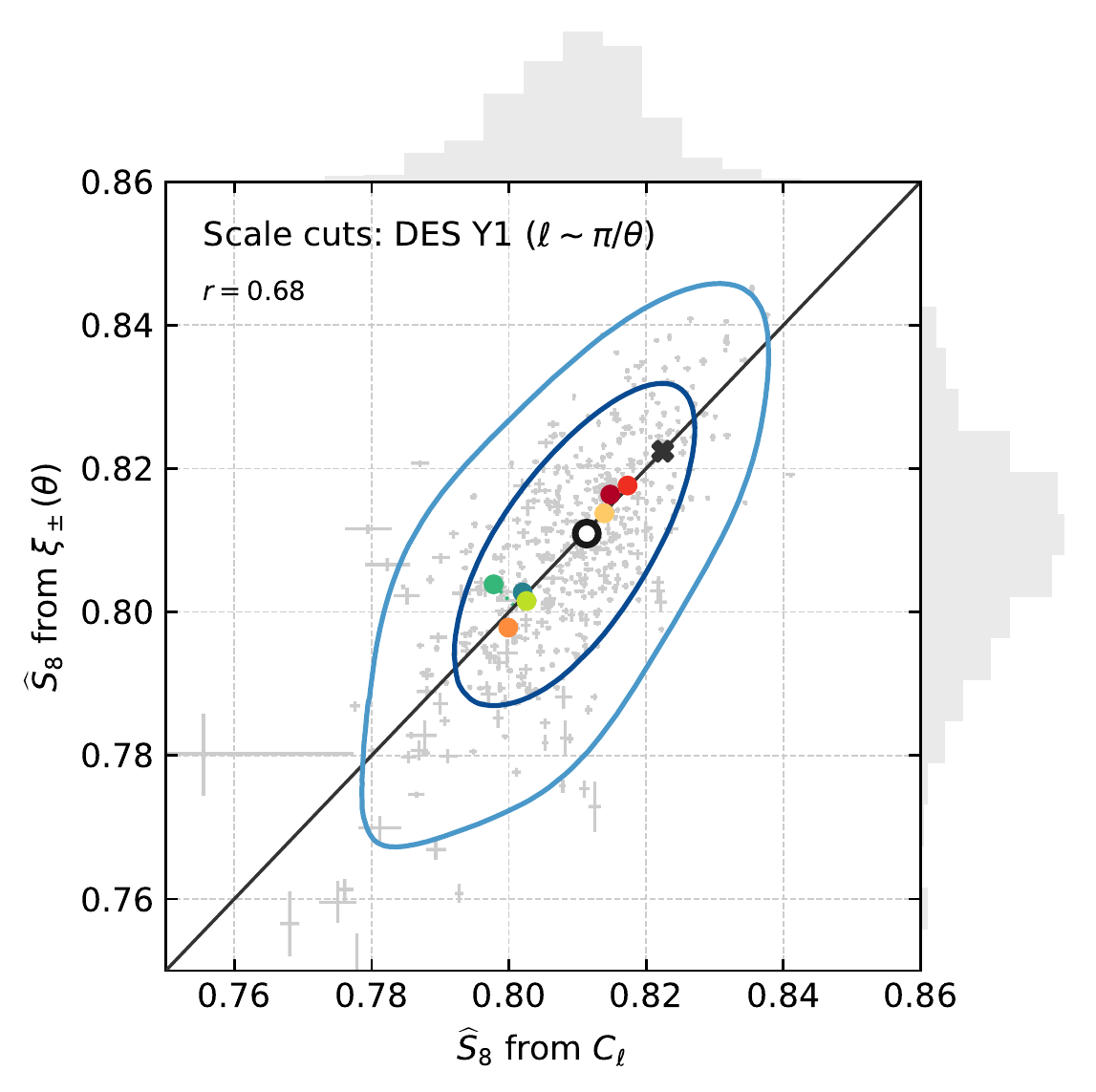}
    };
    \node at (a.south west) [anchor=north west] {
        \includegraphics[scale=0.6, trim=0 0 0 0, clip]{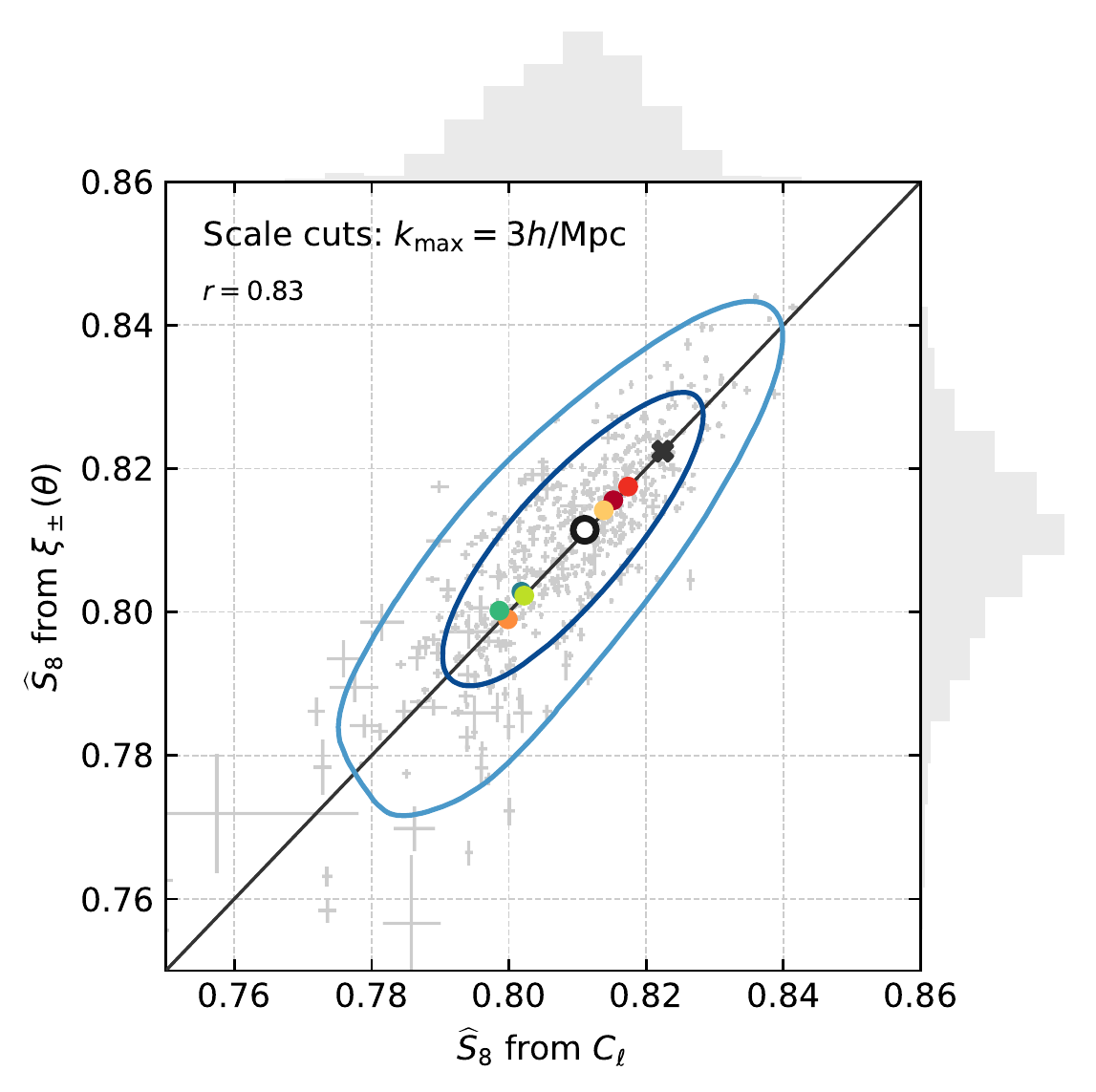}
    };
    \node(d) at (a.south east) [anchor=north west] {
        \includegraphics[scale=0.6, trim=0 0 0 0, clip]{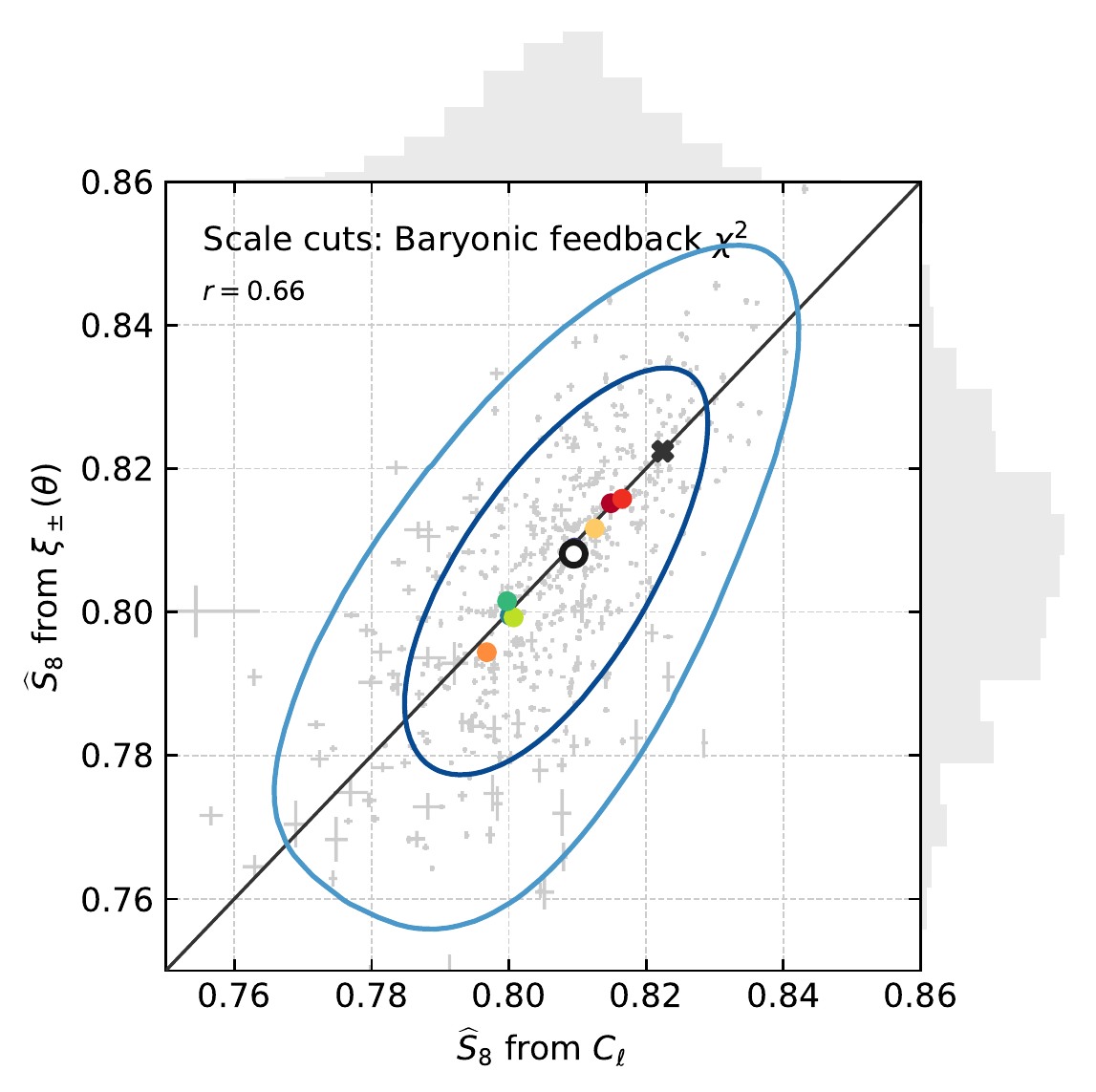}
    };
    \node at (d.north east)[anchor=center, xshift=10mm, yshift=0mm]{
        \includegraphics[scale=0.6]{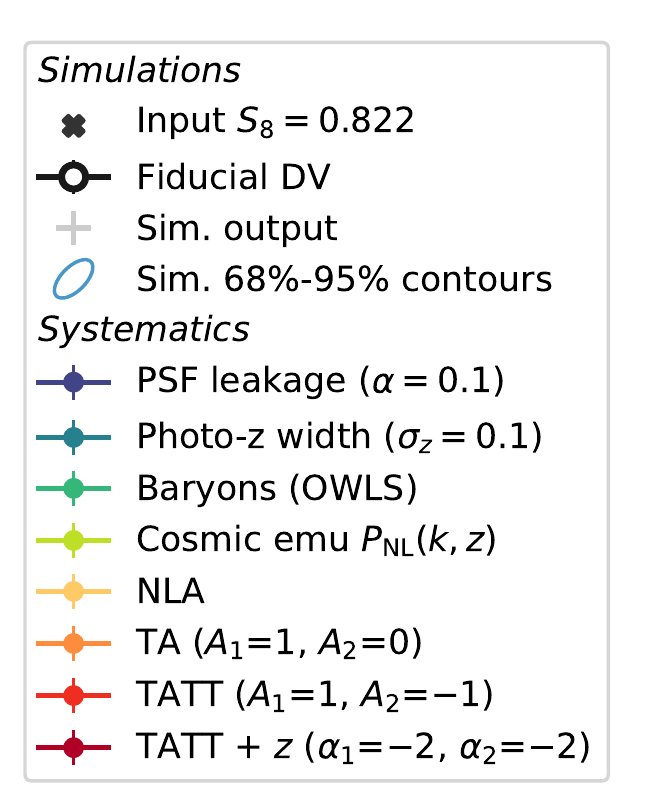}};
    \end{tikzpicture}
    \caption{
    Distribution of estimated $\widehat{S_8}$ from real space (vertical axis) \vs harmonic space (horizontal axis) analyses, and biases induces by unmodeled systematic effects. Each plot corresponds to a different scale cut indicated in the upper left. The scattered gray points show the mean of the posteriors of the Gaussian simulations described in \cref{sec:sims} in both real and harmonic space, as well as the error associated with the IS estimator (mostly indistinguishable for realistic scale cuts). \cd{Note these error bars denote the uncertainty on the mean of the posteriors rather than their width.} Blue contours show the 68\% and 95\% contours of these points. Their Pearson correlation coefficient $r$ is reported in the upper left. Marginal distributions are shown by the gray (unnormalized) histograms above and to the right of each panel.
    The mean of the posterior for the fiducial, noiseless data vector is shown by the white dot encircled in black.
    Data vectors contaminated with various unmodeled systematic effects are analyzed in the same way and results are shown by the colored points.
    The value of $S_8$ used as input is shown by the cross. The result of the analysis of the fiducial data vector provides a reference point for the estimator, which is expected to differ from the input due to projection effects (while remaining consistent in terms of the width of the posterior).}
    \label{fig:S8_r_vs_F}
\end{figure*}

\subsection{Estimated parameter differences between harmonic and real space cosmic shear}
\label{sec:posterior_shifts}

We now apply the importance sampling methodology described in \cref{sec:is} to compute the mean posterior points $\SeightCl$ and $\Seightxi$ for each simulation using harmonic and real-space measurements described in \cref{sec:sims}, for different scale cuts described in \cref{sec:scales}.

\begin{figure*}
    \centering
    \includegraphics[scale=0.6]{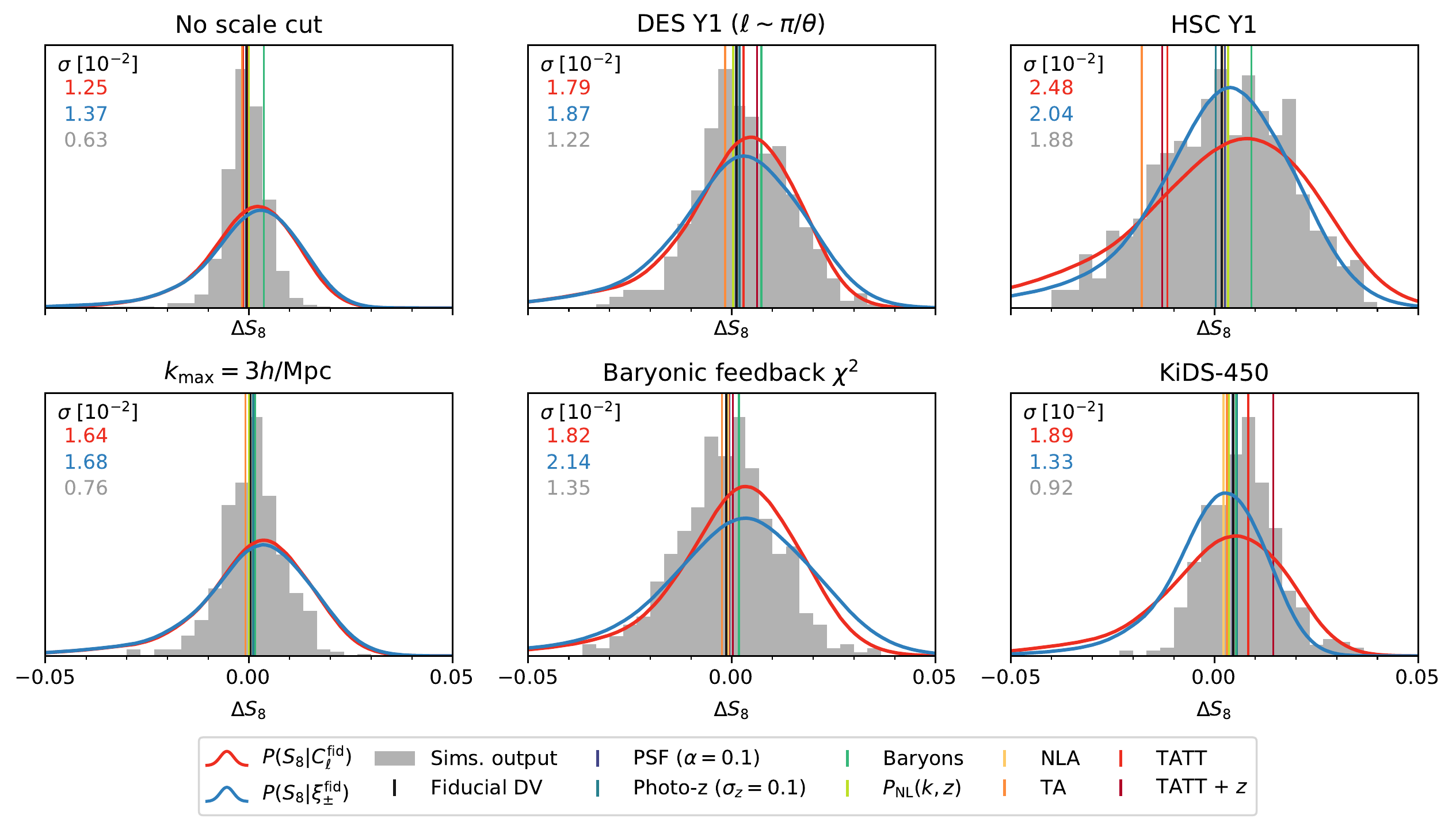}
    \caption{Distribution of differential biases ${\Delta \widehat{S_8} \equiv \Seightxi-\SeightCl}$ between real space and harmonic space analyses.The histograms show the posterior mean difference derived from simulations. They are compared to individual posteriors, $P(S_8|X)$ with $X=\Cl,\xipm$, derived from fiducial data vectors and shifted with respect to their own mean $S_8$. Each panel corresponds to a different scale cut used in this analysis, as indicated above. The standard deviations (multiplied by 100) of the $\Cl$ posterior (red), $\xipm$ posterior (blue) and ${\Delta \widehat{S_8}}$ histograms (gray) are reported in the upper left of each panel. Vertical lines indicate the differential biases computed for data vectors contaminated with unmodeled systematics.}
    \label{fig:deltaS8vsPost}
\end{figure*}

We consider four fiducial scale cuts: \ref{it:sc_i} no scale cuts, \ie using all measured scales from the simulations ($\ellmax=8192$, $\theta_{\min}^+=\SI{1}{\arcminute}$ and $\theta_{\min}^-=\SI{10}{\arcminute}$) \ref{it:sc_ii} using DES Y1 real-space cuts converted to harmonic space with the relation $\ell = \pi/\theta$, \ref{it:sc_iii} a $\kmax$ type cut with a threshold at \SI{3}{\h\per\mega\parsec}, and \ref{it:sc_iv} cuts derived from baryonic feedback contamination. The baryonic feedback cut is the most conservative choice, while the $\kmax$ cut is (relatively) more aggressive, and the DES Y1 ($\ell=\pi/\theta$) lies in between (see \cref{fig:sims_validation_Cl,fig:sims_validation_xi}). Results are shown in \cref{fig:S8_r_vs_F}, where we show with gray error bars the mean of the posterior for each simulation, and in blue the 68\% and 95\% contours of the distribution sampled by these points. Note that these contours do not necessarily reflect the width of the posteriors themselves, but rather how much posteriors may shift for multiple realizations of the cosmic shear measurements at fixed cosmology.
We quantify the statistical discrepancy between harmonic and real-space cosmic shear analyses by characterizing the distribution of the difference between mean posterior points, $\sigma(\Delta\widehat{S_8})$, derived from harmonic and real-space measurements. We show histograms of $\Delta\widehat{S_8}$ in \cref{fig:deltaS8vsPost}, for the four sets of scale cuts used above (as well as HSC Y1 and KiDS-450 scale cuts, discussed in \cref{sec:res_hsc_kids}). We measure its standard deviation as well as the Pearson correlation coefficients, $r$, between $\SeightCl$ and $\Seightxi$, which are reported in \cref{tab:disc}. If the mean posterior shifts from individual statistics have the same spread ($\sigma(\widehat{S_8}\vert_{X})$ for $X=\Cl,\xipm$, also reported for reference), the two metrics measure the same quantity, but when the constraints change, the correlation coefficient $r$ is likely a better measure of the common information used in the two statistics. The spread, however, is useful to determine how likely a certain observed difference between parameters measured by the two statistics is, \ie an observed difference larger than this number would be indicative of a tension between the two analyses.
We proceed similarly for $\Om$ and $\sigma_8$, see \cref{fig:sigma8_r_vs_F,fig:Om_r_vs_F}.

Quantitatively, we find that, using all scales available in simulations, the standard deviation of  $\SeightCl$ and $\Seightxi$, $\sigma(\widehat{S_8}\vert_{X})$ is 0.010 for both statistics ($X=\Cl,\xipm$). The spread of the difference is ${\sigma(\Delta\widehat{S_8})=0.007}$, for the fiducial noise level and DES Y3-like survey characteristics. These are lower bound on the error that can be reached with DES Y3 data.
For realistic scale cuts, we find that $\sigma(\widehat{S_8}\vert_{X})$ increases similarly for both statistics while $\sigma(\Delta\widehat{S_8})$ varies significantly across scale cuts (by a factor of almost 2), as seen in \cref{fig:S8_r_vs_F,fig:deltaS8vsPost}. We find significantly smaller variations for $\sigma(\Delta\widehat{\Om})$ and $\sigma(\Delta\widehat{\sigma_8})$.
In particular, we observe that the $\kmax$ cut provides the highest correlation coefficient at about 83\%, with similar spread in either space. For this cut, we find $\sigma(\SeightCl)\equiv\sigma(\Seightxi)=0.013$ and $\sigma(\SeightCl - \Seightxi)=0.0076$. For the {DES~Y1 ($\ell \sim \pi/\theta$)} cut, which by comparison of $\tmin$ is close to a $\kmax$ cut between 2 and \SI{3}{\h\per\mega\parsec}, we find a lower correlation coefficient of 68\%, indicating that the simplistic conversion fails at capturing the same information in each estimator. Moreover, posterior means are slightly more scattered in real space, consistent with the fact that, for these scale cuts, the fiducial signal-to-noise ratio is lower in real space (see \cref{tab:all_cuts}). Similarly, the baryonic feedback cut that we have applied here yields asymmetric results with a tail for lower values in real space, as expected from the difference in predicted signal-to-noise ratio (47 for $\Cl$ and 40 for $\xipm$), and a correlation coefficient of 68\%.
Overall, we measure $\sigma(\Delta\widehat{S_8})$ to be a fraction $\sim\numrange{0.6}{0.9}$ of the scatter for individual statistics, which is to be compared to $\sqrt{2}\approx1.4$ for fully uncorrelated estimators.
For $\sigma_8$ and $\Om$, we observe a similar trend for the scatter of the difference to increase for stricter cuts. However, the correlation coefficients $r$ are much less sensitive to scale cuts, which we interpret as a consequence of the strong, banana-shaped degeneracy between $\sigma_8$ and $\Om$ visible for all scale cuts---the difference in posterior means between statistics is subdominant to the variance of posterior means across realizations.

\begin{figure}
    \centering
    \includegraphics[scale=0.6]{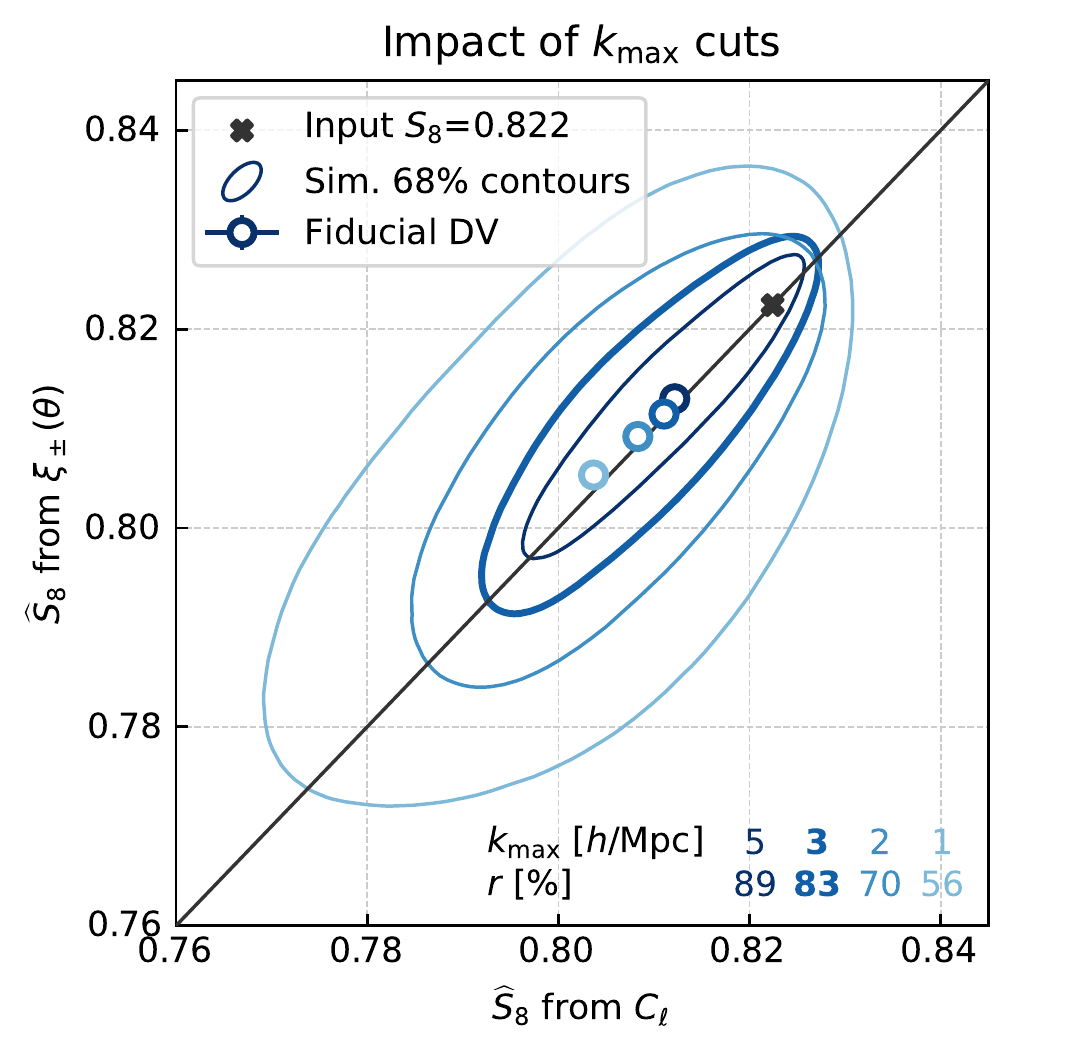}
    \caption{
    Impact of the $\kmax$ cut on the distribution of $\widehat{S_8}$. We show the 68\% contours, similar to the lower left panel of \cref{fig:S8_r_vs_F}, for $\kmax$ scale cuts, varying $\kmax$ from 1 to \SI{5}{\h\per\mega\parsec} (light to dark blue).
    The values of $\widehat{S_8}$ derived from the fiducial data vector are shown by the white dots encircled in blue.
    }
    \label{fig:S8_kmax}
\end{figure}

\begin{figure}
    \centering
    \includegraphics[scale=0.6]{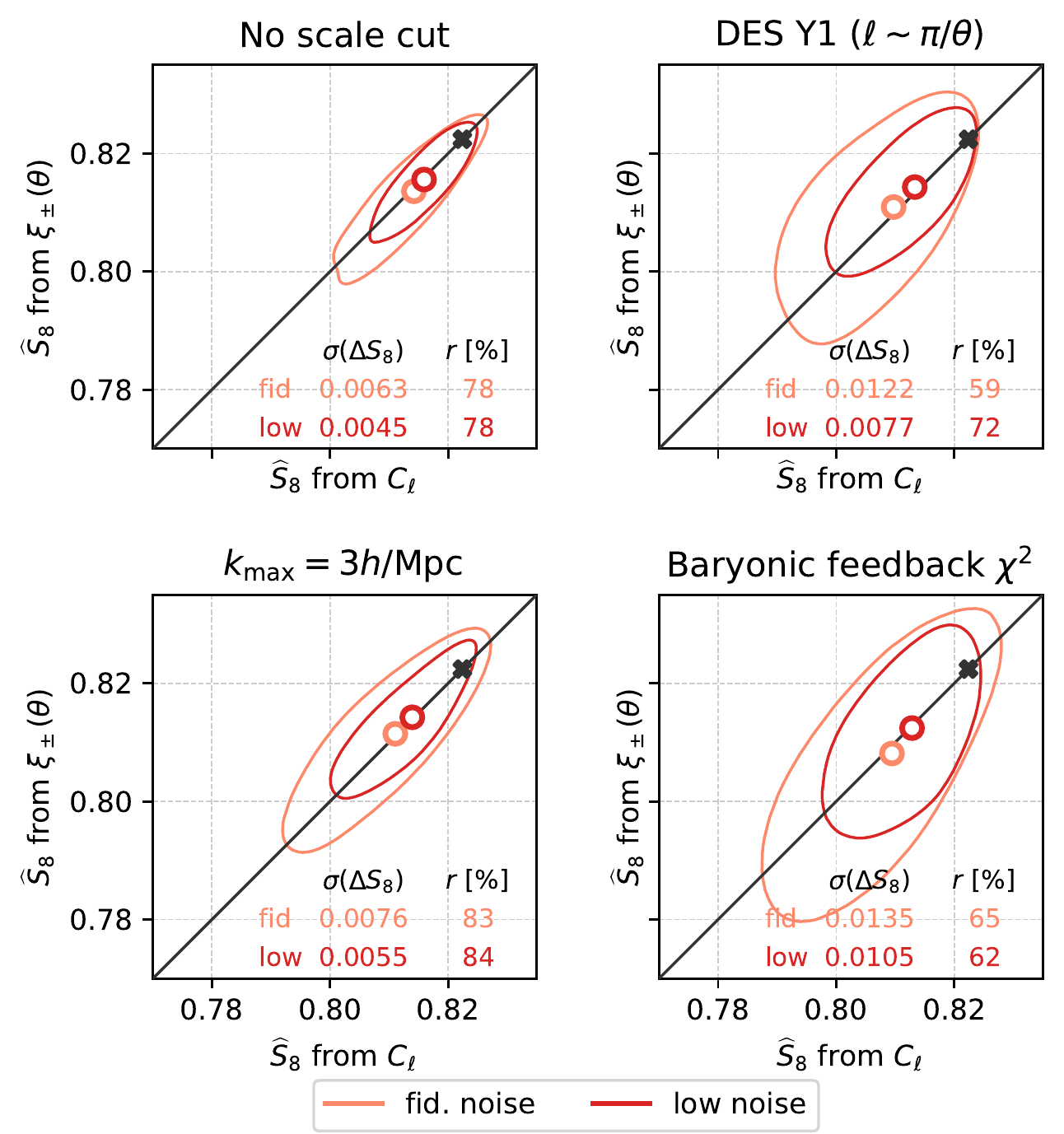}
    \caption{Impact of shape noise on the distribution of $\widehat{S_8}$. The low noise contours are obtained by dividing the variance of intrinsic ellipticities $\sigma_e^2$ by two in the simulations, which is, with respect to the covariance matrix, equivalent to a deeper survey with doubled galaxy density (see also \cref{fig:sims_validation_Cl,fig:sims_validation_xi}). The values of $\widehat{S_8}$ derived from the fiducial data vector are shown by the white dots encircled in red.}
    \label{fig:S8_noise}
\end{figure}

We proceed to further explore the distribution of parameters shifts and attempt to gain insight into the generalization of our results to the next generation of weak lensing surveys. We perform two tests, one where we repeat measurements for $\kmax$-type cuts for different values of $\kmax$ in the range \SIrange{1}{5}{\h\per\mega\parsec}, which is an indication of the confidence in small-scale modeling, and one where we emulate a deeper survey by reducing shape-noise.
In \cref{fig:S8_kmax}, we plot the 68\% contours obtained when varying $\kmax$ from 5 down to \SI{1}{\h\per\mega\parsec}. We first observe that these cuts return very similar constraints in both spaces, indicating that, they are well-performing and physically motivated choices, if consistency is desired. We find that $\sigma(\SeightCl - \Seightxi)$ goes from 0.0049 to 0.019, with a correlation coefficient going from 89\% to 56\%, as expected when decreasing $\kmax$ from 5 to \SI{1}{\h\per\mega\parsec}. We find that the posterior mean derived from the fiducial data vectors shifts towards lower values for decreasing $\kmax$, albeit with negligible differential biases.
In \cref{fig:S8_noise}, we show 68\% contours obtained when decreasing the noise level, which here we achieve by dividing the variance of intrinsic ellipticities $\sigma_e^2$ by two (while keeping the density fixed) in simulations. Contours shrink, as expected, towards the input value. Both the spread of the posterior means and the spread of the difference decrease, though we note that the amount to which they decrease with respect to one another depends on the choice of scale cut, making it difficult to separate the effect of noise and scale cuts on discrepancies in the harmonic and real space analyses. In other words, shape-noise acts as an effective cut-off at small scales where it dominates the signal.

We expect that numerical results presented here will not change dramatically for the real DES Y3 data. At the time of conducting this work, the DES Y3 catalogs and pipeline were not yet finalized, but the number density, footprint and redshift distributions we used do capture the essential properties of the final DES Y3 data.

\begin{table*}
    \centering
    \begin{tabular}{l l c c c c c c}
         & & \multicolumn{6}{c}{Statistics} \\
        \cmidrule(r){3-8}
        Parameter ($\theta$) & Scale cut & $\sigma\left(\theta\left|C_\ell^{\rm fid}\right.\right)$ & $\sigma\left(\theta\left|\xi_\pm^{\rm fid}\right.\right)$ & $\sigma\left(\widehat{\theta}\vert_{\Cl}\right)$ & $\sigma\left(\widehat{\theta}\vert_{\xipm}\right)$ & $\sigma\left(\Delta\widehat{\theta}\right)$ & Pearson $r$ \\
        \midrule
        \multirow{4}{*}{$S_8$}
        & (i) No scale cut & 0.013 & 0.014 & 0.010 & 0.010 & 0.006 & 0.79 \\
        & (ii) DES Y1 ($\ell\sim\pi/\theta$) & 0.016 & 0.019 & 0.012 & 0.014 & 0.011 & 0.68 \\
        & (iii) $\kmax=\SI{3}{\h\per\mega\parsec}$ & 0.016 & 0.017 & 0.013 & 0.013 & 0.008 & 0.83 \\
        & (iv) Baryonic feedback $\chi^2$ & 0.018 & 0.021 & 0.014 & 0.018 & 0.014 & 0.66 \\

        \midrule
        \multirow{4}{*}{$\sigma_8$}
        & (i) No scale cut & 0.058 & 0.060 & 0.050 & 0.047 & 0.031 & 0.79 \\
        & (ii) DES Y1 ($\ell\sim\pi/\theta$) & 0.070 & 0.070 & 0.056 & 0.053 & 0.041 & 0.71 \\
        & (iii) $\kmax=\SI{3}{\h\per\mega\parsec}$ & 0.071 & 0.064 & 0.057 & 0.052 & 0.036 & 0.79 \\
        & (iv) Baryonic feedback $\chi^2$ & 0.077 & 0.077 & 0.063 & 0.061 & 0.046 & 0.73 \\
        
        \midrule
        \multirow{4}{*}{$\Om$}
        & (i) No scale cut & 0.049 & 0.054 & 0.042 & 0.043 & 0.026 & 0.81 \\
        & (ii) DES Y1 ($\ell\sim\pi/\theta$) & 0.063 & 0.063 & 0.049 & 0.046 & 0.033 & 0.76 \\
        & (iii) $\kmax=\SI{3}{\h\per\mega\parsec}$ & 0.063 & 0.059 & 0.050 & 0.046 & 0.029 & 0.82 \\
        & (iv) Baryonic feedback $\chi^2$ & 0.069 & 0.068 & 0.053 & 0.050 & 0.034 & 0.78 \\
    \end{tabular}
    \caption{Comparison of the scatter of parameter estimators between harmonic and real space two-point statistics. For each parameter $\theta=S_8,\sigma_8,\Om$ and two-point statistics $X=\Cl,\xipm$, we report the scatter of the mean of the posterior $\sigma(\widehat{\theta}\vert_{X})$ compared to the width of the posterior for the fiducial data vector $\sigma(\theta|X^{\rm fid})$ for power spectra and correlation functions, the scatter of the difference of posterior means between the two statistics $\sigma(\Delta\widehat{\theta})$ (\textit{penultimate column}) and the Pearson correlation coefficient $r$ of posterior means (\textit{last column}). If $\widehat{\theta}\vert_{\Cl}$ and $\widehat{\theta}\vert_{\xipm}$ were independent random variable with equal variance, we would have $\sigma(\Delta\widehat{\theta})=\sqrt{2}\sigma(\widehat{\theta}|_X)$.}
    \label{tab:disc}
\end{table*}

\subsection{Bias from systematic effects}
\label{sec:sys}

We now compute posterior shifts due to theoretical and astrophysical uncertainties. As argued in the introduction, different effects may impact $\Cl$ and $\xipm$ measurements differently, resulting in non-equal shifts in posterior distributions, thus creating a \textit{differential} bias. Although it is difficult to assess all potential biases and their interplay, we nonetheless propose to measure individual (differential) biases from a selection of systematics, representative of theoretical and observational uncertainties pertaining to the current generation of weak lensing surveys. To do so, we apply the IS analysis pipeline to noiseless theory data vectors computed with varying modeling assumptions,
or contaminated with spurious signals,
as detailed in \cref{sec:sys_th}. Results are overlaid in \cref{fig:S8_r_vs_F,fig:deltaS8vsPost} for comparison with expected shifts from statistical fluctuations described in the previous section. In particular, we show the result of analyzing the following data vectors (we indicate in italic the label used in the plots):
\begin{itemize}
    \item \textit{Fiducial DV.} Noiseless data vector computed from the baseline model detailed in \cref{sec:model} with the fiducial cosmology (see \cref{tab:params}).
    \item \textit{PSF leakage ${\alpha=0.1}$.} Data vector computed from the fiducial data vector with an additive bias measured from PSF elliptiticies and residuals, with a leakage fraction $\alpha=0.1$.
    \item \textit{Photo-z width (${\sigma_z=0.10}$).} Noiseless data vector computed at the fiducial cosmology but with redshift distributions convolved by a Gaussian kernel of width $\sigma_z=0.1$.
    \item \textit{Baryons (OWLS).} Noiseless data vector computed at the fiducial cosmology with a power spectrum including small-scale rescaling due to baryonic feedback, see \cref{eq:baryon_Pk_ratio}.
    \item \textit{Cosmic emu ${P_{\rm NL}(k,z)}$.} Noiseless data vector computed with a non-linear matter power spectrum modelled with the Mira-Titan emulator, as opposed to the \halofit prescription.
    \item \textit{NLA.} Noiseless data vector computed from the baseline model at the fiducial cosmology, but with a non-zero amplitude ($\Aia=1.5$) of the intrinsic alignments (IA), assuming the fiducial NLA model. This is a check that the cosmology is not significantly affected by intrinsic alignments when the model is correct. We note that the small shifts with respect to the fiducial data vector indicate that projection effects---that shift the posterior mean from input parameter values---are somewhat dependent on the input parameters.
    
    \item \textit{TA (${A_1=1}$, ${A_2=0}$).} Noiseless data vector computed at the fiducial cosmology but where we switched the IA model to TATT. For this first TATT data vector, we only include the TA component, which includes the additional density-tidal field contribution, with respect to NLA. Current constraints on TATT amplitudes $A_{1,2}$ and redshift-dependence parameters $\alpha_{1,2}$ remain fairly weak. However, \citet{2019MNRAS.489.5453S} showed that all four parameters are of order unity and that DES Y1 data favor $A_1>0$, $A_2<0$ and show a mild preference for $\alpha_{1,2}<0$. We therefore use $A_1=1$ and $A_2=0$ here, and no redshift dependence, \ie $\alpha_{1,2}=0$.
    
    \item \textit{TATT (${A_1=1}$, ${A_2=-1}$).} Noiseless data vector computed at the fiducial cosmology with the TATT model, incorporating the TT contribution with $A_1=1$ and $A_2=-1$, and no redshift dependence, \ie $\alpha_{1,2}=0$.

    \item \textit{TATT + $z$ (${\alpha_1=-2}$, ${\alpha_2=-2}$).} Noiseless data vector computed at the fiducial cosmology with the TATT model with both TA and TT contributions and redshift dependence with parameters $\alpha_{1,2}=-2$.
    
\end{itemize}

Biases should be compared with the mean posterior point from the analysis of the fiducial data vector ("Fiducial DV") and differences between harmonic and real space should be measured perpendicular to the gray diagonal (and, strictly speaking, multiplied by $\sqrt{2}$).
For the three fiducial scale cuts, and given our DES Y3-like setup, we find most systematics tested here yield estimate $S_8$ well within 68\% contours in the $(\SeightCl,\Seightxi)$ plane and that the differences between harmonic and real space biases are typically within $\sim0.5\sigma(\Delta\widehat{S_8})$, displaced along the diagonal.
The largest bias comes from baryonic feedback, which lies beyond the 68\% contour when no scale cuts are applied and approaches it for the "DES ($\ell\sim\pi/\theta$) cut. Then, smaller biases are found for the TA model and non-linear power spectrum, and, to a lesser extent, data vectors contaminated with photo-$z$ width---all of which result in biases of about 0.01 for each estimator, but generally negligible differential biases.
When comparing TA and TATT models, we find, somewhat counter-intuitively, that the model including TT contributions yield a smaller bias, on both statistics, than the TA-only model. This depends on details of how the NLA model is able to mimic TA and TT contributions and absorb the non-cosmological shear signal. However, we note that the choice of a negative $A_2$ reduces the overall IA contamination, especially on small scales, and that the TT part is likely canceling part of the beyond NLA contributions in TA.
We conclude that the various effects we examined impact the two statistics similarly for the three fiducial scale cuts we show and do not bias one over the other for a DES Y3-like configuration. This indicates that these scale cuts capture sufficiently similar information for those systematics to have basically the same effect on both two-point statistics, with regard to $S_8$ estimation.
For $\sigma_8$ and $\Om$ (see \cref{fig:sigma8_r_vs_F,fig:Om_r_vs_F}), baryons do create a differential bias when no cut is applied, but are well controlled with realistic cuts. On the other hand, intrinsic alignments including TA and TT contributions, in particular with redshift dependence, create a mild differential bias, pushing $\sigma_8$ high and $\Om$ low for harmonic space compared to real space. A possible explanation is that the redshift dependence, once the shear field is projected, becomes a scale-dependent effect which impacts both statistics differently.
The contamination from PSF leakage and NLA (which is included in the model) are found to be negligible for all cuts.

In \cref{fig:S8_kmax}, we vary the $\kmax$ cut and find that, for most systematics, decreasing $\kmax$ does not create significant differential biases. We observe that $\widehat{S_8}$ moves similarly for the fiducial data vector and for most contaminated data vectors, although these are not shown to maintain readiblity.
These shifts are therefore likely due to the lesser information content at $\kmax=\SI{1}{\h\per\mega\parsec}$, combined with projection effects.
When decreasing shape-noise in \cref{fig:S8_noise}, we see two countereffects: with lower noise, the posterior tightens closer to the true input value, but the reweighting of small scales increases some biases---particularly baryons and IA--- beyond the tightened 68\% contours for the most aggressive cuts.

We note that, by testing effects one at a time, we cannot probe all their combinations and how they add up to increase biases, or reversely, cancel each other out.
Finally, we point out that for all systematic effects tested here, the IS pipeline provides accurate estimates, characterized by effective numbers of samples in the range $10^3-10^4$. We ran standard MCMC chains for a limited number of cases and found excellent good agreement for all of them.

\subsection{Comparison with previous work}
\label{sec:res_hsc_kids}

\begin{figure}
    \centering
    \includegraphics[scale=0.6]{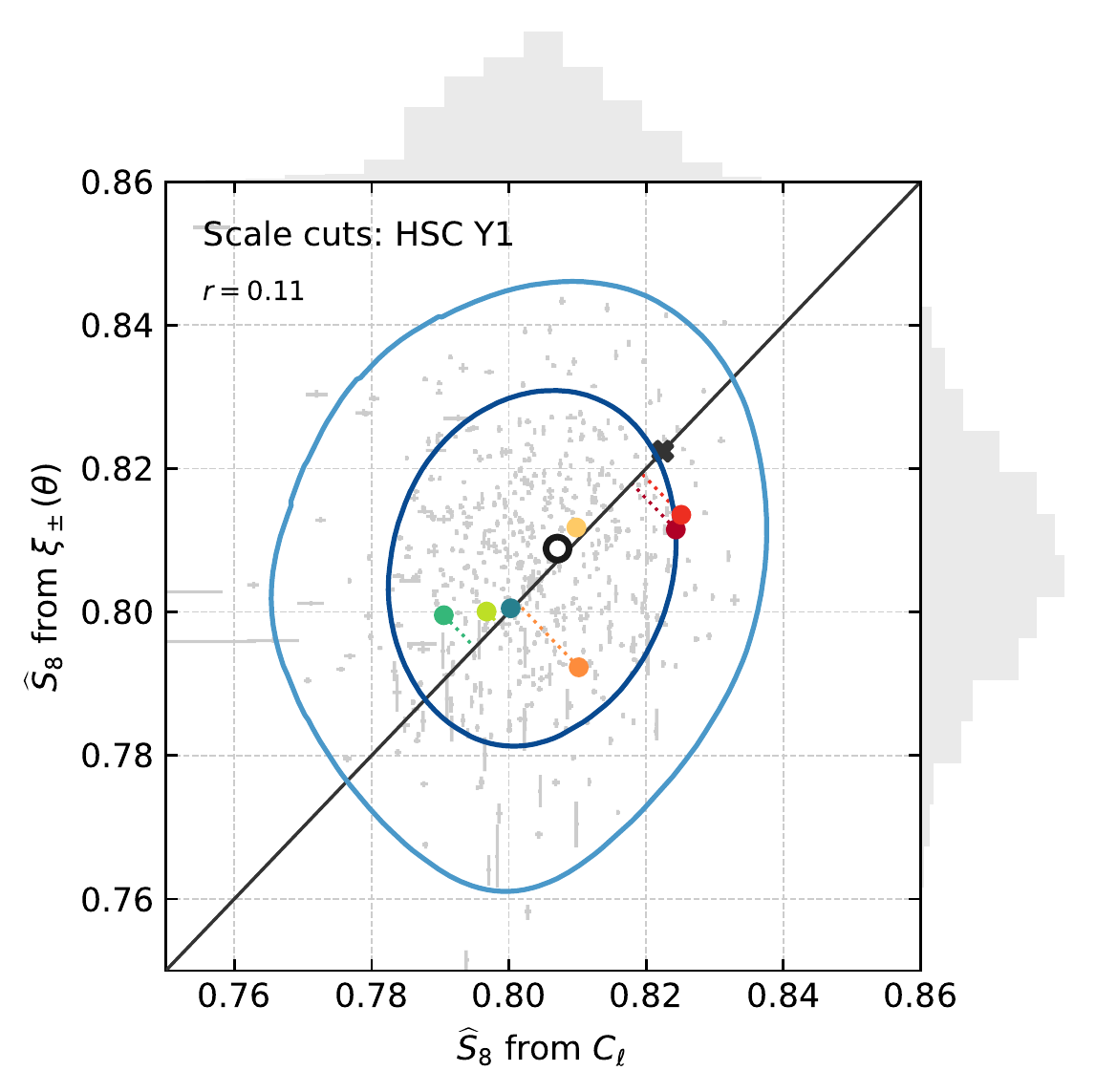}
    \\
    \includegraphics[scale=0.6]{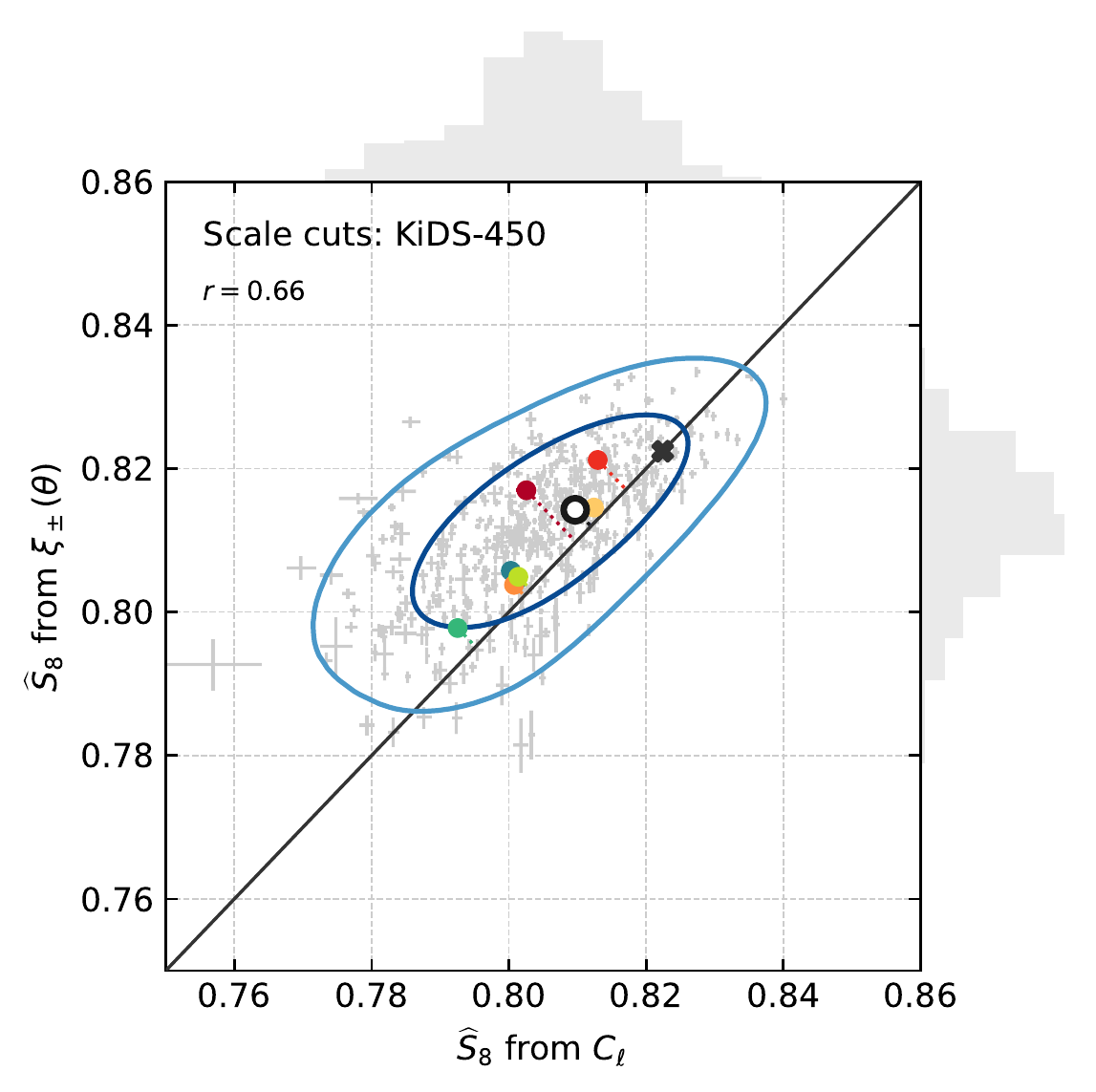}
    \\
    \includegraphics[scale=0.6]{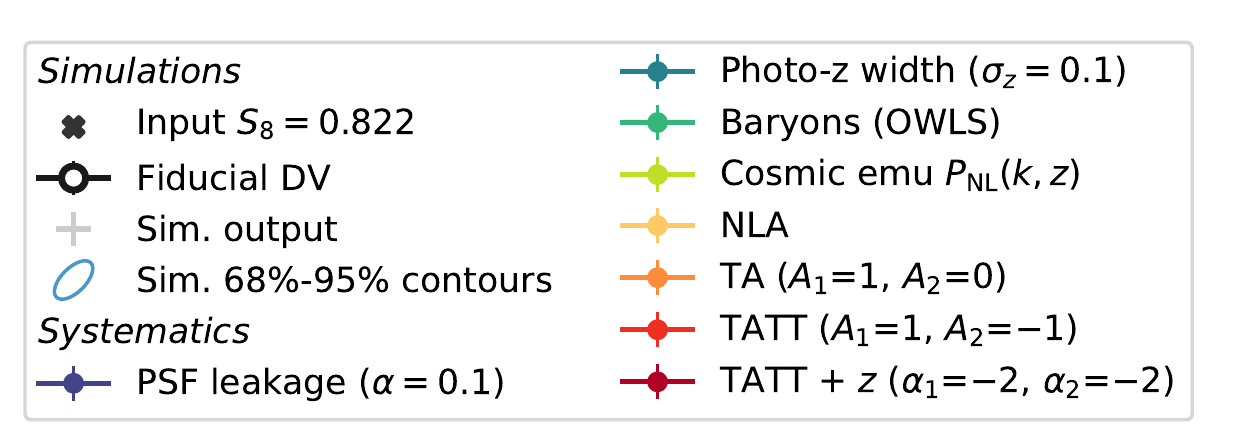}
    \caption{
    Distribution and biases of $\widehat{S_8}$ from real (vertical axis) \vs harmonic (horizontal axis) space for mock DES Y3 surveys analyzed with scale cuts that were used in the published cosmic shear analyses of HSC Y1 data \citep[top,][]{2020PASJ...72...16H,2019PASJ...71...43H} and KiDS-450 \citep[bottom,][]{2017MNRAS.465.1454H,2017MNRAS.471.4412K}. See \cref{fig:S8_r_vs_F} for a description of elements in the plot.}
    \label{fig:HSC_KiDS}
\end{figure}

The HSC \citep{2020PASJ...72...16H,2019PASJ...71...43H} and KiDS \citep{2017MNRAS.465.1454H,2017MNRAS.471.4412K} collaborations have published cosmic shear analyses in both harmonic and real spaces using, respectively, HSC Year 1 and KiDS-450 data, sharing shear catalogues and redshift distributions across analyses, and found discrepancies in their cosmological constraints of order $\numrange[range-phrase=-]{0.5}{1}\sigma$. We now repeat the exercise of analyzing simulations and contaminated data vectors using corresponding scale cuts found in those publications, that is, removing bins outside the corresponding $\tmin/\tmax$ and $\ellmin/\ellmax$ ranges while maintaining the binning we used throughout this work.
We caution the reader that here
we use the same simulations of a DES Y3-like survey as presented in previous sections, and only apply published scale cuts for comparison with cuts shown above. Since the simulations do not capture the depth and geometry of the different surveys, we cannot directly apply the conclusions here to the published KiDS and HSC results. We can, however, gain qualitative insights to how the chosen scale cuts might have resulted in the apparent large scatter between the constraints from real- and harmonic-space statistics.

For the analysis of HSC first-year data in real space \citep{2020PASJ...72...16H}, scale cuts are the same for all redshift bin pairs. The small-scale cut is chosen so that the difference in $\xi_\pm$ from baryons \citep[AGN model from][]{2015MNRAS.450.1212H} is smaller than 5\%. The large-scale cut is chosen from PSF contamination and impacts only $\xip$. They are
$\SI{7.08}{\arcminute} < \theta < \SI{56.2}{\arcminute}$ and $\SI{28.2}{\arcminute} < \theta < \SI{178}{\arcminute}$ for $\xi_+$ and $\xi_-$, respectively.
For the harmonic space analysis \citep{2019PASJ...71...43H}, the cuts are also the same for all redshift bin pairs. The large-scale cut $\ell_{\rm min}=300$ is determined by the detection of B-modes at lower multipoles from PSF leakage. The small-scale cut ${\ell_{\rm max}=1900}$ is chosen to avoid uncertainties from intrinsic alignments at scale smaller than \SIrange{1}{2}{\per\h\mega\parsec}, corresponding to $\ell \gtrsim 2000$. Both analyses share modeling choices, \eg for the non-linear matter power spectrum, intrinsic alignments and the impact of baryons. Although the fiducial model assumes no baryonic feedback, the authors test the analytic model of \cite{2015MNRAS.450.1212H} and find that constraints shift by less than $1\,\sigma$.

For the KiDS-450 analyses, scale cuts are also chosen to be identical across redshift bins. In real space \citep{2017MNRAS.465.1454H}, a large scale cut of \SI{72}{\arcmin} is imposed on $\xip$ because of additive shear biases at larger scales, while $\xim$ is used up to \SI{300}{\arcmin}, only limited by the extent of the KiDS-450 patches. Small scale cuts of \SI{0.5}{\arcmin} and \SI{4.2}{\arcmin} are applied to $\xip$ and $\xim$ due to uncertainties in the model and low signal-to-noise ratio. In harmonic space \citep{2017MNRAS.471.4412K}, the large scale cut is $\ellmin=76$, limited by the extent of KiDS-450 patches, and the small-scale cut of $\ellmax=1310$ is chosen to be in the regime where the quadratic power spectrum estimator is found to recover simulation inputs. We note that authors caution that the real-space analysis uses more non-linear scales information, therefore expecting differences.

In the four aforementioned HSC and KiDS-450 papers, the robustness of the results with respect to model choices were thoroughly explored, in particular the impact of baryonic feedback and intrinsic alignment, as well as to instrumental effects (some of which dictated scale cuts, as mentioned above). In particular, KiDS-450 analyses, which use small-scale measurements, include baryonic effects in their baseline model. Both collaborations also model intrinsic alignments with the NLA model in each analysis.

\Cref{fig:HSC_KiDS} shows the distribution of $\SeightCl$ and $\Seightxi$ and the impact of systematics when using HSC and KiDS-450 published scale cuts on a DES Y3-like survey. We present corresponding results for $\sigma_8$ and $\Om$ in \cref{fig:sigma8_r_vs_F,fig:Om_r_vs_F}, respectively.
For HSC scale cuts, we find that the correlation coefficient is of order 10\% for $S_8$, $\sigma_8$ and $\Om$, somewhat consistent with findings in Section 6.7 and Appendix 5 of \citet{2020PASJ...72...16H} obtained from running the full pipeline on one hundred $N$-body mock survey simulations ($S_8$ was found to show a correlation of 50\%, albeit on the posterior median rather than the mean, though the difference may also arise from survey configurations). Noticeably, all systematic effects we probed point in the direction where $\sigma_8$ as measured from harmonic space is higher than $\sigma_8$ from real space measurements (and lower $\Om$), which coincides with actual observations, although the amplitudes are found to be smaller here than actual discrepancies observed in data, meaning that no single systematic may explain this discrepancy. It is possible that all systematic effects combined could create a more significant bias between the two analyses, though this would require further investigation. However, we also note that baryonic feedback points to a higher $S_8$ from real space measurement---matching observations in directions, but not in amplitude---while intrinsic alignments point in the other direction. Therefore, while we observe significant systematic trends in $\sigma_8$ and $\Om$, we also find that distribution of $\Delta\widehat{S_8}$ is significantly broader here than for other cuts, as shown in the top right panel of \cref{fig:deltaS8vsPost}, and we refrain from attributing the observed difference in $S_8$ to either systematics or statistics.
For KiDS-450 scale cuts, we observe a significant trend for real space measurement to yield higher mean of $S_8$ than harmonic space measurements, the former being typically closer to the input value in our simulations. This is likely due to a combination of the different scale cuts and projection effects. We find a higher correlation coefficient of 66\% in this case, and note that all systematics have higher $S_8$ for real space measurements.
In particular, we find that the bias created by intrinsic alignments from tidal torquing, especially with redshift dependence (and, for HSC, the TA model),
lie close to the boundary of the 68\% region in the $(\SeightCl,\Seightxi)$ plane.
While these observations are not sufficient to make any conclusive claim regarding observed discrepancies (\ie distinguish a statistical fluke from systematic biases), they do shed some light on the interactions between the choice of scale cuts and systematic effects. Moreover, in the case of HSC, we note that a change in area increases error bars in the same proportion at all scales (through the $\fsky$ coefficient in the covariance matrix, see \cref{eq:cov_cl}) and we find that our results still hold when using our "low noise" simulations: biases remain practically unchanged while $(\SeightCl,\Seightxi)$ contours shrink, making those biases all the more concerning. A caveat is that to generate those simulations, we decrease the shape-noise $\sigma_e^2/\bar{n}$ by dividing the ellipticity standard deviation $\sigma_e$ by 2, while HSC would be better matched by increasing the galaxy density $\bar{n}$, though the two should be equivalent with respect to shape-noise alone.

\section{Conclusion}
\label{sec:summary}

In this work, we have investigated the impact of scale cuts and systematic effects on the cosmological constraints derived from the analysis of cosmic shear two-point statistics in harmonic \textit{vs} real space.
As a quantity projected along the line-of-sight, the observed angular two-point statistic at a given scale receives contribution from a broad range of three-dimensional, Fourier $k$-modes, where physics is naturally described. Moreover, these contributions do not align perfectly between the harmonic and real space statistics, which are related through a Bessel integral. Therefore, imposing a hard cut in one space means imposing soft, or oscillatory, cuts in other spaces, making it difficult to find unambiguous correspondence between various analyses. In addition, theoretical uncertainties and observational systematic effects may induce differential biases that need to be disentangled from statistical fluctuations.

Motivated by discrepancies found in the literature on the parameter $S_8$ by the HSC Y1 \citep{2020PASJ...72...16H,2019PASJ...71...43H} and KiDS-450 \citep{2017MNRAS.465.1454H,2017MNRAS.471.4412K} collaborations between their analyses of two-point statistics in harmonic and real space,
we explore the expected consistency of Year 3 cosmic shear data from the Dark Energy Survey and how similar discrepancies can arise.
We suggest several scale cuts and a method to test them, which we apply to the forthcoming analysis of DES Y3.
To do so, we generate 500 mocks of a DES Y3-like survey from Gaussian simulations, which we analyze using a fast importance sampling method with various scale cuts, in order to measure the discrepancies that can be expected from pure statistical fluctuations in parameter space versus that originating in systematic effects.

Our findings are:
\begin{itemize}
    \item We motivate two new methods to determine small-scale cuts from theory, both readily applicable to harmonic and real space two-point statistics; one is based on a three-dimensional $k$-mode cut-off, the other on a $\chi^2$ distance between alternative predictions for the data vectors, here applied to the baryonic feedback model (which is the most conservative cut we test). We also use DES Y1 cuts, converted with ${\ell\sim\pi/\theta}$.
    \item Given our DES Y3 setup, we find that $\sigma(\Delta\widehat{S_8})$, the scatter of the difference $\Delta\widehat{S_8}$ between posterior means from harmonic and real space analyses, is of order \numrange{0.08}{0.14}. Its value is a fraction $\sim\numrange{0.6}{0.9}$ of the scatter for individual statistics (it would be $\sqrt{2}\approx1.4$ for independent estimators). The correlation coefficient between $\SeightCl$ and $\Seightxi$ is highly sensitive to the choice of scale cut, decreasing with more conservative cuts. Among the scale cuts we try, the $\kmax$-based scale cuts yield the best consistency metrics. In particular they lead to symmetric scatter in the two statistics and to a high correlation coefficient (86\% for $\kmax=\SI{3}{\h\per\mega\parsec}$).
    \item We estimate the differential bias in $S_8$ due to a variety of systematics and modeling choices and we do not find, overall, one statistic to be intrinsically more biased than the other. Biases are generally small in our DES Y3-like setup and our proposed cuts, leading to shifts of less than a third of the statistical uncertainty. The partial exceptions are intrinsic alignment mechanism including tidal torquing (and redshift dependence) and baryonic feedback processes, especially for $\sigma_8$ and $\Om$. We conclude that our proposed cuts are immune to systematics tested here and are good candidates for the upcoming analysis for DES Y3, and easily adaptable to other surveys' characteristics.
    \item Our results indicate that with deeper surveys, and lower statistical errors, the biases due to systematics will be more significant---harmonic and real space statistics could then lead to different results in $S_8$ and other parameters. Extrapolating our results to LSST, with an effective number density of \SI{30}{gal \per arcmin\squared} and area of \SI{18000}{deg\squared} \citep[following][]{2013MNRAS.434.2121C}, error bars in the noise-dominated regime would reduce by $\sim\num{4.3}$, so one might expect $\sigma(\Delta\widehat{S_8})\sim{0.002}$.
\end{itemize}

Although the trends observed for various scale cuts match our expectations, the numerical value of the scatter $\sigma(\SeightCl-\Seightxi)$ is a complex function of scale cuts, survey characteristics and modeling, thus requiring simulations to quantify.
We tested the impact of a number of systematic effects and alternatives in the ingredients of our baseline model, including non-linear power spectrum, baryonic feedback, intrinsic alignments, PSF leakage and redshift distribution uncertainty
and found that the
largest discrepancies are due to baryonic feedback, intrinsic alignments sourced by tidal torquing (TATT with $A_2=-1$), particularly when redshift dependence is present but not modeled. This work will serve to guide the choice of small-scale cuts for the forthcoming analysis of DES Y3 cosmic shear data in harmonic space.

We also applied scale cuts used in the published HSC and KiDS-450 cosmic shear analyses and compared statistical and systematic differences in $S_8$, $\sigma_8$ and $\Om$. Although we used simulations with DES Y3-like characteristics, preventing us from drawing conclusions about observed spread, we do highlight some of our findings. We find a very low correlation of $S_8$ estimators when using HSC cuts, consistent with results of \citet{2020PASJ...72...16H}. Therefore the scatter of individual estimators $\widehat{S_8}$ is typically smaller than the scatter of the difference. For KiDS-450 cuts, we find asymmetric results, with the two-point functions measurements yielding higher $S_8$ than power spectra (as a matter of fact, closer to the truth because of a projection effect). It is finally worth noting that all systematic effects we tested point in the direction where ${\widehat{\sigma_8}|_{\Cl}>\widehat{\sigma_8}|_{\xipm}}$ (and ${\widehat{\Om}|_{\Cl}<\widehat{\Om}|_{\xipm}}$) for HSC, consistent with observations, although with modest but non-negligible amplitudes, typically of a fraction of the statistical spread. This indicates that no single systematic effect we test can create a significant differential bias, while a combination of effects could create differences of order $\sim1\sigma$.
Recent KiDS-1000 results compared three different statistics---two-point functions $\xipm$, band powers of $\Cl$, COSEBIs \citep[Complete Orthogonal Sets of E-/B-mode Integrals,][]{2010A&A...520A.116S}---and found them to be consistent \citep{2020arXiv200715633A}, although all of them rely on initial $\xipm$ measurements for thin bins in $\theta$ in the range \SIrange{0.5}{300}{\arcmin}, corresponding to band powers from $\ell=100$ to 1500. Scale cuts were unchanged for $\xipm$ and no cuts were applied to band powers or COSEBIs.

For future surveys, such as LSST, Euclid and Roman, those systematic shifts are expected to remain the same while statistical scatter will decrease with either higher depth or increased area, making control of those effects all the more important.
We also note that we restricted the analysis to a single choice of estimator for each statistic, namely a pseudo-$\Cl$ estimator of the shear power spectrum and the standard, unweighted estimator of the correlation functions $\xipm(\theta)$ \citep{2001PhR...340..291B}. For each statistic, different estimators exist \citep{1998PhRvD..57.2117B,2011MNRAS.412...65H} as well as alternative two-point statistics \citep[\eg the variance of the aperture mass statistic ][]{1996MNRAS.283..837S}, with various trade-off between computational difficulty and sensitivity to systematic effects. However, we have shown that scale cuts appear to be the key factor in terms of consistency between harmonic and real space, and we do not expect results would change in that regard. Nonetheless, some of these estimators were developed along with mitigation strategies to minimize biases from systematic effects \citep[including deprojection of systematic templates for pseudo-$\Cl$,][]{2017MNRAS.465.1847E,2019MNRAS.484.4127A,2020arXiv200714499W}, which would impact this part of our results, although this is beyond the scope of this paper.
One could also imagine combining both harmonic and real space statistics into a joint analysis, provided that one can model the joint likelihood (in particular the cross-covariance). In other words, if constraints derived from the two statistics independently are not fully correlated, a joint analysis could capture extra information with respect to independent analyses. That information likely lies within the particular $k$-modes that are captured by one statistics and  missed by the other, as discussed in the introduction.
An alternative is to use other statistics that exploit information from both spaces,
such as COSEBIs or $\Psi$- and $\Upsilon$-statistics \citep{2020arXiv200407811A},
at the cost of
increased complexity in estimation from data and modeling from theory.

\section*{Data availability}

A general description of DES data releases is available on the survey website at \url{https://www.darkenergysurvey.org/the-des-project/data-access/}. DES Y1 cosmological data is available on the DES Data Management website hosted by the National Center for Supercomputing Applications at \url{https://des.ncsa.illinois.edu/releases/y1a1}. This includes the redshift distributions used in this analysis. DES-Y3 data will be made available at \url{https://des.ncsa.illinois. edu/releases}. The \texttt{CosmoSIS} software \citep{2015A&C....12...45Z} is available at \url{https://bitbucket.org/joezuntz/cosmosis/wiki/Home}.

\section*{Acknowledgements}

This paper has gone through internal review by the DES collaboration.

The authors would like to thank Masahiro Takada and Scott Dodelson for useful discussions that motivated this work. The authors would also like to thank the anonymous referee for helpful comments.

Funding for the DES Projects has been provided by the U.S. Department of Energy, the U.S. National Science Foundation, the Ministry of Science and Education of Spain, the Science and Technology Facilities Council of the United Kingdom, the National Center for Supercomputing Applications at the University of Illinois at Urbana-Champaign, the Kavli Institute for Cosmological Physics at the University of Chicago, Financiadora de Estudos e Projetos, Fundacao Carlos Chagas Filho de Amparo a Pesquisa do Estado do Rio de Janeiro , Conselho Nacional de Desenvolvimento Cientifico e Tecnologico and the Ministerio da Ciencia e Tecnologia, and the Collaborating Institutions in the Dark Energy Survey.

The Collaborating Institutions are Argonne National Laboratories, the University of Cambridge, Centro de Investigaciones Energeticas, Medioambientales y Tecnologicas-Madrid, the University of Chicago, University College London, DES-Brazil, Fermilab, the University of Edinburgh, the University of Illinois at Urbana-Champaign, the Institut de Ciencies de l'Espai (IEEC/CSIC), the Institut de Fisica d'Altes Energies, the Lawrence Berkeley National Laboratory, the University of Michigan, the National Optical Astronomy Observatory, the Ohio State University, the University of Pennsylvania, the University of Portsmouth, and the University of Sussex.

The analysis made use of the \cosmosis software \citep{2015A&C....12...45Z} for the baseline model computations, the \numcosmo library\footnote{\url{https://numcosmo.github.io/}} \citep{2014ascl.soft08013D,2018MNRAS.480.5386D} to compute $\kmax$ cuts and \cref{fig:Cl_xi_vs_lnk}, \multinest \citep{2009MNRAS.398.1601F} for nested sampling chains, {\sc SMT} \citep{SMT2019} to generate optimized Latin Hypercube samples and Python packages including {\sc SciPy} \citep{Jones:2001}, {\sc NumPy} \citep{Oliphant:2006}, {\sc Matplotlib} \citep{Hunter:2007}, {\sc GetDist} \citep{Lewis:2019}, {\sc Numba} \citep{10.1145/2833157.2833162} and {\sc Seaborn} \citep{michael_waskom_2014_12710}.




\bibliographystyle{mnras}
\bibliography{papers_standard} 




\appendix

\section{Additional plots on $\sigma_8$ and $\Om$}
\label{sec:app_s8_Om}

\begin{figure*}
    \centering
    \includegraphics[scale=0.6]{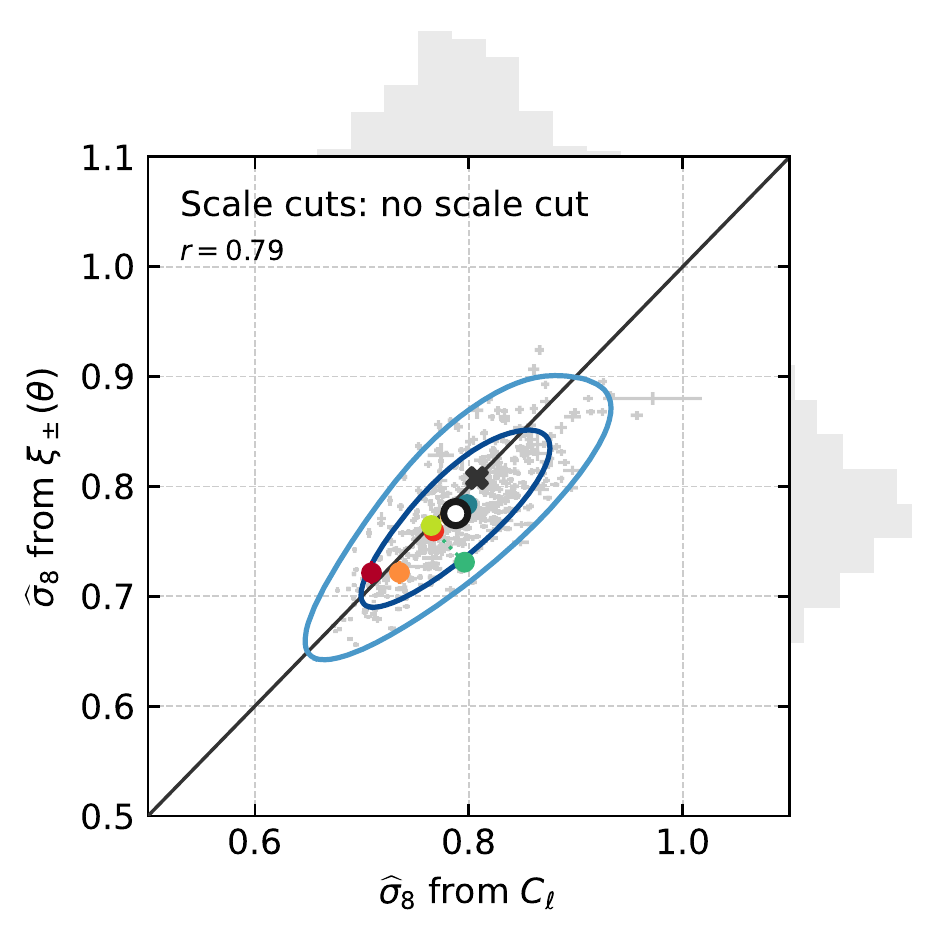}
    \includegraphics[scale=0.6]{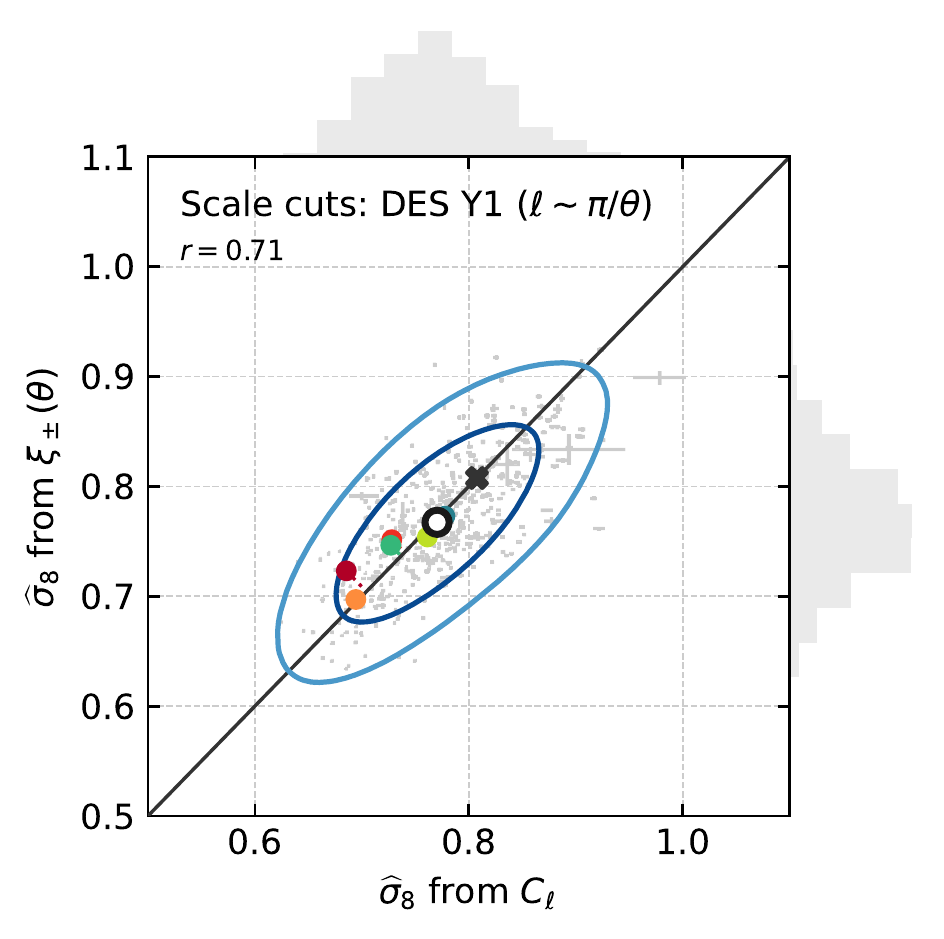}
    \includegraphics[scale=0.6]{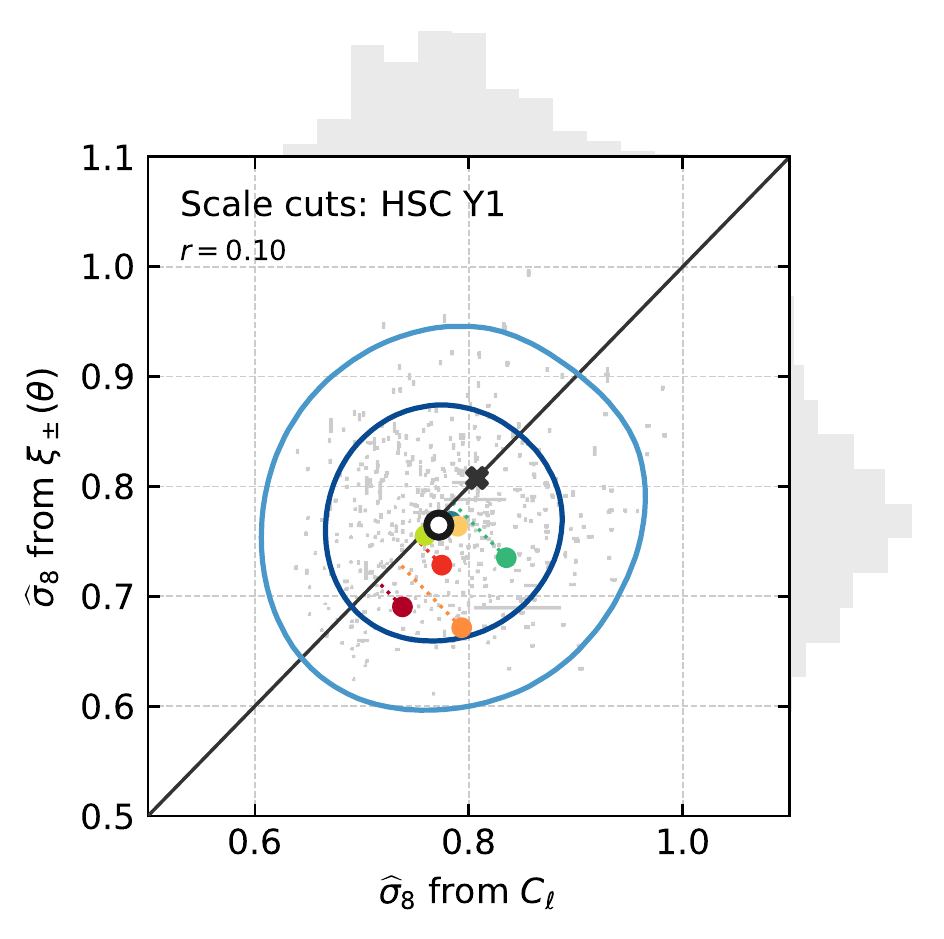}
    \\
    \includegraphics[scale=0.6]{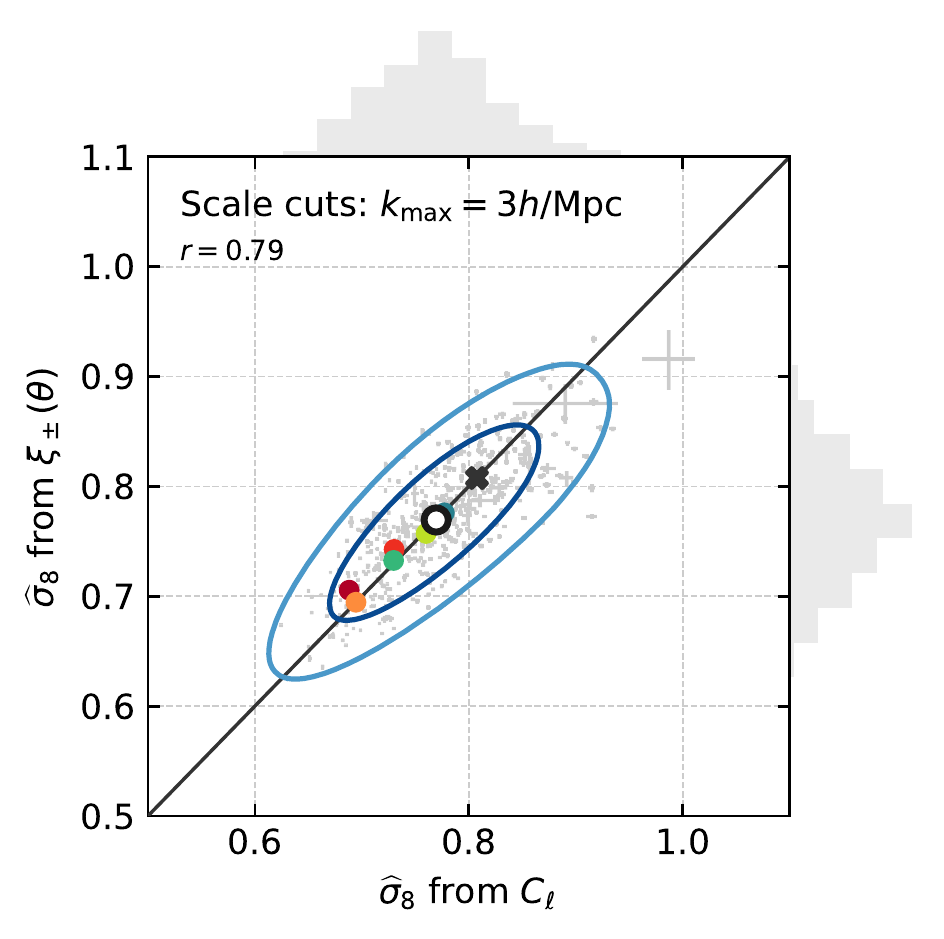}
    \includegraphics[scale=0.6]{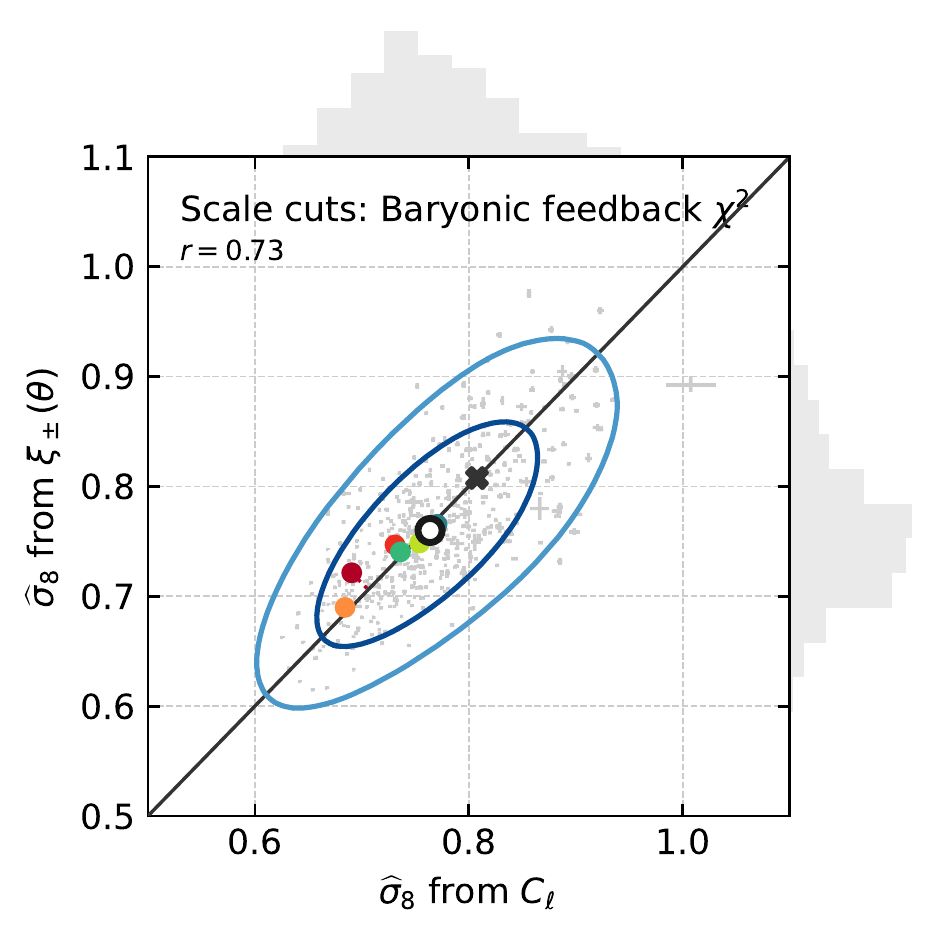}
    \includegraphics[scale=0.6]{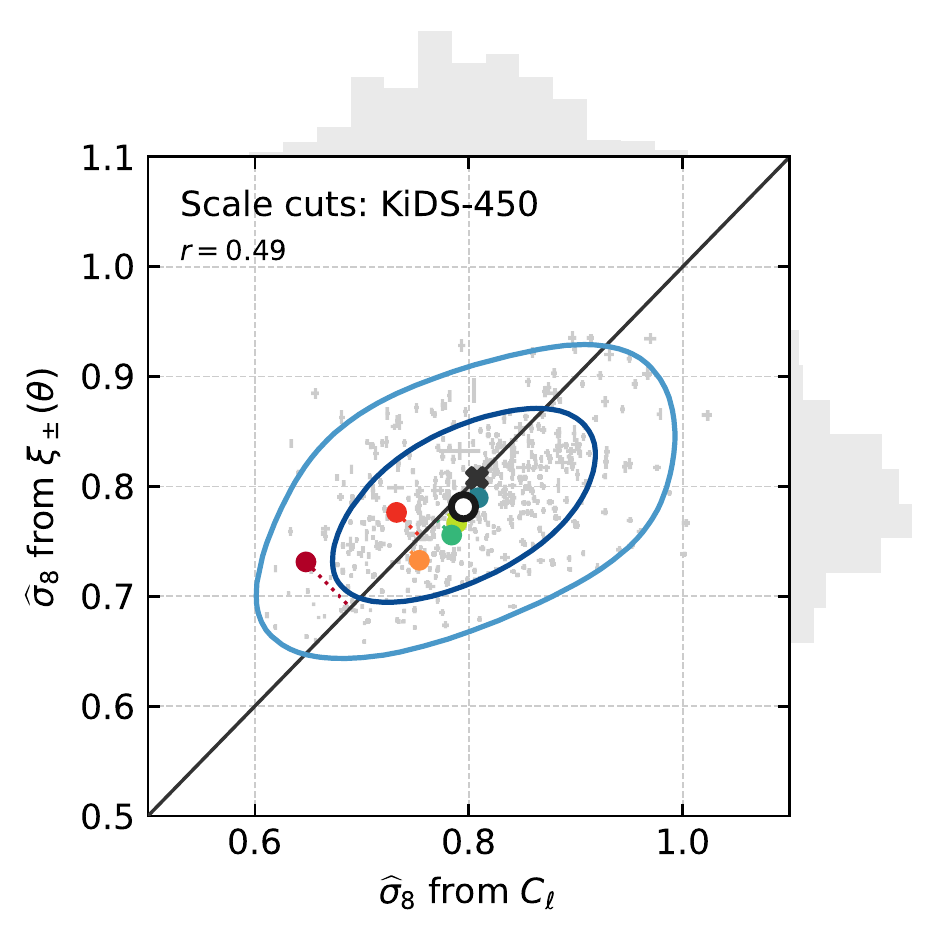}
    \\
    \includegraphics[scale=0.6]{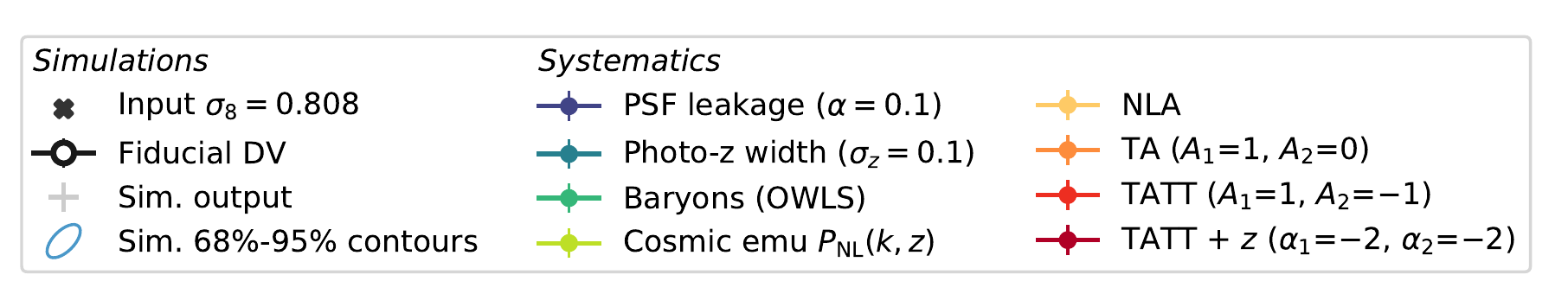}
    \caption{Same as \cref{fig:S8_r_vs_F,fig:HSC_KiDS} for $\sigma_8$.}
    \label{fig:sigma8_r_vs_F}
\end{figure*}

\begin{figure*}
    \centering
    \includegraphics[scale=0.6]{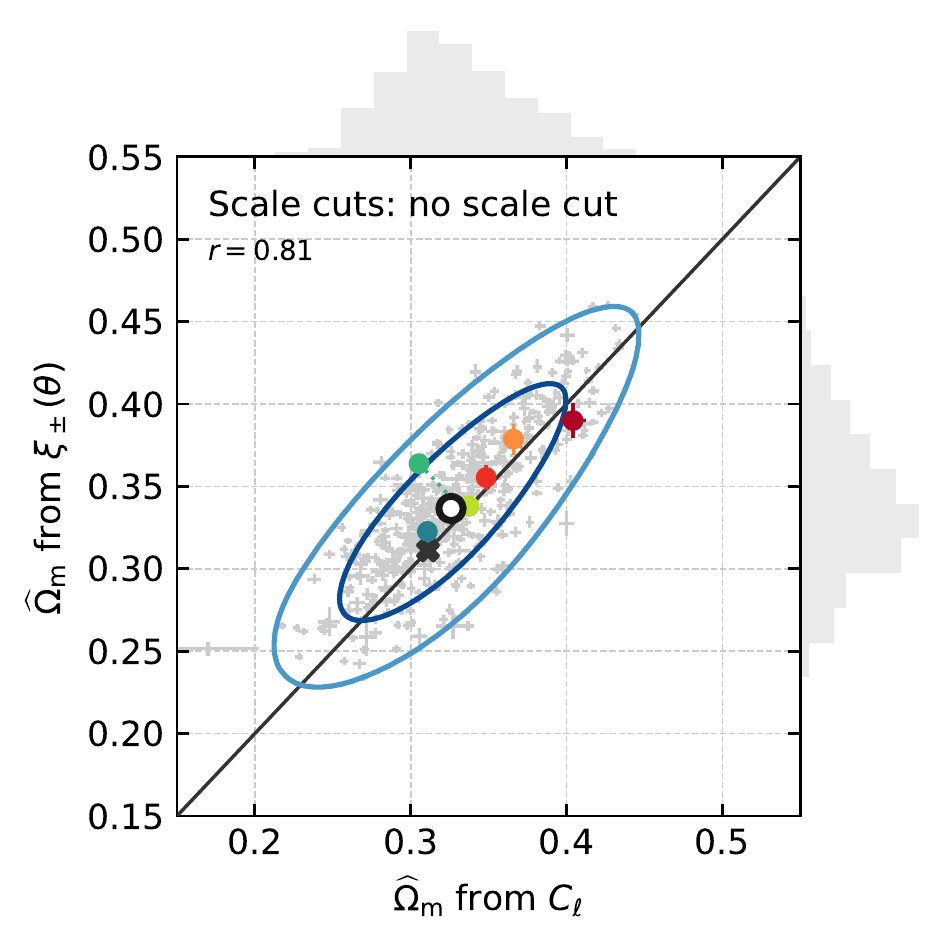}
    \includegraphics[scale=0.6]{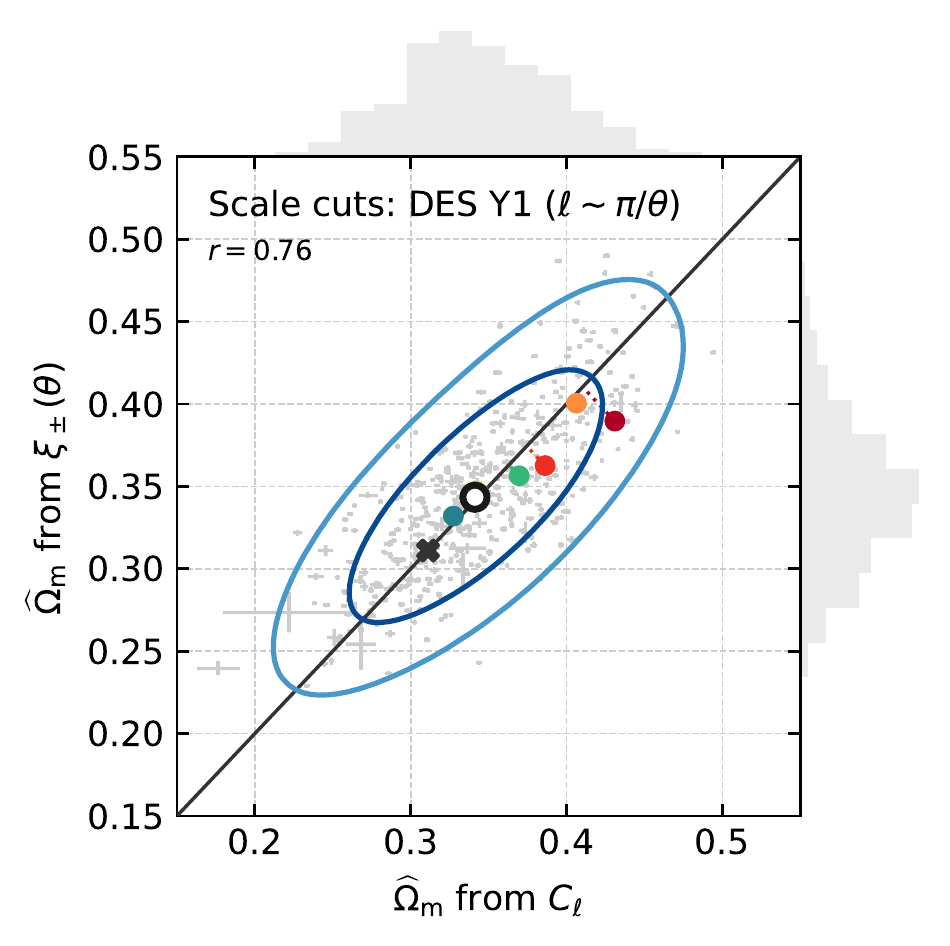}
    \includegraphics[scale=0.6]{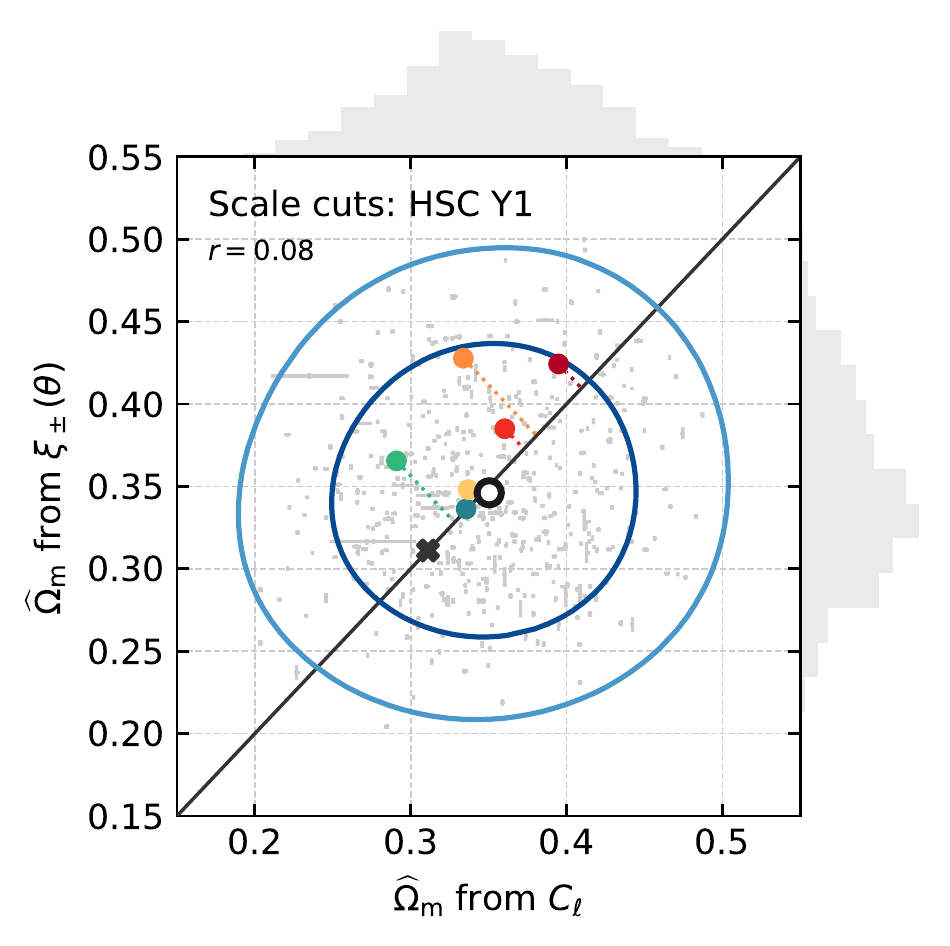}
    \\
    \includegraphics[scale=0.6]{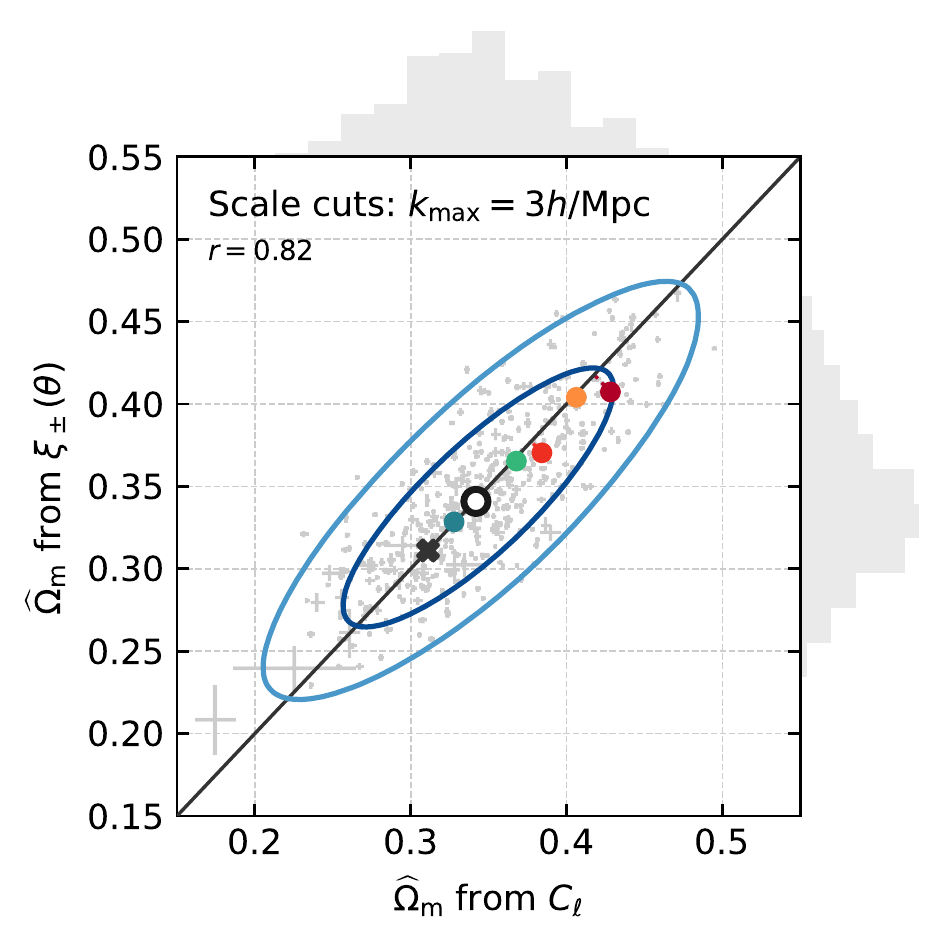}
    \includegraphics[scale=0.6]{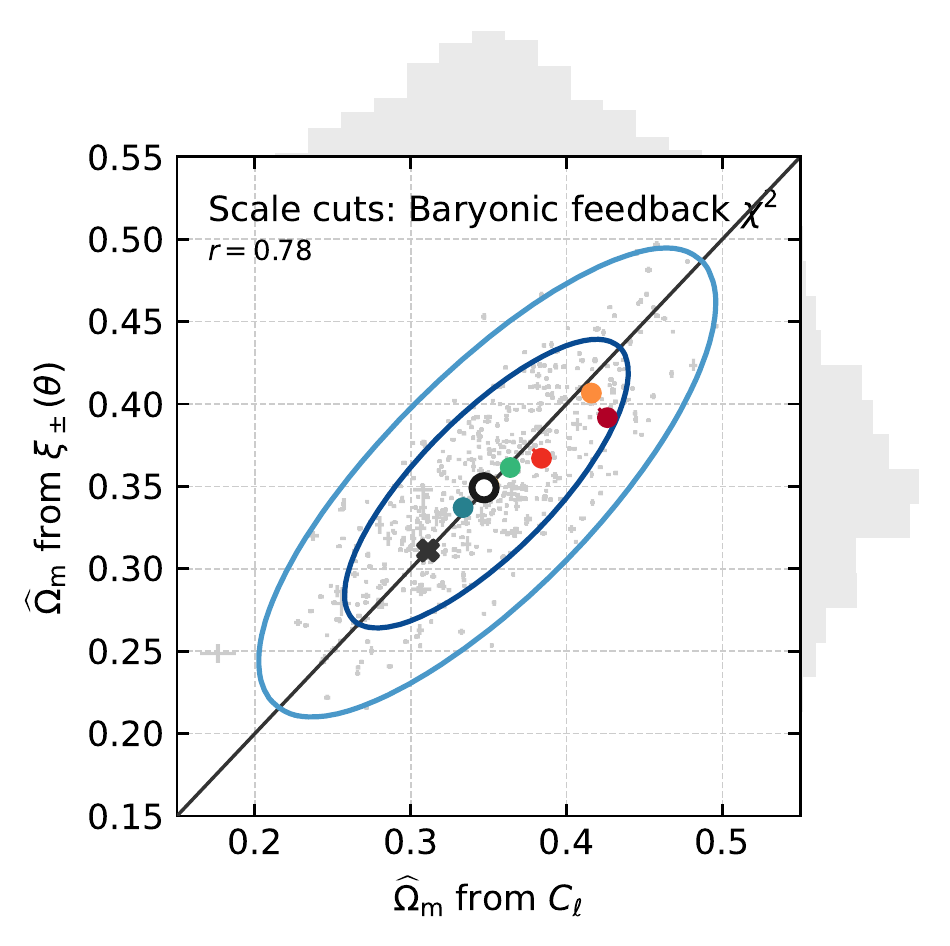}
    \includegraphics[scale=0.6]{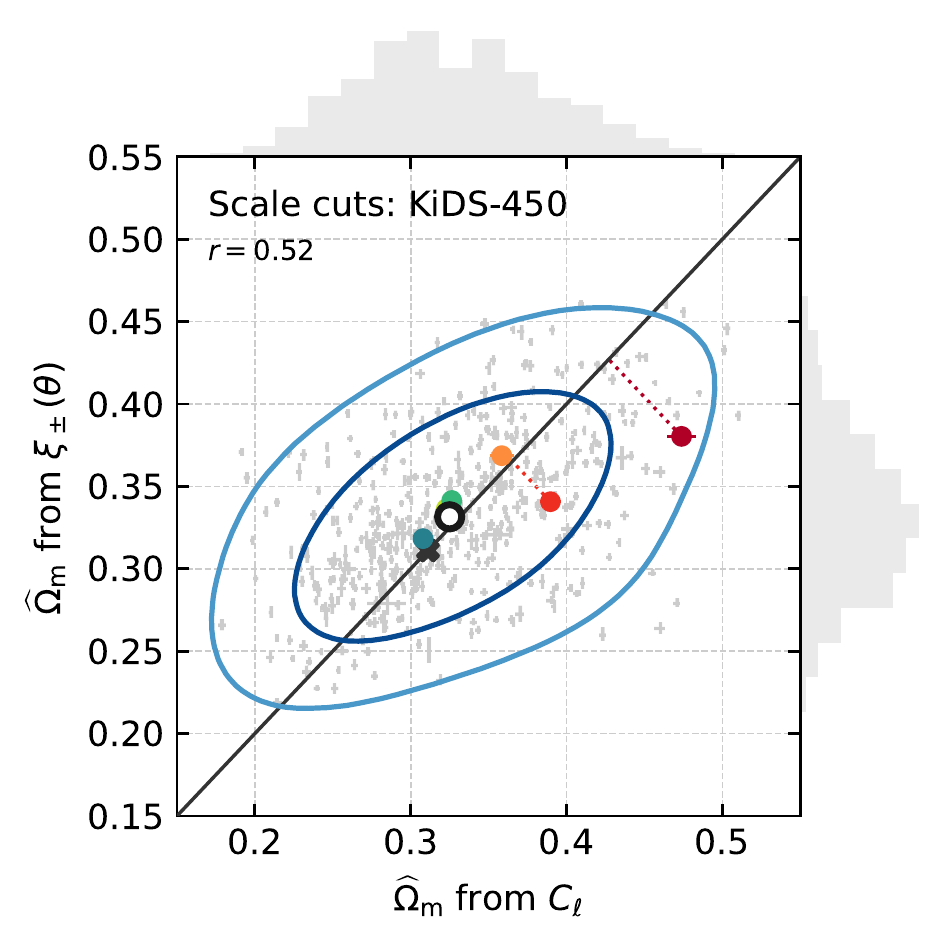}
    \\
    \includegraphics[scale=0.6]{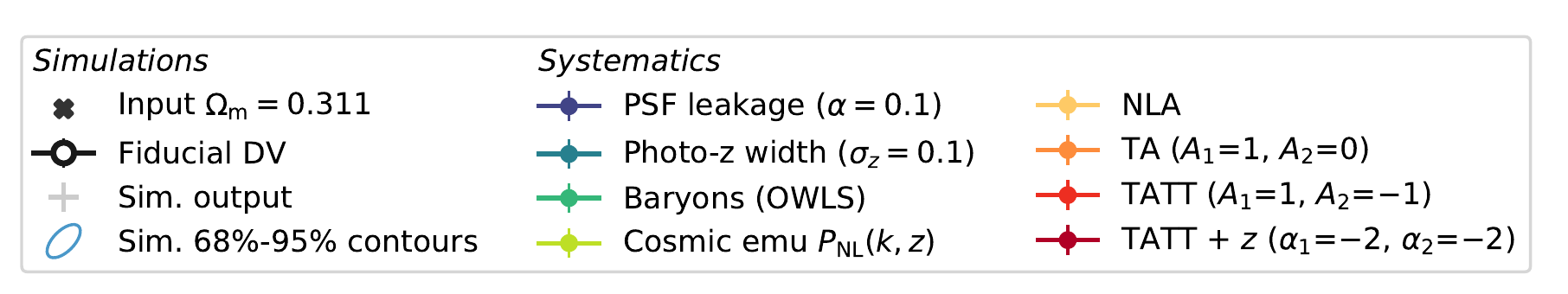}
    \caption{Same as \cref{fig:S8_r_vs_F,fig:HSC_KiDS} for $\Om$.}
    \label{fig:Om_r_vs_F}
\end{figure*}

\Cref{fig:sigma8_r_vs_F,fig:Om_r_vs_F} show the equivalents of \cref{fig:S8_r_vs_F,fig:HSC_KiDS} for $\sigma_8$ and $\Om$, respectively.

\section*{Affiliations}

$^{1}$ Department of Physics and Astronomy, University of Pennsylvania, Philadelphia, PA 19104, USA\\
$^{2}$ Department of Astronomy and Astrophysics, University of Chicago, Chicago, IL 60637, USA\\
$^{3}$ Kavli Institute for Cosmological Physics, University of Chicago, Chicago, IL 60637, USA\\
$^{4}$ Center for Cosmology and Astro-Particle Physics, The Ohio State University, Columbus, OH 43210, USA\\
$^{5}$ Institute of Physics, Laboratory of Astrophysics, \'Ecole Polytechnique F\'ed\'erale de Lausanne (EPFL), Observatoire de Sauverny, 1290 Versoix, Switzerland\\
$^{6}$ Instituto de F\'{i}sica Te\'orica, Universidade Estadual Paulista, S\~ao Paulo, Brazil\\
$^{7}$ Laborat\'orio Interinstitucional de e-Astronomia - LIneA, Rua Gal. Jos\'e Cristino 77, Rio de Janeiro, RJ - 20921-400, Brazil\\
$^{8}$ Department of Astronomy/Steward Observatory, University of Arizona, 933 North Cherry Avenue, Tucson, AZ 85721-0065, USA\\
$^{9}$ Institut de F\'{\i}sica d'Altes Energies (IFAE), The Barcelona Institute of Science and Technology, Campus UAB, 08193 Bellaterra (Barcelona) Spain\\
$^{10}$ Department of Applied Mathematics and Theoretical Physics, University of Cambridge, Cambridge CB3 0WA, UK\\
$^{11}$ Department of Physics, Carnegie Mellon University, Pittsburgh, Pennsylvania 15312, USA\\
$^{12}$ Department of Physics, Duke University Durham, NC 27708, USA\\
$^{13}$ Institute for Astronomy, University of Edinburgh, Edinburgh EH9 3HJ, UK\\
$^{14}$ Departamento de F\'isica Matem\'atica, Instituto de F\'isica, Universidade de S\~ao Paulo, CP 66318, S\~ao Paulo, SP, 05314-970, Brazil\\
$^{15}$ Fermi National Accelerator Laboratory, P. O. Box 500, Batavia, IL 60510, USA\\
$^{16}$ Kavli Institute for Particle Astrophysics \& Cosmology, P. O. Box 2450, Stanford University, Stanford, CA 94305, USA\\
$^{17}$ Instituto de Fisica Teorica UAM/CSIC, Universidad Autonoma de Madrid, 28049 Madrid, Spain\\
$^{18}$ Institute of Cosmology and Gravitation, University of Portsmouth, Portsmouth, PO1 3FX, UK\\
$^{19}$ CNRS, UMR 7095, Institut d'Astrophysique de Paris, F-75014, Paris, France\\
$^{20}$ Sorbonne Universit\'es, UPMC Univ Paris 06, UMR 7095, Institut d'Astrophysique de Paris, F-75014, Paris, France\\
$^{21}$ Department of Physics \& Astronomy, University College London, Gower Street, London, WC1E 6BT, UK\\
$^{22}$ SLAC National Accelerator Laboratory, Menlo Park, CA 94025, USA\\
$^{23}$ Instituto de Astrofisica de Canarias, E-38205 La Laguna, Tenerife, Spain\\
$^{24}$ Universidad de La Laguna, Dpto. Astrofísica, E-38206 La Laguna, Tenerife, Spain\\
$^{25}$ Center for Astrophysical Surveys, National Center for Supercomputing Applications, 1205 West Clark St., Urbana, IL 61801, USA\\
$^{26}$ Department of Astronomy, University of Illinois at Urbana-Champaign, 1002 W. Green Street, Urbana, IL 61801, USA\\
$^{27}$ Astronomy Unit, Department of Physics, University of Trieste, via Tiepolo 11, I-34131 Trieste, Italy\\
$^{28}$ INAF-Osservatorio Astronomico di Trieste, via G. B. Tiepolo 11, I-34143 Trieste, Italy\\
$^{29}$ Institute for Fundamental Physics of the Universe, Via Beirut 2, 34014 Trieste, Italy\\
$^{30}$ Institut d'Estudis Espacials de Catalunya (IEEC), 08034 Barcelona, Spain\\
$^{31}$ Institute of Space Sciences (ICE, CSIC),  Campus UAB, Carrer de Can Magrans, s/n,  08193 Barcelona, Spain\\
$^{32}$ Observat\'orio Nacional, Rua Gal. Jos\'e Cristino 77, Rio de Janeiro, RJ - 20921-400, Brazil\\
$^{33}$ Department of Physics, University of Michigan, Ann Arbor, MI 48109, USA\\
$^{34}$ School of Mathematics and Physics, University of Queensland,  Brisbane, QLD 4072, Australia\\
$^{35}$ Faculty of Physics, Ludwig-Maximilians-Universit\"at, Scheinerstr. 1, 81679 Munich, Germany\\
$^{36}$ Institute of Theoretical Astrophysics, University of Oslo. P.O. Box 1029 Blindern, NO-0315 Oslo, Norway\\
$^{37}$ Jet Propulsion Laboratory, California Institute of Technology, 4800 Oak Grove Dr., Pasadena, CA 91109, USA\\
$^{38}$ Department of Astronomy, University of Michigan, Ann Arbor, MI 48109, USA\\
$^{39}$ Department of Physics, Stanford University, 382 Via Pueblo Mall, Stanford, CA 94305, USA\\
$^{40}$ D\'{e}partement de Physique Th\'{e}orique and Center for Astroparticle Physics, Universit\'{e} de Gen\`{e}ve, 24 quai Ernest Ansermet, CH-1211 Geneva, Switzerland\\
$^{41}$ Santa Cruz Institute for Particle Physics, Santa Cruz, CA 95064, USA\\
$^{42}$ Center for Astrophysics $\vert$ Harvard \& Smithsonian, 60 Garden Street, Cambridge, MA 02138, USA\\
$^{43}$ Australian Astronomical Optics, Macquarie University, North Ryde, NSW 2113, Australia\\
$^{44}$ Lowell Observatory, 1400 Mars Hill Rd, Flagstaff, AZ 86001, USA\\
$^{45}$ George P. and Cynthia Woods Mitchell Institute for Fundamental Physics and Astronomy, and Department of Physics and Astronomy, Texas A\&M University, College Station, TX 77843,  USA\\
$^{46}$ Instituci\'o Catalana de Recerca i Estudis Avan\c{c}ats, E-08010 Barcelona, Spain\\
$^{47}$ Physics Department, 2320 Chamberlin Hall, University of Wisconsin-Madison, 1150 University Avenue Madison, WI  53706-1390\\
$^{48}$ Institute of Astronomy, University of Cambridge, Madingley Road, Cambridge CB3 0HA, UK\\
$^{49}$ Department of Astrophysical Sciences, Princeton University, Peyton Hall, Princeton, NJ 08544, USA\\
$^{50}$ Centro de Investigaciones Energ\'eticas, Medioambientales y Tecnol\'ogicas (CIEMAT), Madrid, Spain\\
$^{51}$ School of Physics and Astronomy, University of Southampton,  Southampton, SO17 1BJ, UK\\
$^{52}$ Computer Science and Mathematics Division, Oak Ridge National Laboratory, Oak Ridge, TN 37831\\
$^{53}$ Max Planck Institute for Extraterrestrial Physics, Giessenbachstrasse, 85748 Garching, Germany\\
$^{54}$ Universit\"ats-Sternwarte, Fakult\"at f\"ur Physik, Ludwig-Maximilians Universit\"at M\"unchen, Scheinerstr. 1, 81679 M\"unchen, Germany\\
$^{55}$ Department of Physics and Astronomy, Pevensey Building, University of Sussex, Brighton, BN1 9QH, UK\\



\bsp	
\label{lastpage}
\end{document}